\documentclass[journal,A4]{IEEEtran}

\usepackage{cite}
\usepackage{graphicx}
\usepackage{subfigure}
\usepackage{amssymb,amsmath} 
\usepackage{multirow}
\usepackage{booktabs}

\begin{document}

\title{Validation of cross sections for Monte Carlo simulation of the photoelectric effect}

\author{Min Cheol Han, Han Sung Kim, Maria Grazia Pia, Tullio Basaglia, Matej Bati\v{c},  Gabriela Hoff, \\
Chan Hyeong Kim,  and Paolo Saracco
\thanks{Manuscript received 14 September 2015.}
\thanks{This work has been partly funded by CAPES BEX6460/10-0 and PVE-71/2013 grants, Brazil.}
\thanks{T. Basaglia is with CERN, CH-1211, Geneva, Switzerland (e-mail: Tullio.Basaglia@cern.ch).}
\thanks{M. Bati\v{c} was with INFN Sezione di Genova, Genova, Italy 
             (e-mail: Batic.Matej@gmail.com); he is now with  
             Sinergise, 1000 Ljubljana, Slovenia.}
\thanks{M. C. Han, C. H. Kim and H. S. Kim are with are with 
	the Department of Nuclear Engineering, Hanyang University, 
        Seoul 133-791, Korea 
	(e-mail: mchan@hanyang.ac.kr, chkim@hanyang.ac.kr, hskim87@hanyang.ac.kr).}
\thanks{G. Hoff is with  
             CAPES, Brasilia, Brazil (e-mail:ghoff.gesic@gmail.com).}
\thanks{M. G. Pia and P. Saracco are  with INFN Sezione di Genova, Via Dodecaneso 33, I-16146 Genova, Italy 
	(phone: +39 010 3536328, fax: +39 010 313358, e-mail:
	MariaGrazia.Pia@ge.infn.it, Paolo.Saracco@ge.infn.it).}
}

\maketitle

\begin{abstract}
Several total and partial photoionization cross section calculations, based on
both theoretical and empirical approaches, are quantitatively evaluated with
statistical analyses using a large collection of experimental
data retrieved from the literature to identify the state of the art for modeling 
the photoelectric effect in Monte Carlo particle transport.
Some of the examined cross section models are available in general purpose Monte
Carlo systems, while others have been implemented and subjected to validation tests 
for the first time to estimate whether they could improve the 
accuracy of particle transport codes.
The validation process identifies Scofield's 1973
non-relativistic calculations, tabulated in the Evaluated Photon Data Library
(EPDL), as the one best reproducing experimental measurements of total cross
sections.
Specialized total cross section models, some of which derive from more recent
calculations, do not provide significant improvements.
Scofield's non-relativistic calculations are not surpassed regarding the
compatibility with experiment of K and L shell photoionization cross sections either,
although in a few test cases Ebel's parameterization produces more accurate
results close to absorption edges.
Modifications to Biggs and Lighthill's  parameterization implemented in Geant4 
significantly reduce the accuracy of total cross sections at low energies with 
respect to its original formulation.
The scarcity of suitable experimental data hinders a similar extensive analysis
for the simulation of the photoelectron angular distribution, which is limited
to a qualitative appraisal.

\end{abstract}


\section{Introduction}
\label{sec_intro}
\PARstart{P}{hotoionization} 
is important in various experimental domains, such as material analysis
applications, astrophysics, photon science and bio-medical physics.
As one of the interactions photons undergo in matter, it is relevant in
experimental methods concerned with the energy deposition resulting from photons
as primary or secondary particles.
Photoionization is also experimentally relevant for the secondary atomic processes that it
induces, X-ray fluorescence and Auger electron emission, which play a role
in many physics research areas and technological applications.
Extensive reviews, which cover both the theoretical and experimental aspects of
this process, can be found in the literature: 
\cite{pratt_1973, pratt_1973_err, samson_1976, kelly_1990, starace_1983, starace_2006,
amusia_1990, berkowitz_2002} are some notable examples among them.

This paper is concerned with modeling the physics of photoionization in a
pragmatic way: the simulation of this process in general purpose Monte
Carlo codes for particle transport.

Calculations for the simulation of the photoelectric effect are implemented in
these codes \cite{hubbell_2006}, nevertheless a comprehensive, quantitative
appraisal of their validity is not yet documented in the literature.
Assessments 
previously reported in the
literature concern comparisons of cross sections with NIST reference values,
such as \cite{demarco_2002, tns_nist}, or comparisons of cross section libraries
used in Monte Carlo codes \cite{zaidi_2000}, or involve complex observables
resulting from several physics processes in the full simulation of an
experimental set-up, such as \cite{chica_2009}.
Comparisons with experimental data of basic modeling features of photoionization
in Monte Carlo codes, such as those shown in \cite{egs5}, are usually limited to
qualitative visual appraisal and to a restricted sample of photon energies and
target materials.
It is worthwhile to note that the validation of simulation
models implies their comparison with experimental measurements
\cite{trucano_what}; comparisons with tabulations of theoretical calculations or
analytical parameterizations, such as those reported in
\cite{cirrone2010} as validation of Geant4 photon interaction cross
sections, do not constitute a validation of the simulation software.

The analysis documented  here evaluates the methods adopted in
widely known Monte Carlo systems  for the calculation of
photoelectric cross sections for the elements of the periodic table, as well as other
modeling approaches not yet implemented in these codes.
This investigation aims to assess the capabilities of Monte Carlo
codes for particle transport in this respect, and identify the state of the art for the
simulation of the photoelectric effect.

Special emphasis is devoted to the validation and possible improvement of
photoionization simulation in Geant4 \cite{g4nim,g4tns}; nevertheless, the
results documented in this paper provide information relevant to other Monte
Carlo systems as well.

The simulation of the atomic relaxation following the ionization of an atom has
been treated in previous publications \cite{tns_relax, tns_relax_nist,
tns_relax_prob, tns_binding}, therefore it is not addressed in this paper.


\section{Strategy of This Study}

This study concerns an extensive set of models for the simulation of
photoionization, which are representative of the variety of theoretical and
empirical methods documented in the literature.
The validation test concerns single ionisation of neutral atoms by 
non-polarized photons, as this is the context handled by general purpose 
Monte Carlo codes.

Computational performance imposes constraints on the complexity of physics
calculations to be performed in the course of simulation; hence the analysis in
this paper is limited to theoretical cross sections for which tabulations of
pre-calculated values are available and to empirical models that are
expressed by means of simple analytical formulations.
To be relevant for general purpose Monte Carlo systems, tabulated data should
cover the whole periodic table of elements and an extended energy range.

The evaluation mainly concerns total and partial photoelectric cross sections:
in particle transport, the former are relevant to determine the occurrence of
the photoionization process, while the latter determine which shell is ionized.
Calculated cross sections are quantitatively compared with a wide set of
experimental data collected from the literature.
The compatibility with experiment for each model, and the differences in
compatibility with experiment across the various models, are quantified by means
of statistical methods.

In addition, methods for the determination of the photoelectron angular
distribution are examined; nevertheless, due to the scarcity of pertinent
experimental data, their analysis is limited to qualitative considerations.


Computational algorithms pertaining to how basic physics modeling features are
used in the transport environment, such as methods for dealing with the
macroscopic cross sections for compounds or mixtures
\cite{egs_compound1,egs_compound2}, are outside the scope of this paper.





\section{Physics Overview}
\label{sec_physics}

Photoionization has been the object of theoretical and experimental interest for
several decades; only a brief overview of the physics relevant to the simulation
of the photoelectric effect in general purpose Monte Carlo codes is included
here to facilitate the understanding of the validation tests reported in 
this paper.

In the photoelectric effect a photon disappears and an electron is ejected from
an atom.
The energy of the photoelectron corresponds to the difference between the energy
of the absorbed photon and the energy binding the electron to the atom.

General purpose Monte Carlo codes for particle transport consider single photon
interactions with isolated atoms in their ground state; they neglect
interactions with ions and excited states, and multiple ionizations.
Photon interactions are treated regardless of the environment of the target
medium; this assumption neglects solid state effects and other features related
to the molecular structure of the medium.
The environment can have a significant effect on the cross sections near the
photoionization thresholds of both inner and outer shell electrons; due to the
limitations of their underlying physics assumptions, current general purpose
Monte Carlo codes are not usually exploited for the simulation of
experimental scenarios involving EXAFS (Extended X-ray Absorption Fine
Structure), XANES (X-ray Absorption Near Edge Structure) and other techniques
for which detailed accounting of material structure is required.

\subsection{Total and Partial Cross Sections}

The photoelectric cross section as a function of energy exhibits a
characteristic sawtooth behavior corresponding to absorption edges, as the
binding energy of each electron subshell is attained and corresponding
photoionization is allowed to occur.

Early theoretical calculations of photoionization cross sections were
limited to the K shell; they are represented by the papers of Pratt
\cite{pratt_1960}, providing the asymptotic behavior for arbitrarily high
energies, and Pratt et al. \cite{pratt_1964}, reporting calculations in the
energy range between 200~keV and 2~MeV.
More extensive calculations became available only at a later stage: 
Rakavy and Ron \cite{rakavy_1967} calculated cross sections for all subshells of
five elements over the energy range 1~keV to 2~MeV, and
Schmickley and Pratt \cite{schmickley_1967} reported cross sections for K to M
shells for three elements from 412 to 1332~keV.

Scofield's non-relativistic calculations \cite{scofield_1973} in a
Hartree-Slater framework represented a major advancement in the field, as they
covered systematically all subshells over the whole periodic table of the
elements.
More recent calculations were performed by Chantler
\cite{chantler_1995,chantler_2000} in a self-consistent relativistic
Dirac-Hartree-Fock framework.

These theoretical calculations provide the basis for the tabulated data libraries 
listed in section \ref{sec_compilations}.

Various empirical formulations of photoionization cross sections are reported in
the literature, e.g. in \cite{hubbell_1969}.
They derive from fits to experimental data, parameterizations of theoretical
calculations and semi-empirical methods involving both measured data and
theoretical considerations.

\subsection{Photoelectron Angular Distribution}

Fischer's non-relativistic theory \cite{fischer_1931} addresses the
calculation of differential cross sections in the low energy region.
The first relativistic treatment of the photoelectric effect was given by Sauter
\cite{sauter_1931, sauter_1931a}, who calculated the K-shell cross section in the Born
approximation; it concerns the lowest order in Z$\alpha/\beta$ (where
Z is the atomic number of the target, $\alpha$ is the fine structure constant and $\beta$ is $v/c$).
A comparison of these theories is discussed in \cite{davisson_evans_1952},
which showed that Sauter's theory applies even in the non-relativistic realm, 
despite being derived for relativistic electrons.

Gavrila \cite{gavrila_1959} and Nagel \cite{nagel_1963} extended Sauter's results
to the next order in Z$\alpha/\beta$.
Further calculations by Gavrila are available for the L shell \cite{gavrila_1961}.

Monte Carlo codes generate the photoelectron angular distribution based on these
differential cross section calculations; their respective approaches are
documented in section \ref{sec_mc}.



\section{Photoionization Cross Section Compilations}
\label{sec_compilations}

The photoelectric cross section compilations considered in this study are
listed in Table~\ref{tab_models},
which reports the label by which they are identified in the validation analysis,
the corresponding references and their coverage in energy, atomic number and atomic shells.
A brief overview of their features 
is summarized in the following subsections, where the compilations appear
in  chronological order of publication; more details can be found
the related references.

\begin{table}[htbp]
  \centering
  \caption{Compilations of photoionization cross sections considered in this study}
    \begin{tabular}{lrrrrc}
   \toprule
    \textbf{Identifier} & \multicolumn{2}{c}{\textbf{Energy range}} & \multicolumn{2}{c}{\textbf{Z range}} & \textbf{Shell} \\
    \midrule
    Biggs	\cite{biggs3}							&  10 eV 	& 100 GeV  	&  1	& 100 	& no \\
    BiggsG4	\cite{biggs3,grishin_1994}					&  10 eV 	& 100 GeV  	&  1	& 100 	& no \\
    Brennan\cite{brennan_cowan_1992}				&  30 eV   	& 700 keV 	&  3	& 92 		& no \\
    Chantler \cite{chantler_1995,chantler_2000}			&  10 eV 	& 433 keV 	&  1	& 92 		& K \\
    Ebel  	\cite{ebel_2003}							&  1 keV 	&  300 keV 	&  1	& 92 		& all \\
    Elam  	\cite{elam_2002}						&  100 eV 	& 1 MeV 		&  1	& 98 		& no \\
    EPDL97 	\cite{epdl97}							&  10 eV 	& 100 GeV  	&  1	& 100 	& all \\
    Henke 	\cite{henke_1993}						&  10 eV 	&  30 keV 		&  1	& 92 		& no \\
    McMaster 	\cite{mcmaster_1969, shaltout_2006} 		&  1 keV 	&  700 keV 	&  1	& 94 		& no \\
    PHOTX 	\cite{photx}							&  1 keV 	& 100 MeV 	&  1	& 100 	& no \\
    RTAB  	\cite{rtab}								&  10 eV 	&  300 keV 	&  1	& 99 		& all \\
    Scofield \cite{scofield_1973}						&  1 keV 	& 1.5 MeV 		&  1	& 100 	& all \\
    StormIsrael 	\cite{storm_israel_photon}				&  1 keV 	& 100 MeV 	&  1	& 100 	& no \\
    Veigele 	\cite{veigele_1973}						&  100 eV 	& 1 MeV 		&  1	& 94 		& no \\
    XCOM-DB  	\cite{xcom}						&  1 keV 	& 500 keV 	&  1	& 100 	& no \\
    XCOM	  	\cite{xcom}						&  1 keV 	& 100 GeV 	&  1	& 100 	& no \\

    \bottomrule
    \end{tabular}%
  \label{tab_models}%
\end{table}%
\subsection{McMaster's Tables}
\label{sec_mcmaster}

The compilation by McMaster et al. \cite{mcmaster_1969,mcmaster_news} reports
coherent and incoherent photon scattering cross sections, photoelectric and
total cross sections between 1~keV and 1~MeV.
The data concern elements with atomic numbers 1 to 99, with the exception of
atomic numbers 84, 85, 87, 88, 89, 91, and 93.
The photoelectric cross sections were obtained by a combination of methods: by
least-squares fitting of semi-empirical data, when they were available, by
theoretical calculations from theory in regions where the data were insufficient
or were considered unreliable, and by interpolation between semi-empirical and
theoretical values, where no data of either kind were available.
The semi-empirical photoelectric cross sections were derived by subtracting
theoretical scattering cross sections from experimental total photon attenuation
data.
The theoretical cross sections are based on Schmickley and Pratt's
\cite{schmickley_1967} calculations.

The original compilation  was updated in 2006 \cite{shaltout_2006}.

\subsection{Storm and Israel's Tables}
\label{sec_storm}

Storm and Israel's tables \cite{storm_israel_photon} encompass photon
interaction cross sections for atoms with atomic numbers between 1 and 100 and
photon energy from 1~keV to~100 MeV.
The photoelectric cross sections are derived from several theoretical
references: calculations by Brysk and Zerby \cite{brysk_1968}, Rakavy and Ron
\cite{rakavy_1967}, and Schmickley and Pratt \cite{schmickley_1967}.

\subsection{Veigele's Tables}
\label{sec_veigele}

The photoelectric cross sections reported in Veigele's compilation
\cite{veigele_1973} are of theoretical origin in the lower energy range (from
100~eV to energies varying between 1 and 10~keV, depending on the element) and
were calculated with semi-empirical methods at higher energies (up to 1~MeV).

The low energy cross sections derive from non-relativistic, self-consistent
field calculations, which are based on independent particle approximation.
At higher energies, theoretical scattering cross sections, calculated by
relativistic self-consistent field methods, were subtracted from experimental
total attenuation data; the resulting calculated photoelectric cross sections
and available photoelectric cross section measurements from 1 keV to 1 MeV were
then fitted by a least-squares procedure to produce recommended values.

Veigele's compilation concerns elements with atomic numbers from 1 to 94.


\subsection{Scofield's Calculations}
\label{sec_scofield}

Scofield's 1973 \cite{scofield_1973} compilation reports photoeffect cross
sections for all subshells, for all elements with atomic numbers from 1 to 101,
over the photon energy range between 1~keV and 1.5~MeV. 
The calculations were later extended down to 100~eV \cite{saloman_1988}.


These cross sections derive from non-relativistic calculations based on a
solution of the Dirac equation for the orbital electrons moving in a static
Hartree-Slater central potential.
In this approximation the electron-electron interaction term is replaced with an
average value, thus making the calculation less computationally intensive, but
in principle also less accurate, than the full Hartree-Fock model, which requires 
the calculation of the self-consistent field for each term.
In the same reference \cite{scofield_1973} Scofield reports renormalization
factors for atomic numbers 2 to 54 to convert the cross sections calculated in
the Dirac-Hartree-Slater approximation to values expected from a relativistic
Hartree-Fock model.

Comparisons with experimental photon mass attenuation coefficients
\cite{saloman_1986,saloman_1988} tend to favour Scofield's unrenormalized values
over the renormalized ones.
Nevertheless, it is worth noting that these evaluations of Scofield's
photoeffect cross sections do not compare the two sets of calculations with
direct photoelectric cross sections measurements: the reference data in these
comparisons involve the subtraction of theoretically calculated photon
scattering contributions from measured photon attenuation coefficients.
It is also unclear whether these evaluations rest on a qualitative appraisal only
or are based on objective statistical methods.


\subsection{Biggs and Lighthill's Parameterisation}
\label{sec_biggs}

Biggs and Lighthill \cite{biggs1,biggs2,biggs3} expressed the total
photoelectric cross sections as an empirical parameterization of data deriving
from various semi-empirical and theoretical sources, which include, among
others, Henke's 1982 compilation \cite{henke_1982} and the 1978 version of EPDL
\cite{epdl78}, the latter in turn based on Storm and Israel's, McMaster's,
Scofield's and Veigele's compilations.

For element $i$ and energy range $j$ the cross section is represented by the
formula:
\begin{equation}
\sigma_{ij} = \frac{A_{ij1}}{E} + \frac{A_{ij2}}{E^2} + \frac{A_{ij3}}{E^3} + \frac{A_{ij4}}{E^4} 
\label{eq_biggs}
\end{equation}
The tabulations in \cite{biggs3} list the most recent compilation of the
$A_{ij}$ coefficients of equation~\ref{eq_biggs} assembled by the original
authors of the parameterization.

Modified $A_{ij}$ coefficients for some gases
are documented in \cite{grishin_1994}.

\subsection{Henke's Compilation}
\label{sec_henke}

The compilation of photoabsorption cross sections 
by Henke et al. \cite{henke_1993} covers the energy range from
10~eV to 30~keV, for elements with atomic numbers up to 92.

Photoabsorption cross sections for energies below 10~keV are based on both
theoretical calculations and experimental data, interpolating across the atomic
number Z for elements where experimental data were scarce.
Above 10~keV photoabsorption cross sections derive from the semi-empirical
parameterizations by Biggs and Lighthill \cite{biggs3}.

An earlier compilation by Henke \cite{henke_1982}, concerning a narrower energy
range (from 30~eV to 10 keV), is based on similar criteria; cross sections from
1.5 to 10 keV are taken from \cite{biggs2}.

\subsection{PHOTX and XCOM}
\label{sec_xcom}

The PHOTX \cite{photx, photx_rsicc} data library was developed as a basis for 
the photon cross section file of ENDF/B-VI \cite{ENDFBVI}.
It provides cross sections for coherent and incoherent scattering, photoelectric
absorption, and pair production in the field of the nucleus and in the field of
the atomic electrons.
The data concern elements with atomic numbers between 1 and 100, and photon
energies from 1~keV to 100~MeV.

XCOM \cite{xcom} is a photon cross section database compiled by the NIST
(National Institute of Standards and Technology of the United States of America).
It concerns elements with atomic numbers up to 100 and photon energies from
1~keV to 100~GeV.
Two sets of tabulations, identified as XCOM and XCOM-DB, are evaluated in
this paper.
XCOM corresponds to the standard energy grid available from the XCOM web
site managed by NIST.
XCOM-DB is encompassed in the DABAX database, which is part of the X-ray Optics
Software Toolkit (XOP) \cite{xop_1997}; it is mainly addressed to the photon science
experimental community and is limited to photon energies up to 509.5 keV.

Regarding the photoelectric effect, both PHOTX and XCOM include Scofield's 1973
unrenormalized cross sections \cite{scofield_1973} up to 1.5~MeV.
At higher energies a semi-empirical formula from \cite{hubbell_1969} connects
Scofield's values at 1.5 MeV to the asymptotic high energy limit calculated by
Pratt \cite{pratt_1960}.

\subsection{Brennan and Cowan's Calculations}
\label{sec_brennan}

Brennan and Cowan developed a collection of software programs
\cite{brennan_cowan_1992} for the calculation of photoabsorption cross sections,
atomic scattering factors and other quantities relevant to photon science.
The photoelectric cross sections are based on Cromer and Liberman's
\cite{cromer_liberman_1970} theory.

Tabulations of photoelectric cross sections derived from Brennan and Cowan's
calculations are included in the XOP software system
\cite{xop_1997}.

\subsection{EPDL} 
\label{sec_epdl}

The EPDL (Evaluated Photon Data Library) includes photon interaction data
concerning photoionization, photoexcitation, coherent and incoherent scattering,
and pair and triplet production.
The latest version at the time of writing this paper was released in 1997, and
is commonly identified as EPDL97 \cite{epdl97}.
It is part of the ENDF/B-VII.1 \cite{ENDFBVII} evaluated nuclear data file.

EPDL97 includes total and partial cross sections for elements with atomic numbers
between 1 and 100, and for photon energies from 1~eV to 100~GeV. 
Partial cross sections are tabulated for all subshells.

From the edge to 1~MeV, subshell ionization cross sections are based on
Scofield's data as in \cite{saloman_1988}; total photoionization cross sections
are summed over all subshells.
From 1~MeV to 100~GeV the total cross sections are based on Hubbell's data
reported in \cite{hubbell_1980}.
Scofield's subshell cross sections have been extended up to 100 GeV by ensuring
that the sum of the subshell cross sections is equal to HubbeIl's total, and
maintaining the same ratio between subshell cross sections over the entire
energy range from 1 MeV to 100 GeV.
At 1~MeV the total photoionization cross section is identical from both sources,
therefore the two sets of calculations could be combined in a consistent manner.

EPDL97 documentation reports rough estimates and qualitative comments about the
accuracy of the tabulated data, but it does not document how these estimates
were produced.
To the best of our knowledge systematic, quantitative validation of EPDL97
photoionization data is not documented in the literature.

EPDL97 is extensively used in Monte Carlo simulation; details are given in section 
\ref{sec_mc}.

\subsection{Chantler's Calculations}
\label{sec_chantler}

Chantler calculated photoelectric cross sections in a self-consistent relativistic
Dirac-Hartree-Fock framework \cite{chantler_1995,chantler_2000}.
The exchange potential of Chantler's approach follows that of Cromer and
Liberman \cite{cromer_liberman_1970, cromer_liberman_1981} and Brennan and Cowan
\cite{brennan_cowan_1992}, and is different from the approach used by Scofield
\cite{scofield_1973}.

The calculations are carried out in isolated atom and independent particle
approximations: each atom is treated as a standalone system, not influenced by
other atoms or particles, and each electron is considered to move in an
effective potential of the nucleus with the average repulsive force of the
electrons.
This effective screening neglects some correlation; it also neglects the fact
that the potential for one electron is not identical to that of a different
electron.

Chantler's published tabulations report total and K-shell cross sections for
elements with atomic number from 1 to 92; they cover an energy range comprised
between 1-10~eV and 0.4-1~MeV (the lower and upper bounds vary with the atomic
number).


%
%

\subsection{RTAB}
\label{sec_rtab}

The RTAB \cite{rtab} database encompasses three sets of photoionization cross
sections along with an extensive set of tabulations concerning photon elastic
scattering: a set of original calculations (identified in the following as RTAB
cross sections), an extension of these cross sections based on EPDL97
(identified as RTABX) and Scofield's 1973 cross sections.
They are listed for atomic numbers from 1 to 100.

All the data in the RTAB database (apart from those reported from other sources)
have been consistently computed in the same Dirac-Slater potential.
Scofield's 1973 \cite{scofield_1973} cross sections were also computed in a
Dirac-Slater potential, although differently from those included in the RTAB
database.
According to \cite{rtab}, subshell cross sections are calculated in the Dirac-Slater potential to obtain
total atom cross sections from threshold up to several hundred keV.
For higher energies, these values are smoothly joined to the total-atom values
in the EPDL97, thus extending the cross sections in the RTABX collection up to
100 GeV.
RTAB photoelectric cross sections are tabulated for energies up to 300 keV.

The RTAB photoelectric database has not been exploited yet in general purpose
Monte Carlo systems.

\subsection{Elam's Database}

Elam's database \cite{elam_2002} provides photon cross section data for elements
with atomic numbers between 1 and 98, and for photon energy from 100~eV to 1~MeV.

Above 1~keV the photoabsorption cross sections derive from the XCOM ones of
\cite{xcom_1987}, which were parameterized with a cubic spline algorithm.
Below 1~keV they are based on the 1981 version of EPDL \cite{epdl81}.
Appropriate algorithms were applied to the data to ensure a smooth connection
between the two sets of cross sections.
Edge discontinuities were adjusted to be consistent with Williams' atomic
binding energies \cite{williams}.

\subsection{Ebel's Parameterizations}
\label{sec_elam}

Ebel et al. \cite{ebel_2003} developed parameterizations of total photoelectric
absorption coefficients and of subshell absorption coefficients in the energy
range from 1~keV to 300~keV, for elements with atomic number up to 92.
They are based on fitting fifth order polynomials in the logarithm of the photon
energy to Scofield's 1973 \cite{scofield_1973} cross section data.
The coefficients for the parameterization are tabulated for each energy interval 
identified by the absorption edges of a given element.
Photoelectric cross sections can be calculated using the tabulations.

This compilation was developed especially for application in fundamental
parameter programs for quantitative X-ray analysis.


\section{Photoionization in Monte Carlo Codes}
\label{sec_mc}

General purpose Monte Carlo codes for particle transport include algorithms for
the simulation of the photoelectric effect.

The original version of EGS4 \cite{egs4} calculated photoelectric total cross
sections based on Storm and Israel's tables \cite{storm_israel_photon} and
generated the photoelectron with the same direction as the incident photon.
Later evolutions introduced the use of PHOTX cross sections
\cite{sakamoto} and the generation of the photoelectron angular distribution
\cite{egs4_angular} based on Sauter's theory \cite{sauter_1931}.
These features are currently implemented in EGS5 \cite{egs5}.
EGSnrc \cite{egsnrc} provides the option of calculating total photoelectric
cross sections based on Storm and Israel's tables as originally in EGS4 or on a
fit to XCOM \cite{xcom} cross sections, while it uses subshell cross sections
based on EPDL \cite{epdl97}.
It samples the photoelectron angular distribution according to the method
described in \cite{egs4_angular} based on Sauter's theory.

ETRAN \cite{etran} uses Scofield's 1973 \cite{scofield_1973} cross sections for
energies from 1 keV to 1.5 MeV and extends them to higher energies by exploiting
Hubbell's method \cite{hubbell_1969} to connect the values at 1.5 MeV to the
asymptotic high energy limit calculated by Pratt \cite{pratt_1960}.
It samples the direction of the photoelectron from Fischer's
\cite{fischer_1931} distribution for electron energies below 50~keV and from
Sauter's \cite{sauter_1931} distribution for higher energies.

FLUKA \cite{fluka1,fluka2} calculates photoelectric cross sections based on
EPDL97 and samples the photoelectron direction according to Sauter's theory
\cite{sauter_1931}.

ITS \cite{its5} calculates photoelectric cross sections based on 
Scofield's 1973 non-renormalized values.
The angle of the photoelectron with respect to the parent photon is described by
Fischer's distribution \cite{fischer_1931} at lower energies and by Sauter's
\cite{sauter_1931} formula at higher energies.

MCNP5 \cite{mcnp5} and MCNPX \cite{mcnpx27e} provided different options of data
libraries for the calculation of photoelectric cross sections: two version of
EPDL (EPDL97 \cite{epdl97} and EPDL89 \cite{epdl89}), and ENDF/B-IV
\cite{ENDFB-IV} data complemented by Storm and Israel's tables
\cite{storm_israel_photon} for atomic numbers greater than 83.
MCNP6 \cite{mcnp6} has extended the minimum energy cut-off for photon transport
down to 1~eV; the necessary photo-atomic cross sections derive from ENDF/B-VI
version 8, which in turn is based on EPDL97 regarding photon interactions.

In the first version of Penelope including photon transport \cite{sempau_1997}
photoelectric cross sections were interpolated from XCOM; in more recent
versions \cite{penelope2008,penelope2011} they are interpolated
from EPDL97 tabulations.
The photoelectron angular distribution is sampled from Sauter's differential
cross section for the K shell \cite{sauter_1931}.

GEANT 3 \cite{geant3} calculated total photoionization cross sections based on
Biggs and Lighthill's \cite{biggs3} parameterizations; the probability of
ionization of the K shell and L subshells was estimated by parameterizations of
the jump ratios deriving from Veigele's \cite{veigele_1973} tables.
The angular distribution of the photoelectron was sampled for the K shell and
for the L$_1$, L$_2$ and L$_3$ subshells based on Sauter's
\cite{sauter_1931,sauter_1931a} and Gavrila's \cite{gavrila_1959,gavrila_1961}
calculations.

The Geant4 toolkit encompasses various implementations of the photoelectric
effect.
The overview summarized here concerns the latest version at the time of writing
this paper: Geant4 10.1, complemented by two correction patches.

The model implemented in Geant4 \textit{standard} electromagnetic package calculates
cross sections based on the analytical formula of Biggs and Lighthill,  but it
uses modified coefficients deriving from a fit to experimental data.
The related reference cited in Geant4 Physics Reference Manual \cite{g4physmanual}
as the source of these modifications does not appear to be consistent, presumably
due to a mismatch between the Russian and English versions of the 
periodical where it was published.
The modifications appear to derive from \cite{grishin_1994}, which reports
fits to experimental data concerning noble gases, hydrogen, carbon, fluorine,
oxygen and silicon; they concern cross sections at low energies.
The atomic cross section calculation according to Biggs-Lighthill
parameterization is implemented in the \textit{G4SandiaTable} and
\textit{G4StaticSandiaData} classes of Geant4 \textit{materials} package.
The same calculation of photoelectric cross sections is also used by the PAI
(PhotoAbsorption-Ionisation) model \cite{g4pai}.
The energy of the emitted photoelectron is determined as the difference between
the energy of the interacting photon and the binding energy of the ionized shell
defined in the \emph{G4AtomicShells} class \cite{tns_binding}, and the
photoelectron angle is calculated according to the Sauter-Gavrila distribution
for K shell \cite{sauter_1931,gavrila_1959}.

Geant4 \textit{low energy} electromagnetic package \cite{lowe_chep,lowe_nss}
encompasses two implementations of the photoelectric effect, one identified as 
``Livermore'' \cite{lowe_e} and one reengineered from the 2008 version of
the Penelope code \cite{penelope2008}; both models calculate total and
partial cross sections based on EPDL97.
The so-called ``Livermore'' model provides three options of computing the
angular distribution of the emitted photoelectron: in the same direction as the
incident photon, based on Gavrila's distribution of the polar angle
\cite{gavrila_1959} for the K shell and the L$_1$ subshell, and
based on a double differential cross section derived from Gavrila's
\cite{gavrila_1959,gavrila_1961} calculations, which can also handle polarized
photons.

In addition, the Geant4 toolkit encompasses two models for the simulation of the
photoelectric effect concerning polarized photons: one for circularly polarised
photons in the \textit{polarisation} package and one in the \textit{low energy}
electromagnetic package, identified as ``Livermore polarized''.
Their evaluation is outside the scope of this paper.


\section{The Validation Process}

The validation process adopts the same methodology used in \cite{tns_rayleigh}
for the validation of photon elastic scattering.
The main concepts and most relevant features of the validation method are
summarized below to facilitate understanding of the results reported in
sections~\ref{sec_resulttot} and \ref{sec_resultshell}.


\subsection{Simulation Models}
\label{sec_sw}

The validation process concerns the methods for the calculation of total and
subshell cross sections summarized in Table \ref{tab_models}.
Some of these simulation models represent novel approaches with respect to those
so far available in Geant4 and in other general purpose Monte Carlo codes.


All the physics models subject to evaluation have been implemented in a
consistent software design, compatible with the Geant4 toolkit, which minimizes
external dependencies to ensure unbiased appraisal of their intrinsic behaviour.
The software adopts a policy-based class design \cite{alexandrescu}, which 
supports the provision of a wide variety of physics models without imposing the 
burden of inheritance from a predefined interface.

A single policy class calculates cross sections 
that exploit tabulations; alternative tabulations, corresponding to different 
physics models, are managed through the file system.
Dedicated policy classes implement cross section calculations
based on analytical formulae.
The same scheme is adopted for total and partial cross section calculation;
the latter is involved in the algorithm that creates a vacancy, which 
drives the subsequent atomic relaxation process.
Alternative modeling of photoelectron
angular distributions is handled through a Strategy pattern \cite{gang4}. 

A photoionization process, derived from the \textit{G4VDiscreteProcess} class of
Geant4 kernel, which in turn is derived from \textit{G4VProcess}, acts as a host
class for the policy classes; they can be interchanged \cite{tns_dna,
em_nss2009, em_chep2009} to determine its behaviour.
The simulation of photoionization according to this software
design is consistent with Geant4 kernel, since Geant4 tracking handles all
processes polymorphically through the \textit{G4VProcess} base class.

Since policy classes are characterized by a single responsibility and have
minimal dependencies on other parts of the software, the adopted programming
paradigm enables independent modeling and test
of all physics options.
Their validation can be performed through simple unit tests.
This strategy ensures greater modeling flexibility and testing agility than the one
adopted in the Geant4 electromagnetic package, where total cross
section, vacancy creation and final state generation are bundled into one object:
that software design choice requires full-scale simulation applications to test
basic physics entities, due to extensive dependencies imposed by the
\textit{G4VEmModel} base class, from which all photoelectric models derive.

Existing physics models in Geant4 have been refactored \cite{refactoring}
consistently with this software design; other models not yet available in Geant4
have been implemented for the first time.
The correctness of the implementation has been verified prior to the validation 
process to ensure that the software reproduces the physical features
of each model consistently.






\begin{table*}[htbp]
  \centering
  \caption{Summary of the experimental total cross section data used in the validation analysis}
 \begin{tabular}{ccccl}
\toprule
Atomic Number 	& Element 	& Energy range (keV) 	& Sample size  	& References \\ 
    \midrule
1 & H & 0.0136 - 0.020 & 27 & \cite{Beynon1965, Kohl1978, Palenius1976} \\
2 & He & 0.025 - 0.277 & 320 & \cite{Henke1967, Lowry1965, Lukirskii1964, Marr1976, Samson1964 ,Samson2002, Watson1972,West1976} \\
3 & Li & 0.046 - 0.400 & 93 & \cite{Baker1962, Huang1999, Mehlman1982, Peterson1963} \\
7 & N & 0.015 - 0.4 & 73 & \cite{Denne1970, Samson1990} \\
8 & O & 0.013 - 0.28 & 215 & \cite{Kohl1978}, \cite{Angel1988, Cairns1965, Samson1985} \\
10 & Ne & 0.022 - 2.952 & 448 & \cite{Marr1976},  \cite{Samson2002, Watson1972, West1976},  \cite{Denne1970},\cite{Suzuki2003} \\
11 & Na & 0.046 - 0.246 & 17 & \cite{Codling1977} \\
13 & Al & 145.4 & 1 & \cite{Gowda1973} \\
17 & Cl & 0.016 - 0.078 & 25 & \cite{Samson1986} \\
18 & Ar & 0.016 - 6 & 487 & \cite{ Henke1967},  \cite{Samson2002, Watson1972, West1976}, \cite{Denne1970}, \cite{Lukirskii1963, Suzuki2005, Yang1987, Zheng2006, Kato2007} \\
19 & K & 0.004 - 0.005 & 12 & \cite{Sander1981} \\
22 & Ti & 59.54 & 1 & \cite{Sogut2002} \\
23 & V & 59.54 & 1 & \cite{Sogut2002} \\
24 & Cr & 59.54 & 1 & \cite{Sogut2002} \\
25 & Mn & 59.54 & 1 & \cite{Sogut2002} \\
26 & Fe & 0.008 - 59.54 & 25 &  \cite{Sogut2002, DelGrande1986, Lombardi1978}\\
27 & Co & 59.54 & 1 & \cite{Sogut2002} \\
28 & Ni & 1.487 - 59.54 & 17 & \cite{Sogut2002, DelGrande1986}  \\
29 & Cu & 59.54 - 661.6 & 9 & \cite{Gowda1973}, \cite{Sogut2002}, \cite{Gowda1974a, Gowda1975, Ranganathayah1977, Titus1959} \\
30 & Zn & 59.54 & 1 & \cite{Sogut2002} \\
33 & As & 59.54 & 1 & \cite{Sogut2002} \\
34 & Se & 59.54 & 1 & \cite{Sogut2002} \\
36 & Kr & 0.015 - 1.626 & 357 & \cite{Henke1967}, \cite{Samson2002}, \cite{West1976}, \cite{Lang1975}, \cite{Suzuki2002} \\
37 & Rb & 0.004 - 0.010 & 4 & \cite{Marr1968} \\
38 & Sr & 59.54 & 1 & \cite{Sogut2002} \\
39 & Y & 279.2 - 661.6 & 2 & \cite{Manna1985} \\
40 & Zr & 59.54 - 661.6 & 7 &\cite{Gowda1973}, \cite{Gowda1974a, Gowda1975, Ranganathayah1977},   \cite{Budak1999a} \\
41 & Nb & 59.54 & 2 & 
\cite{Budak1999a} \\
42 & Mo & 59.54 - 661.6 & 7 & \cite{Sogut2002},  \cite{Titus1959}, \cite{Budak1999a} \\
43 & Tc & 59.54 & 1 & \cite{Budak1999a} \\
44 & Ru & 59.54 & 1 & \cite{Budak1999a} \\
45 & Rh & 59.54 & 1 & \cite{Budak1999a} \\
46 & Pd & 59.54 & 1 & \cite{Budak1999a} \\
47 & Ag & 1.487 - 661.6 & 13 &  \cite{Gowda1973}, \cite{Gowda1974a, Gowda1975, Ranganathayah1977,vTitus1959}, \cite{ Budak1999a, Jahagirdar1992, Lurio1975, Parthasaradhi1966} \\
48 & Cd & 59.54 & 2 &  \cite{Sogut2002,Budak1999a}\\
49 & In & 59.54 & 2 &   \cite{Budak1999a} \\
50 & Sn & 1 - 661.6 & 31 & \cite{Gowda1973}, \cite{DelGrande1986},  \cite{Gowda1974a, Gowda1975, Ranganathayah1977, Budak1999a, Jahagirdar1992}, \cite{Parthasaradhi1966}\\
51 & Sb & 59.54 & 1 & \cite{Budak1999a} \\
52 & Te & 59.54 & 2 &  \cite{Budak1999a} \\
54 & Xe & 0.013 - 6 & 657 & \cite{Samson2002},  \cite{Zheng2006, Suzuki2003, Kato2007, Saito2001} \\
55 & Cs & 0.004 - 59.54 & 14 &  \cite{Sogut2002}, \cite{Marr1968},\cite{Cook1977}\\
57 & La & 59.54 & 1 & \cite{Ertugrul2003} \\
58 & Ce & 59.54 - 661.6 & 6 &  \cite{Manna1985},  \cite{Ertugrul2003,Anasuya1998,Jahagirdar1996,Karabulut1999}\\
59 & Pr & 59.54 - 661.6 & 3 & \cite{Anasuya1998, Jahagirdar1996, Karabulut1999} \\
60 & Nd & 59.54 - 661.6 & 2 & \cite{Anasuya1998}, \cite{Karabulut1999} \\
61 & Pm & 59.54 & 1 & \cite{Karabulut1999} \\
62 & Sm & 59.54 - 145.4 & 4 & \cite{Ertugrul2003}, \cite{Jahagirdar1996, Karabulut1999} \\
63 & Eu & 59.54 & 2 & \cite{Ertugrul2003}, \cite{Karabulut1999} \\
64 & Gd & 59.54 - 661.6 & 5 & \cite{Anasuya1998, Jahagirdar1996, Karabulut1999, Ertugrul1996} \\
65 & Tb & 59.54 & 3 &  \cite{Ertugrul2003},\cite{Karabulut1999}, \cite{Ertugrul1996}\\
66 & Dy & 59.54 - 661.6 & 8 & \cite{Manna1985},   \cite{Ertugrul2003, Anasuya1998, Jahagirdar1996, Karabulut1999, Ertugrul1996} \\
67 & Ho & 59.54 & 1 & \cite{Karabulut1999} \\
68 & Er & 59.54 & 3 &  \cite{Ertugrul2003}, \cite{Karabulut1999}, \cite{Ertugrul1996} \\
70 & Yb & 279.2 - 661.6 & 2 & \cite{Manna1985} \\
72 & Hf & 65.839-68.547 & 25 & \cite{Nayak2006a} \\
73 & Ta & 1.487 - 661.6 & 31 &\cite{Gowda1973},  \cite{DelGrande1986}, \cite{Gowda1974a, Gowda1975, Ranganathayah1977, Jahagirdar1992, Titus1959} \\
78 & Pt & 1 - 40 & 25 & \cite{DelGrande1986} \\
79 & Au & 0.93 - 661.6 & 33 & \cite{DelGrande1986}, \cite{Gowda1974a, Gowda1975, Ranganathayah1977}, \cite{Lurio1975}\\
80 & Hg & 1173 & 1 & \cite{Ghose1966} \\
82 & Pb & 1.487 - 1173 & 36 &  \cite{Gowda1973}, \cite{DelGrande1986},\cite{Gowda1974a, Gowda1975, Ranganathayah1977},\cite{Jahagirdar1992, Lurio1975, Parthasaradhi1966},  \cite{Ghose1966}, \cite{Anasuya1994}  \\
92 & U & 1.487 - 3 & 4 & \cite{DelGrande1986} \\
   \bottomrule
    \end{tabular}%
  \label{tab_exptot1}%
\end{table*}

\tabcolsep=4pt
\begin{table*}[htbp]
  \centering
  \caption{Summary of the experimental subshell cross section data used in the validation analysis}
 \begin{tabular}{cccccl}
\toprule
Z &Element 	& Shell 	& Energy range (keV) 	& Sample size  	& References \\ 
    \midrule
    3     & Li    & K     & 0.069-1.487 & 10    	& \cite{Mehlman1982} \\
    10    & Ne    & L$_1$    & 0.006-0.127 & 14    	& \cite{Codling1976} \\
    18    & Ar    & M$_1$    & 0.001-0.1 & 147   					& \cite{Lynch1973, Houlgate1974, Mobus1993, Samson1974} \\
    20    & Ca    &M$_1$, N$_1$ & 0.038-0.121 & 24     			& \cite{Bizau1987} \\
    25    & Mn    & M$_1$ & 0.113-0.27 & 7 	& \cite{Jimenez1989} \\
    26    & Fe    & K & 59.54 & 1		     		& \cite{Arora1981} \\
    28    & Ni    & K & 59.54 & 1     			& \cite{ Arora1981} \\
    29    & Cu    & K & 59.54-441.8 & 3    	 	& \cite{Arora1981, Gowda1974b} \\
    30    & Zn    & K &59.54 & 1     					& \cite{ Arora1981} \\
    33    & As    & K & 59.54 & 1     				& \cite{ Arora1981} \\
    34    & Se    & K & 59.54 & 1     				& \cite{Arora1981} \\
    35    & Br    & K & 59.54 & 1     			& \cite{ Arora1981} \\
    36    & Kr    & N$_1$    & 0.028-0.041 & 5     				& \cite{Samson1974} \\
    40    & Zr    & K,  L$_1$-L$_3$ & 59.54-441.8 & 15    	& \cite{ Arora1981, Karabulut2005, Gowda1973, Gowda1974b} \\
    41    & Nb    & K, L, L$_1$-L$_3$ & 59.54 & 6    	& \cite{Arora1981, Karabulut2005} 
\\
    42    & Mo    & K, L, L$_1$-L$_3$ & 5.96-74.4 & 6    					& \cite{Arora1981, Karabulut2005} \\
    43    & Tc    & K, L$_1$-L$_3$ & 59.54 & 4     							& \cite{Karabulut2005} \\
    44    & Ru    & K, L$_1$-L$_3$ & 59.54 & 4     		& \cite{ Karabulut2005} \\
    45    & Rh    & K, L$_1$-L$_3$ & 59.54 & 4     	& \cite{Karabulut2005} \\
    46    & Pd    & K, L$_1$-L$_3$ & 59.54 & 4     	& \cite{Karabulut2005} \\
    47    & Ag    & K, L, L$_1$-L$_3$ & 59.54-441.8 & 8	& \cite{Arora1981, Karabulut2005,  Gowda1973, Gowda1974b} \\
    48    & Cd    & K, L, L$_1$-L$_3$ & 59.54 & 4   	& \cite{Karabulut2005} \\
    49    & In    & K, L, L$_1$-L$_3$ & 59.54 & 6     			& \cite{Arora1981, Karabulut2005} \\
    50    & Sn    & K, L, L$_1$-L$_3$ & 59.54-1330 & 19    		& \cite{Arora1981, Karabulut2005, Gowda1973, Gowda1974b}, \cite{Ranganathaiah1979, Ranganath1978}, \cite{Ranganathaiah1981} \\
    51    & Sb    & K, L$_1$-L$_3$ & 59.54 & 4     			& \cite{Karabulut2005} \\
    52    & Te    & K, L$_1$-L$_3$ & 59.54 & 4     			& \cite{Karabulut2005} \\
    53    & I     & K & 59.54 & 2    				& \cite{Arora1981} \\
    54    & Xe    & M$_4$, M$_5$, O$_1$ & 0.0236-1 & 66    					& \cite{Samson1974, Gustafsson1977, Becker1987} \\
    56    & Ba    &  O$_1$ & 0.069-0.119 & 28     				& \cite{Bizau1989} \\

    58    & Ce    & K, L, L$_1$-L$_3$ & 59.54 & 4    			& \cite{ Karabulut2005} \\
    59    & Pr    & K, L$_1$-L$_3$ & 59.54 & 4    		& \cite{Karabulut2005} 
\\
    60    & Nd    & K, L$_1$-L$_3$ & 59.54 & 4    		& \cite{Karabulut2005} 
\\
    61    & Pm    & K, L$_1$-L$_3$ & 59.54 & 4     							& \cite{Karabulut2005} \\
    62    & Sm    & K, L, L$_1$-L$_3$ & 59.54 & 4    			& \cite{Karabulut2005} \\
    63    & Eu    & K, L, L$_1$-L$_3$ & 59.54 & 4     	& \cite{Karabulut2005} \\
    64    & Gd    & K, L$_1$-L$_3$ & 50.3-59.54 & 5    		& \cite{Nayak2007, Karabulut2005} \\
    65    & Tb    & K, L, L$_1$-L$_3$ & 59.54 & 4    			& \cite{Karabulut2005} \\
    66    & Dy    & K, L, L$_1$-L$_3$ & 53.8-59.54 & 5    	& \cite{Hosur2011, Karabulut2005} \\
    67    & Ho    & K, L$_1$-L$_3$ & 59.54 & 4    		& \cite{Karabulut2005} \\
    68    & Er    & K, L, L$_1$-L$_3$ & 59.54 & 4    	& \cite{Karabulut2005} \\
    70    & Yb    & K & 61.34 & 1    	& \cite{Hosur2011} \\
    72    & Hf    & K & 65.29-65.44 & 2    			& \cite{Nayak2007, Nayak2006} \\
    73    & Ta    & K & 67.36-1330 & 15    			& \cite{Nayak2007, Nayak2006,  Gowda1973, Gowda1974b, Ranganathaiah1979, Ranganath1978, Ranganathaiah1981} \\
    74    & W     & K & 59.57-69.64 & 2    						& \cite{Allawadhi1978, Hosur2011} \\
    78    & Pt    & K & 1330 & 1    	& \cite{Ghose1966} \\
    79    & Au    & K,  N$_6$-N$_7$, O$_1$-O$_3$,  P$_1$ & 5-1330 & 60   & \cite{pratt_1964}, \cite{Gowda1974b}, \cite{Ranganathaiah1979},\cite{Ranganathaiah1981}, \cite{Nayak2007}, \cite{Nayak2006}, \cite{Kunz2005}   \\
    80    & Hg    & K, L,  & 59.54-1250 & 2    	& \cite{Allawadhi1978, Ghose1966} \\
    81    & Tl    &  L & 59.54-74.4 & 3    	& \cite{Allawadhi1978, Allawadhi1977} \\
    82    & Pb    & K, L, L$_3$ & 13.6-2750 & 28   	& \cite{pratt_1964}, \cite{Gowda1973}, \cite{Ghose1966},  \cite{Gowda1974b}, \cite{Ranganathaiah1979, Ranganath1978, Ranganathaiah1981},  \cite{Nayak2007}, \cite{Nayak2006}, \cite{Allawadhi1978}, \cite{Allawadhi1977, Arora1981c, Bleeker1962} \\
    83    & Bi    & K, L,  & 59.54-1330 & 4    		& \cite{Allawadhi1978, Allawadhi1977,  Ghose1966} \\
    90    & Th    & K, L$_3$ & 16.9-1330 & 9    	& \cite{Arora1981c, Ranganathaiah1979, Ranganathaiah1981} \\
    92    & U     & K, L, L$_3$ & 17.8-1330 & 10    	& \cite{Arora1981c, Allawadhi1977,  Allawadhi1978, Boyd1965,  Ghose1966} \\
  \bottomrule
    \end{tabular}%
  \label{tab_expshell1}%
\end{table*}
\tabcolsep=6pt

\subsection{Experimental data}
\label{sec_exp}

Experimental data \cite{Beynon1965} -\cite{Boyd1965} for the validation 
of the simulation models were collected from a survey of the literature.
The sample of experimental total cross sections consists of approximately
3000 measurements, which
concern 61 target atoms and span energies approximately from  5~eV to 1.2 MeV.
It includes measurements 
at energies below 100~eV, mostly concerning gaseous targets:
these data are relevant to evaluate the accuracy of calculations performed in
independent particle approximation at very low energies, e.g. the EPDL97 data
library, which extend down to 1~eV.

The sample of subshell cross sections encompasses approximately
600 measurements, which
concern 52 target atoms and span energies approximately from  1~eV to 2.75~MeV.

An overview of the experimental data sample is summarized in Tables \ref{tab_exptot1}-\ref{tab_expshell1}.

The photoionization cross sections reported in the literature as experimentally
measured often derive from measurements of total photon attenuation, from which
theoretically calculated contributions from photon scattering were subtracted.
These semi-empirical values are not appropriate to an
epistemologically correct validation process, which requires the comparison of
simulation models with truly experimental data.
An evaluation of the possible systematic effects induced by using semi-empirical
data \cite{Babu1984} -\cite{Singh1984} as a reference for testing cross section calculations is reported in
Section \ref{sec_exptheo}; it concerns a sample of approximately 1500 total cross sections, 
spanning energies between approximately 50 eV and 6 MeV.

Some experimental measurements have been published only in graphical form;
numerical values were extracted from the plots by means of the PlotDigitizer
\cite{plotdigitizer} digitizing software.
The error introduced by the digitization process was estimated by digitizing a
few experimental data samples, which are reported in the related publications
both in graphical and numerical format.
The reliability of the digitized values is hindered by the difficulty of
appraising the experimental points and their error bars in figures that may span
several orders of magnitude in logarithmic scale, or that appear of questionable
graphical quality in the original publication.
Caution was exercised in using these digitized data in the validation analysis;
they were discarded, if  incompatible with other measurements 
reported in the literature in numerical form.

Large discrepancies are evident in some of the experimental data; systematic
effects are probably present in some cases, where sequences of positive or
negative differences between data samples originating from different
experimental groups are qualitiatively visible, and confirmed by the
Wald-Wolfowitz test \cite{wald} to be incompatible with randomness.
Experimental data exhibiting large discrepancies with respect to other
measurements in similar configurations, which hint at the presence of systematic
effects, are excluded from the validation tests.

The validation process is hindered by physical effects related to the structure
of the target material.
Accuracy of edge position is limited by chemical shifts and the detailed
structure of the experimental material observed.
Usually an accuracy of absolute energies below 1-3~eV is unattainable for this
reason.
At low energies (less than 200-500 eV) the occurrence of collective valence
effects and dipole resonances can lead to much larger deviations (e.g. up to
50~eV or 10\%).
General purpose Monte Carlo codes do not take into account such material
structure effects; the cross sections they use for the simulation of
the photoionization process, briefly outlined in section \ref{sec_compilations},
are not intended to model these features.
This limitation should be taken into account in the evaluation of the results
of the validation process.

Correct estimate of experimental errors is a concern in the validation of
physics models, since unrealistic estimation of the experimental errors may lead
to incorrect conclusions regarding the rejection of the null hypothesis in tests
whose statistic takes experimental uncertainties explicitly into account.
Although technological developments have contributed to improved precision of
measurement, some estimates of experimental uncertainties reported in the
literature may be excessively optimistic, especially when they appear inconsistent with
other measurements exploiting similar experimental techniques.
Experimental measurements claiming much smaller uncertainties than similar ones
have been critically evaluated in the analysis process.



\subsection{Data analysis}
\label{sec_analysis}

The evaluation of the simulation models performed in this study has two
objectives: to validate them quantitatively, and to compare their relative capabilities.

The scope of the software validation process is defined according to the
guidelines of the IEEE Standard devoted to software verification and
validation\cite{ieee_vv}, which conforms to the life cycle process standard
defined in ISO/IEC Standard 12207 \cite{iso12207}.
For the problem domain considered in this paper, the validation process provides
evidence that the software models photoionization consistently with
experiment.

A quantitative analysis, based on statistical methods, is practically possible
only for the validation of cross sections, for which a large sample
of experimental data is available.
The scarcity of angular distribution data in the literature hinders the
validation of simulation models through similar statistical analysis methods:
only qualitative general considerations can be made.

The statistical analysis of photoionization cross sections is articulated over
two stages: the first determines the compatibility between the cross sections
calculated by each simulation model and experimental data, while the second
determines whether the various models exhibit any significant difference in
compatibility with experiment.
The Statistical Toolkit \cite{gof1,gof2} and R \cite{R} are used in the
statistical analysis.
The level of significance of the tests is 0.01, unless stated otherwise.

The first stage of the analysis encompasses a number of test cases, each one
corresponding to a configuration (characterized by photon energy, target element,
experimental source
and, if appropriate, subshell) for which experimental data are available.
The inclusion of the experimental source in the definition of a test case
facilitates the identification of possible systematic effects related
to the experimental environment of the measurements.
For each test case, cross sections calculated by the software are compared with
measured ones by means of goodness-of-fit tests.
The null hypothesis is defined as the equivalence of the simulated and
experimental data distributions subject to comparison, as being drawn from the 
same parent distribution.
The goodness-of-fit analysis is primarily based on the $\chi^2$ test \cite{bock}.
Among goodness-of-fit tests, this test has the property of taking experimental 
uncertainties explicitly into account; consequently, the test statistic is sensitive
to their correct appraisal.

The ``efficiency'' of a physics model is defined as the fraction of test cases
in which the $\chi^2$ test does not reject the null hypothesis.
This variable quantifies the capability of that simulation model to produce
results statistically consistent with experiment over the whole set of test
cases, which in physical terms means over the whole range of photon energies and
target elements involved in the validation process.
Two methods were applied to calculate the uncertainties on the efficiencies: 
the conventional method involving the binomial distribution, described in many
introductory statistics textbooks (e.g. \cite{frodesen}), and a method 
based on Bayes' theorem \cite{paterno_2004}. 
The two methods deliver identical results within the number of significant
digits reported in the following tables; the method based on Bayes' theorem
delivers meaningful results, which are reported in the following sections, 
also in limiting cases, i.e. for efficiencies very close to 0 or to 1, where the
conventional method based on the binomial distribution produces unreasonable
values.

The second stage of the statistical analysis quantifies the differences of the
simulation models in compatibility with experiment.
It consists of a categorical analysis based on contingency tables, which derive
from the results of the $\chi^2$ test: the outcome of this test is classified as
``fail'' or ``pass'', according  to whether the null hypothesis is rejected or not, respectively.
The simulation model exhibiting the largest efficiency is considered as a
reference in the categorical analysis; the other models are compared to it, to
determine whether their difference in compatibility with measurements is
statistically significant.

The null hypothesis in the analysis of a contingency table assumes 
equivalent compatibility with experiment of the cross section models it compares.

A variety of tests is applied to determine the statistical significance of the
difference between the data categories (i.e. cross section models) subject to
evaluation: Pearson's $\chi^2$ test \cite{pearson} (when the number of entries
in each cell of the table is greater than 5), Fisher's exact test \cite{fisher},
Boschloo's \cite{boschloo} test, the test based on Suissa and Schuster's Z-pooled statistic \cite{suissa},
Santner and Snell's test \cite{santner} and Barnard's test \cite{barnard}. 
As some contingency tables contain cells with a large number of entries ($>$100),
Barnard's test was calculated according to the approximate CSM statistic \cite{barnard_csm} to 
reduce the computational burden.
The use of a variety of tests mitigates the risk of introducing systematic effects
in the validation results due to peculiarities in the mathematical formulation
of the test statistic.

Fisher's test is widely used in the analysis of contingency tables.
It is based on a model in which both the row and column sums are fixed in
advance, which seldom occurs in experimental practice; it remains applicable to
cases in which the row or column totals, or both, are not fixed, but in these
cases it tends to be conservative, yielding a larger p-value than the true
significance of the test \cite{agresti}.

Unconditional tests, such as Barnard's test \cite{barnard}, Boschloo's test
\cite{boschloo} and Suissa and Shuster's \cite{suissa} calculation of a
Z-pooled statistic, are deemed more powerful than Fisher's exact test in some
configurations of 2$\times$2 contingency tables \cite{andres_1994,andres_2004},
but they are computationally more intensive.



%
%



\section{Results of Total Cross Section Validation}
\label{sec_resulttot}

\begin{figure}
\centerline{\includegraphics[angle=0,width=8.5cm]{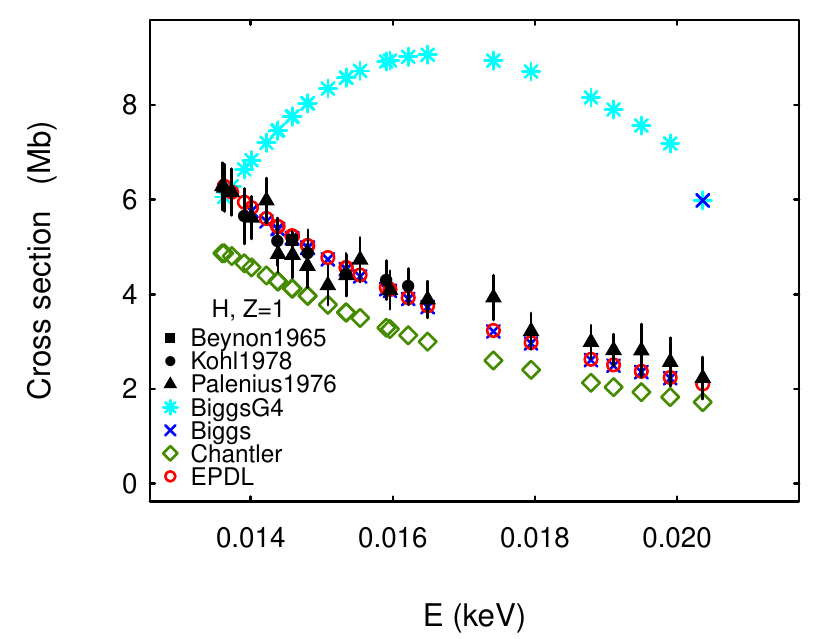}}
\caption{Total photoionization cross section for Z=1 as a function of photon energy.}
\label{fig_tot1}
\end{figure}

\begin{figure}
\centerline{\includegraphics[angle=0,width=8.5cm]{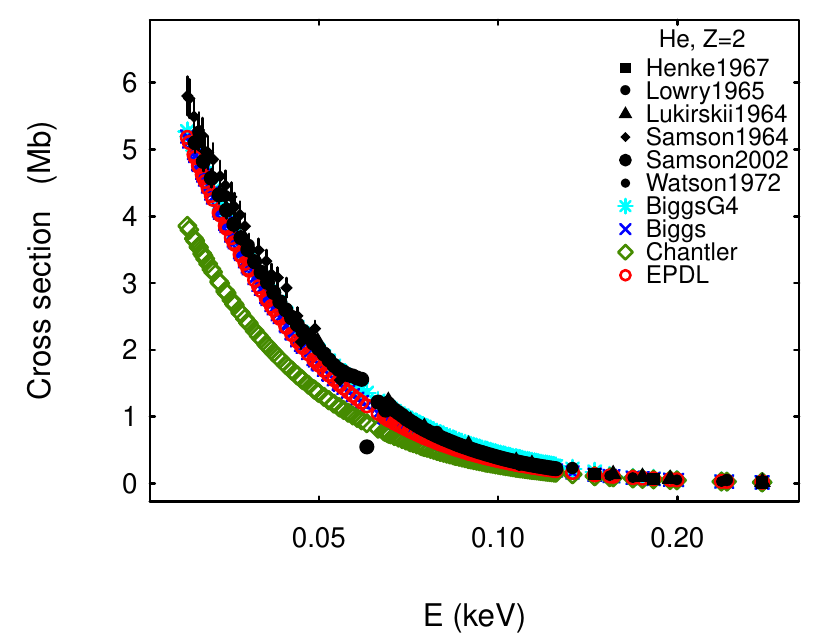}}
\caption{Total photoionization cross section for Z=2 as a function of photon energy.}
\label{fig_tot2}
\end{figure}

\begin{figure}
\centerline{\includegraphics[angle=0,width=8.5cm]{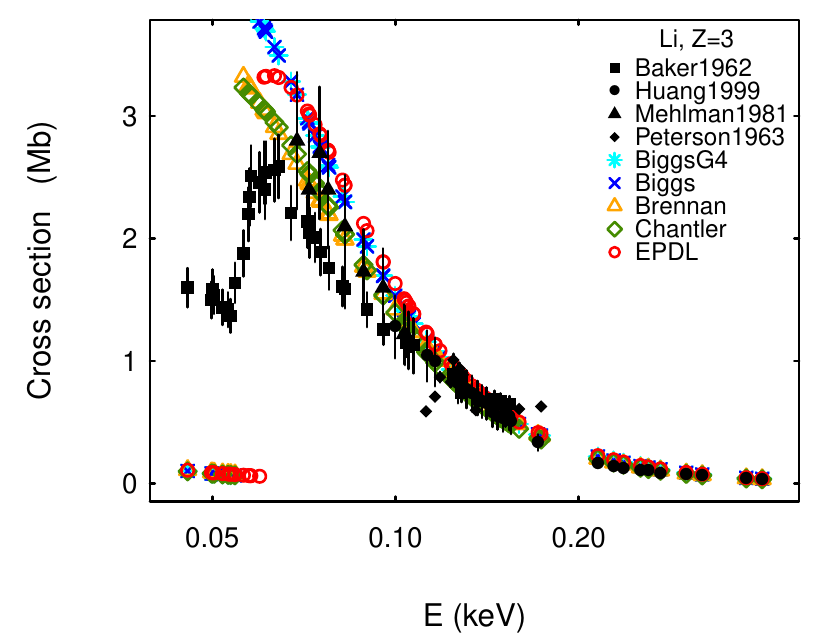}}
\caption{Total photoionization cross section for Z=3 as a function of photon energy.}
\label{fig_tot3}
\end{figure}

\begin{figure}
\centerline{\includegraphics[angle=0,width=8.5cm]{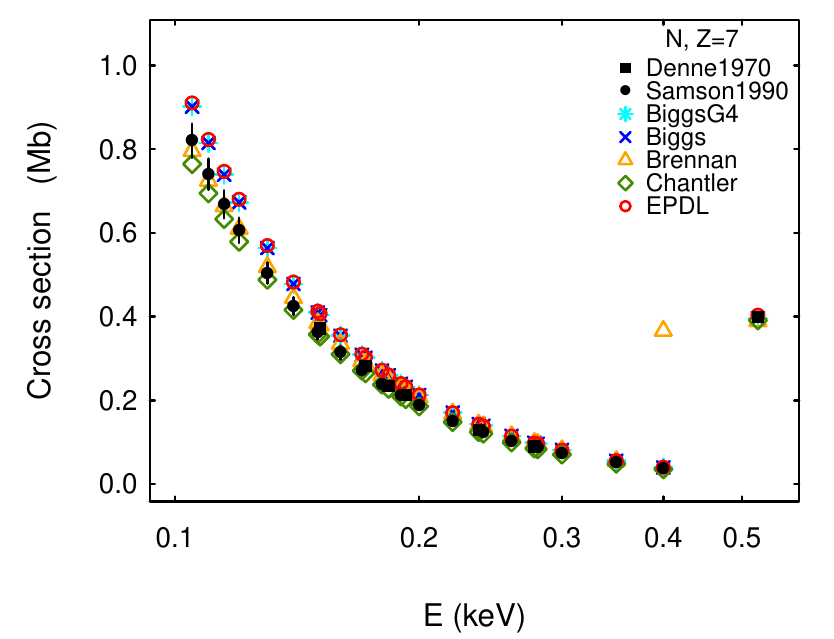}}
\caption{Total photoionization cross section for Z=7 as a function of photon energy.}
\label{fig_tot7}
\end{figure}

\begin{figure}
\centerline{\includegraphics[angle=0,width=8.5cm]{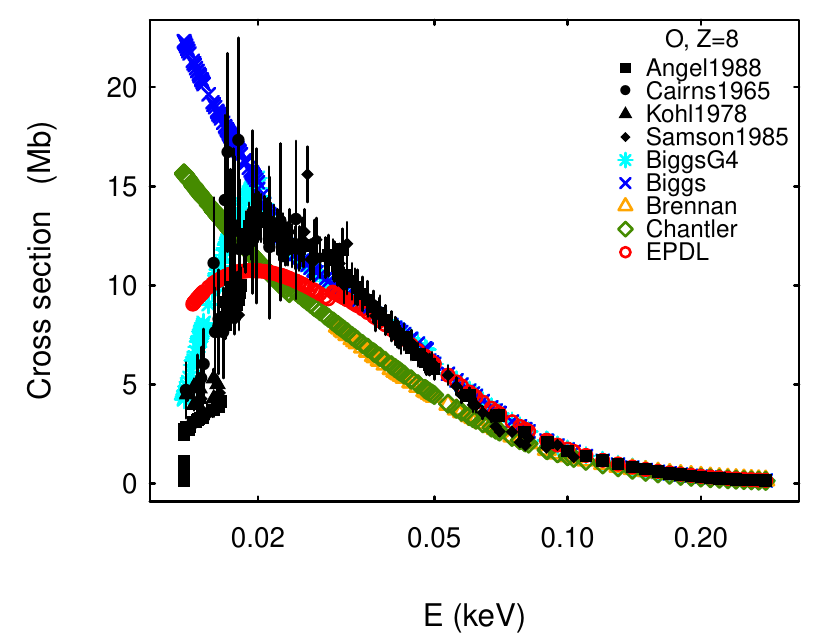}}
\caption{Total photoionization cross section for Z=8 as a function of photon energy.}
\label{fig_tot8}
\end{figure}

\begin{figure}
\centerline{\includegraphics[angle=0,width=8.5cm]{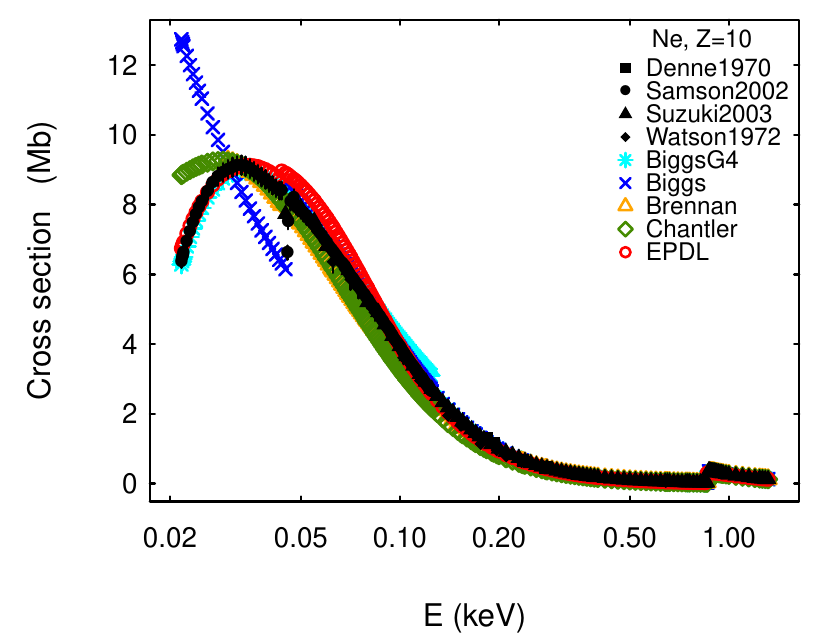}}
\caption{Total photoionization cross section for Z=10 as a function of photon energy.}
\label{fig_tot10}
\end{figure}

\begin{figure}
\centerline{\includegraphics[angle=0,width=8.5cm]{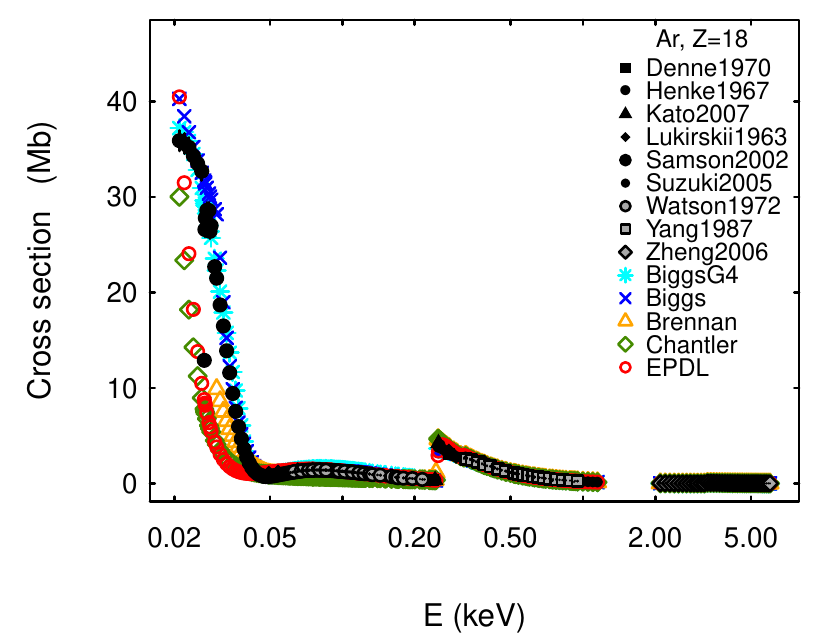}}
\caption{Total photoionization cross section for Z=18 as a function of photon energy.}
\label{fig_tot18}
\end{figure}

\begin{figure}
\centerline{\includegraphics[angle=0,width=8.5cm]{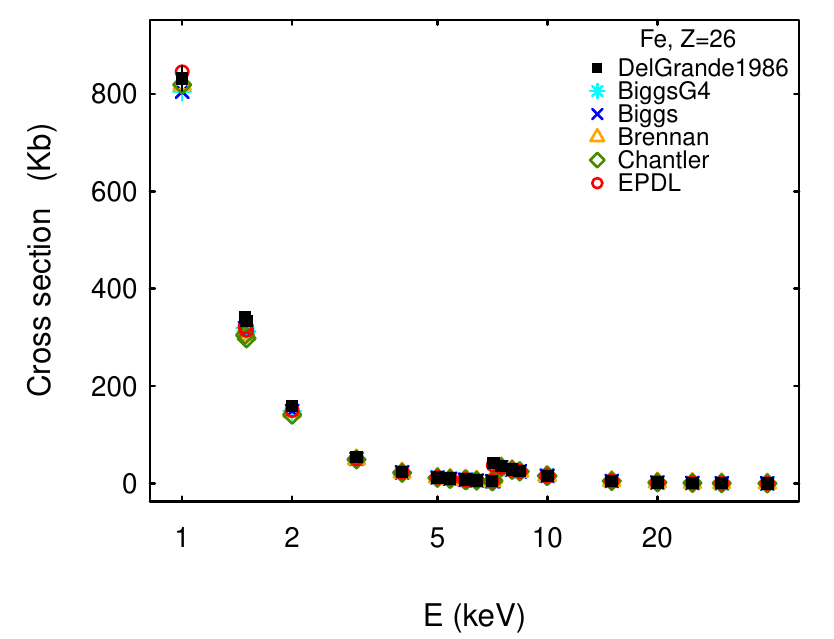}}
\caption{Total photoionization cross section for Z=26 as a function of photon energy.}
\label{fig_tot26}
\end{figure}

\begin{figure}
\centerline{\includegraphics[angle=0,width=8.5cm]{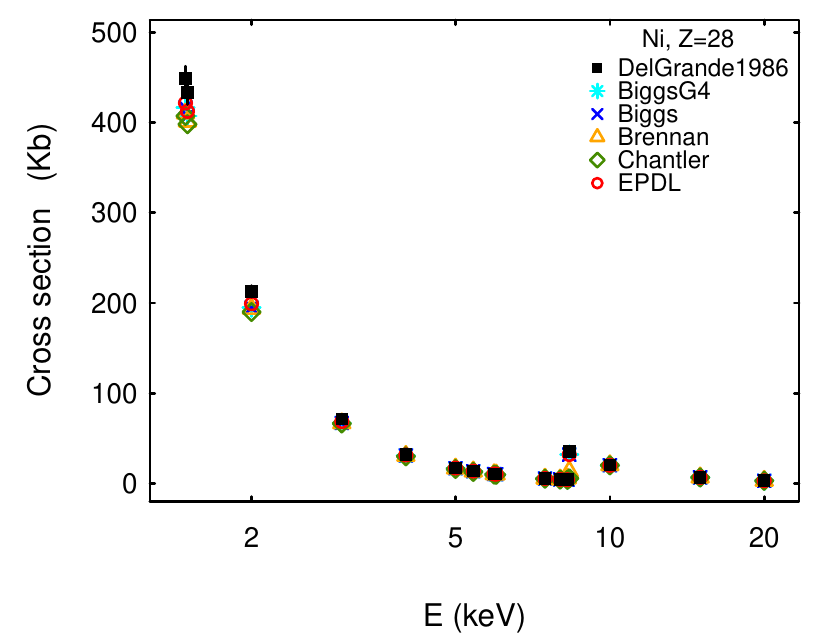}}
\caption{Total photoionization cross section for Z=28 as a function of photon energy.}
\label{fig_tot28}
\end{figure}

\begin{figure}
\centerline{\includegraphics[angle=0,width=8.5cm]{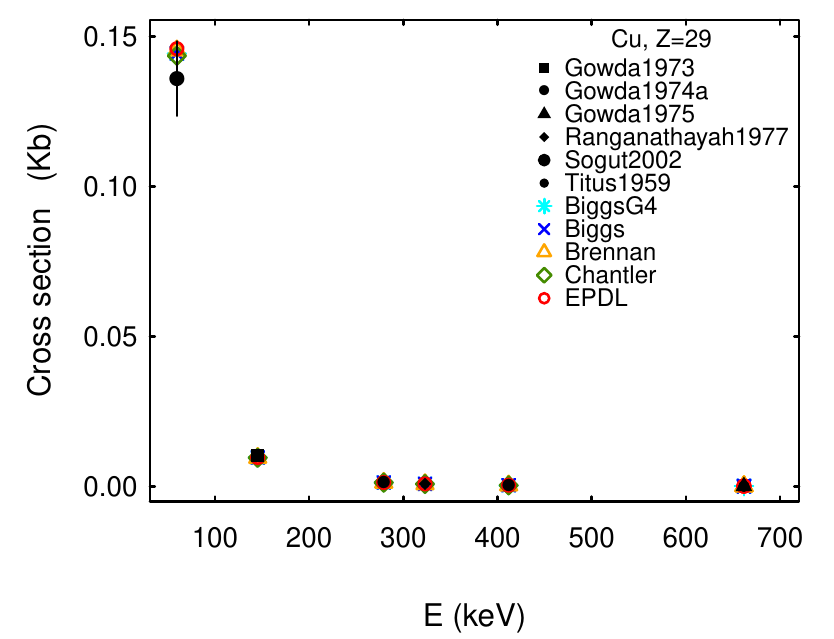}}
\caption{Total photoionization cross section for Z=29 as a function of photon energy.}
\label{fig_tot29}
\end{figure}

\begin{figure}
\centerline{\includegraphics[angle=0,width=8.5cm]{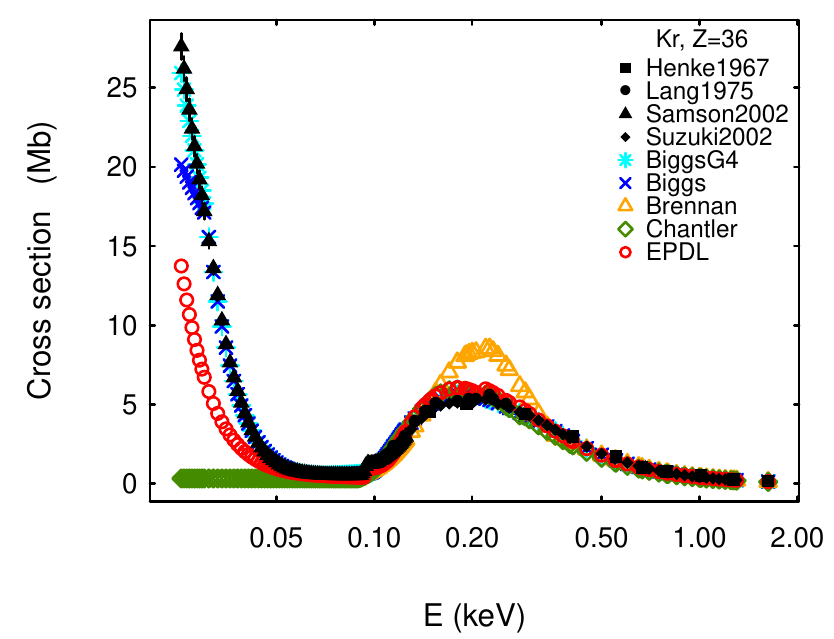}}
\caption{Total photoionization cross section for Z=36 as a function of photon energy.}
\label{fig_tot36}
\end{figure}

\begin{figure}
\centerline{\includegraphics[angle=0,width=8.5cm]{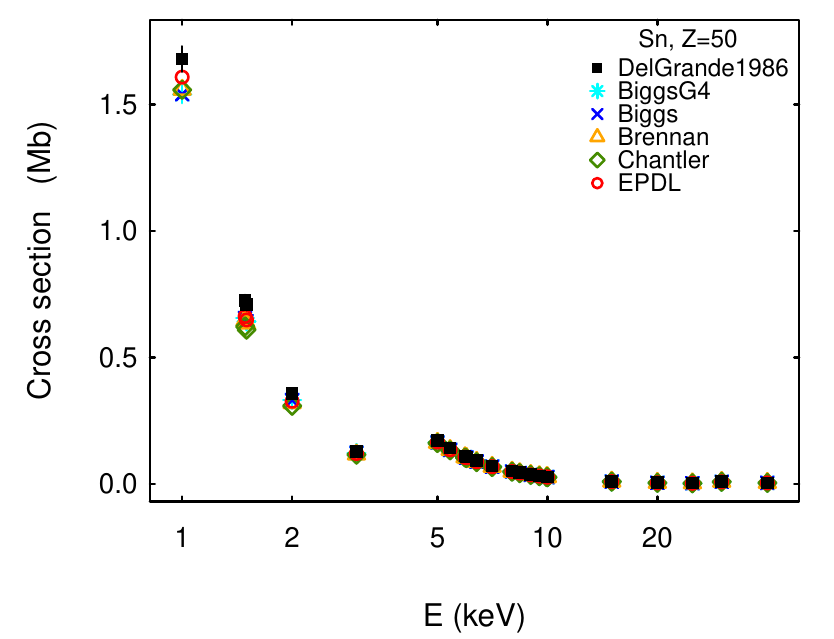}}
\caption{Total photoionization cross section for Z=50 as a function of photon energy.}
\label{fig_tot50}
\end{figure}

\begin{figure}
\centerline{\includegraphics[angle=0,width=8.5cm]{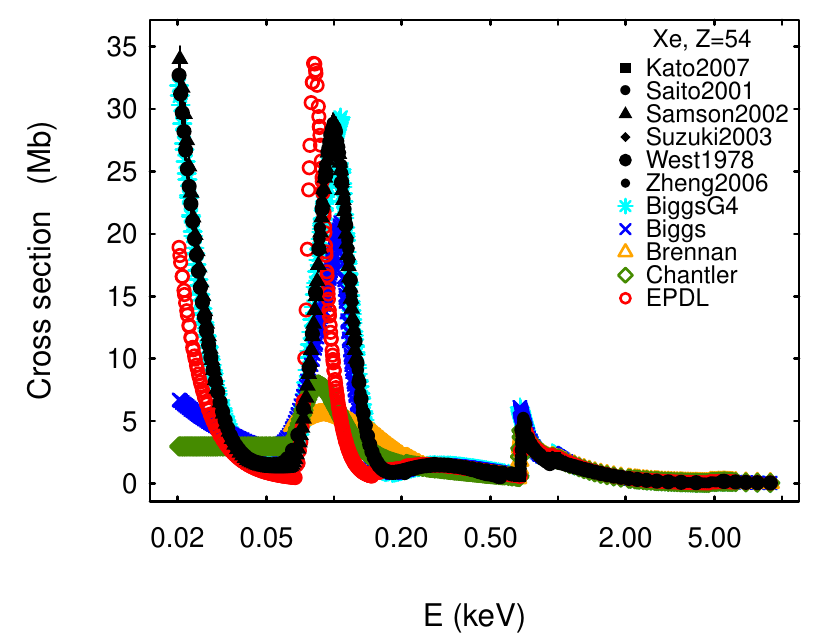}}
\caption{Total photoionization cross section for Z=54 as a function of photon energy.}
\label{fig_tot54}
\end{figure}

\begin{figure}
\centerline{\includegraphics[angle=0,width=8.5cm]{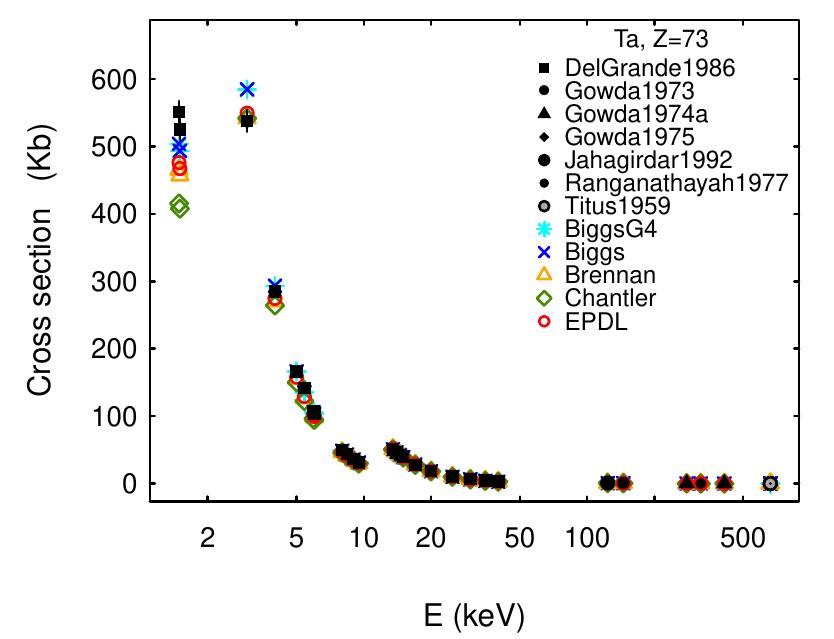}}
\caption{Total photoionization cross section for Z=73 as a function of photon energy.}
\label{fig_tot73}
\end{figure}

\begin{figure}
\centerline{\includegraphics[angle=0,width=8.5cm]{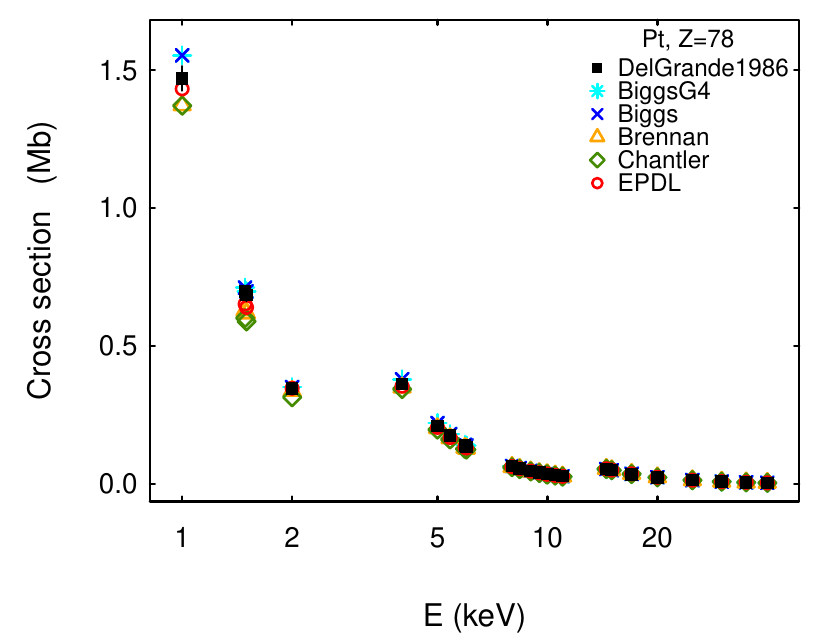}}
\caption{Total photoionization cross section for Z=78 as a function of photon energy.}
\label{fig_tot78}
\end{figure}

\begin{figure}
\centerline{\includegraphics[angle=0,width=8.5cm]{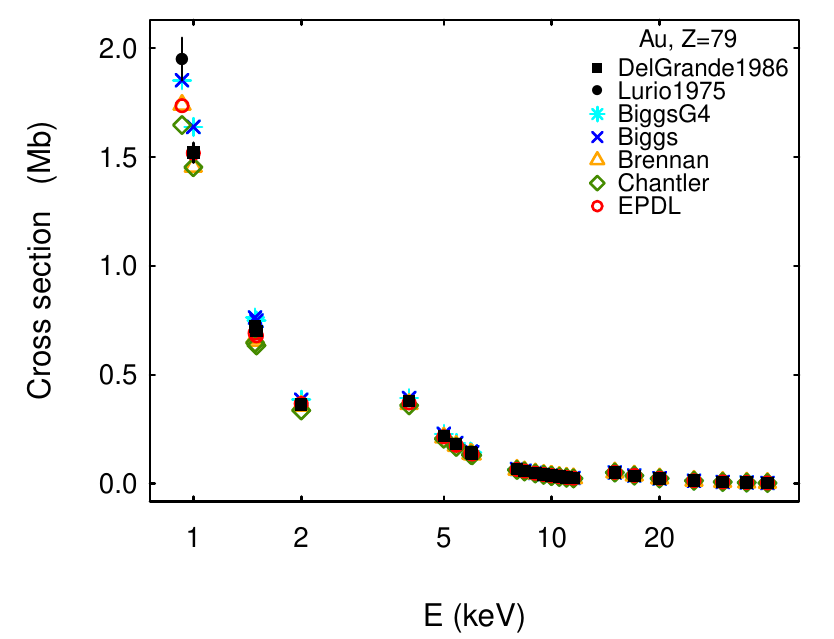}}
\caption{Total photoionization cross section for Z=79 as a function of photon energy.}
\label{fig_tot79}
\end{figure}

\begin{figure}
\centerline{\includegraphics[angle=0,width=8.5cm]{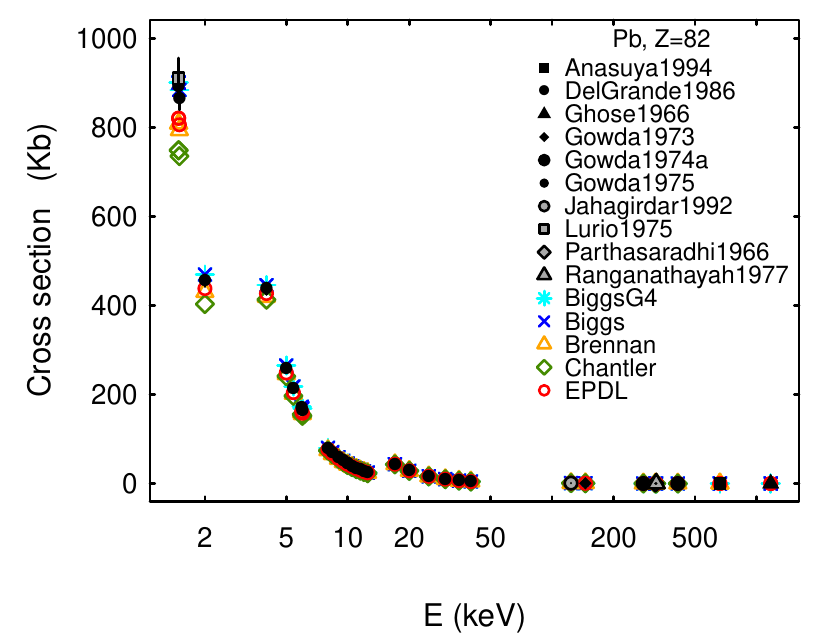}}
\caption{Total photoionization cross section for Z=82 as a function of photon energy.}
\label{fig_tot82}
\end{figure}

\begin{figure}
\centerline{\includegraphics[angle=0,width=8.5cm]{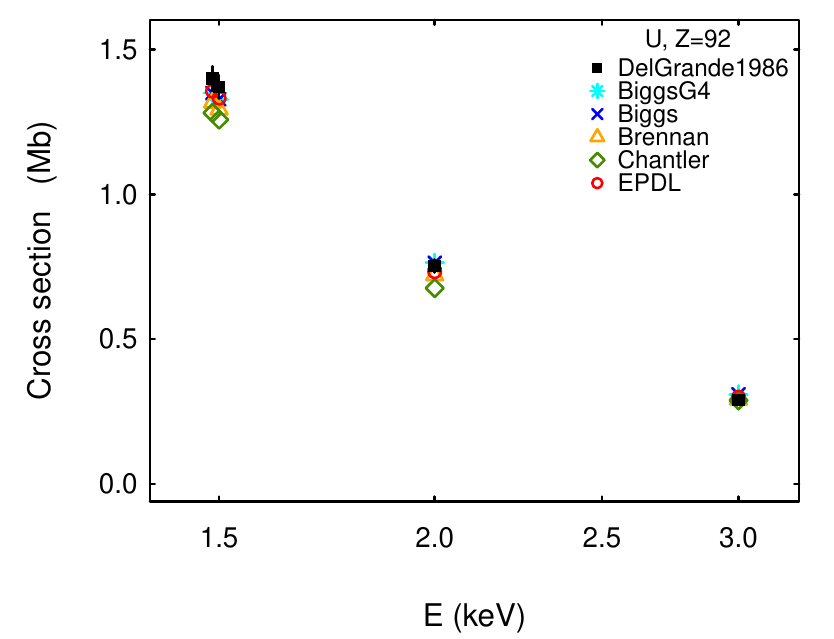}}
\caption{Total photoionization cross section for Z=92 as a function of photon energy.}
\label{fig_tot92}
\end{figure}

\begin{figure}
\centerline{\includegraphics[angle=0,width=8.5cm]{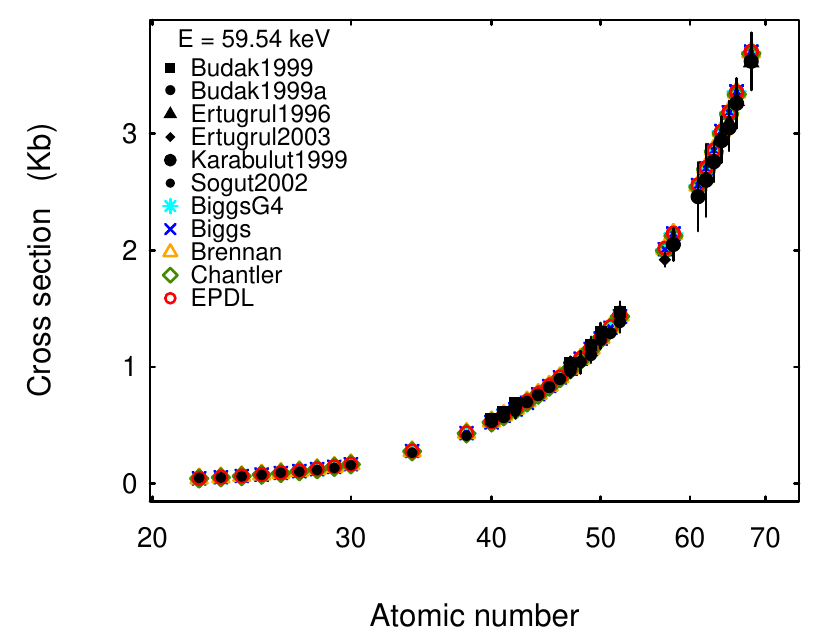}}
\caption{Total photoionization cross section at 59.54~keV as a function of the atomic number Z.}
\label{fig_tote59}
\end{figure}

\begin{figure}
\centerline{\includegraphics[angle=0,width=8.5cm]{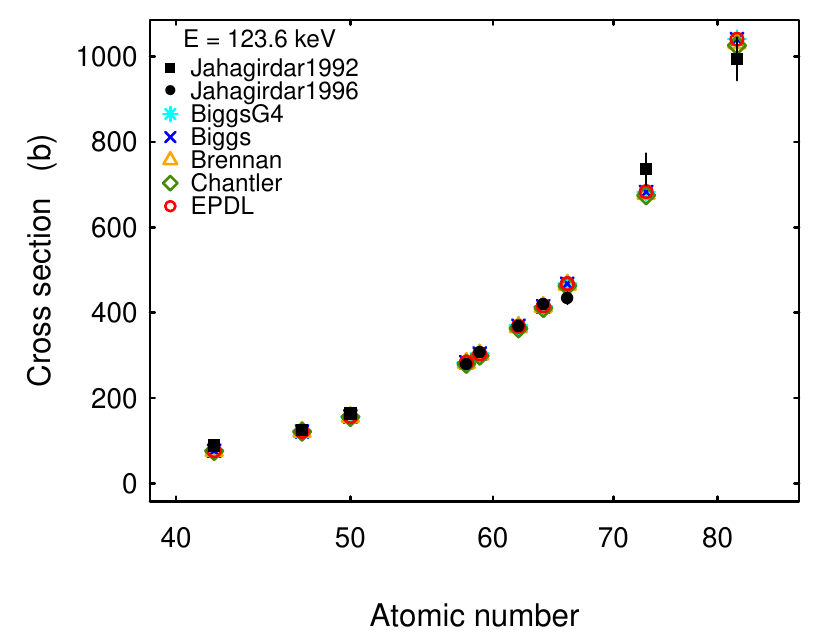}}
\caption{Total photoionization cross section at 123.6~keV as a function of the atomic number Z.}
\label{fig_tote123}
\end{figure}

\begin{figure}
\centerline{\includegraphics[angle=0,width=8.5cm]{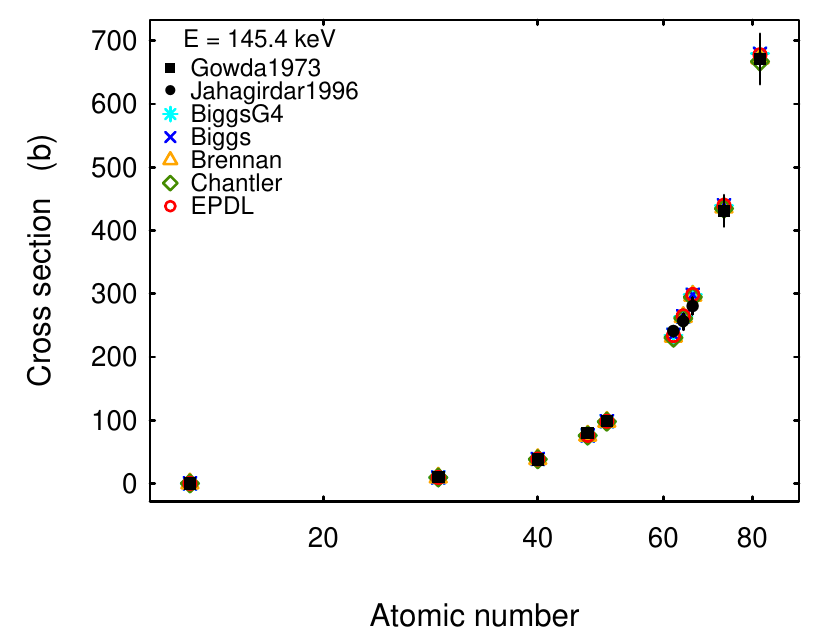}}
\caption{Total photoionization cross section at 145.4~keV as a function of the atomic number Z.}
\label{fig_tot145}
\end{figure}

\begin{figure}
\centerline{\includegraphics[angle=0,width=8.5cm]{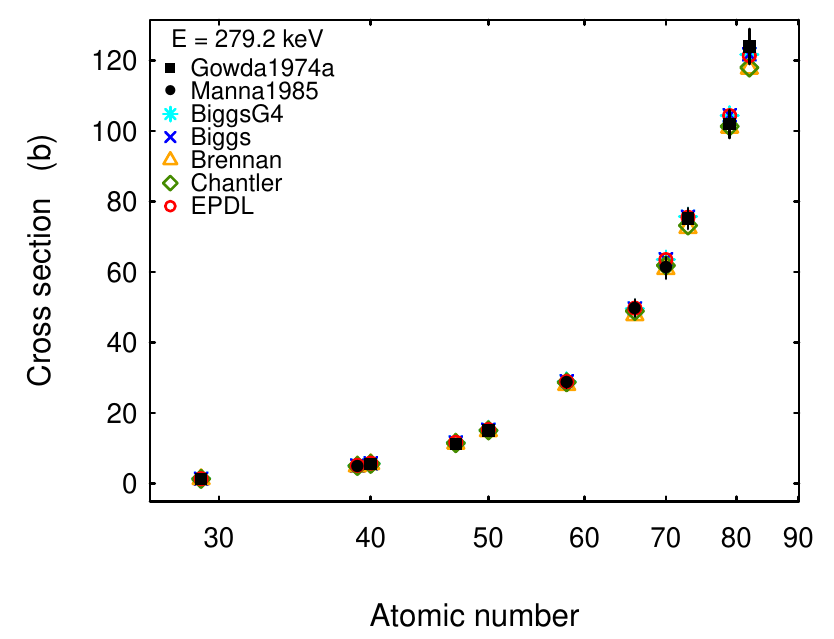}}
\caption{Total photoionization cross section at 279.2~keV as a function of the atomic number Z.}
\label{fig_tote279}
\end{figure}

\begin{figure}
\centerline{\includegraphics[angle=0,width=8.5cm]{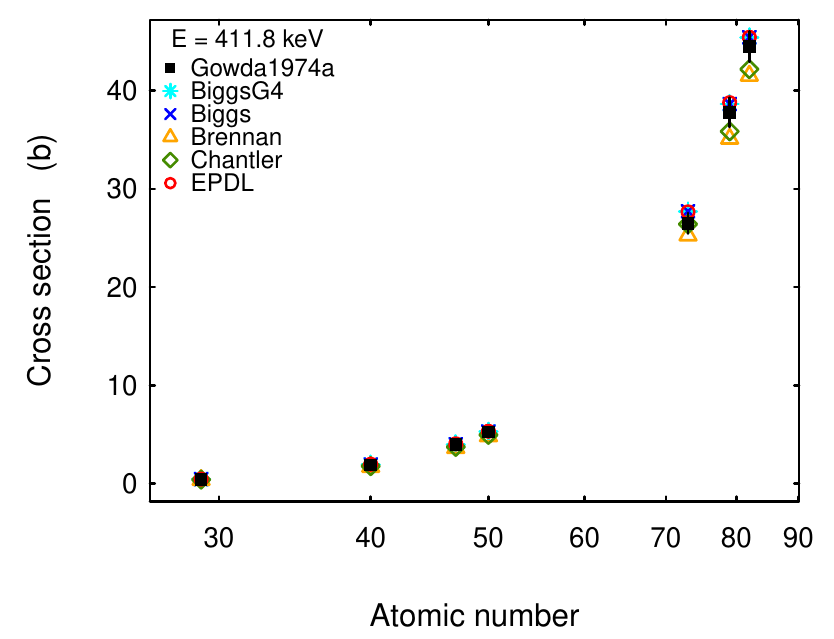}}
\caption{Total photoionization cross section at 411.8~keV as a function of the atomic number Z.}
\label{fig_tote411}
\end{figure}

\begin{figure}
\centerline{\includegraphics[angle=0,width=8.5cm]{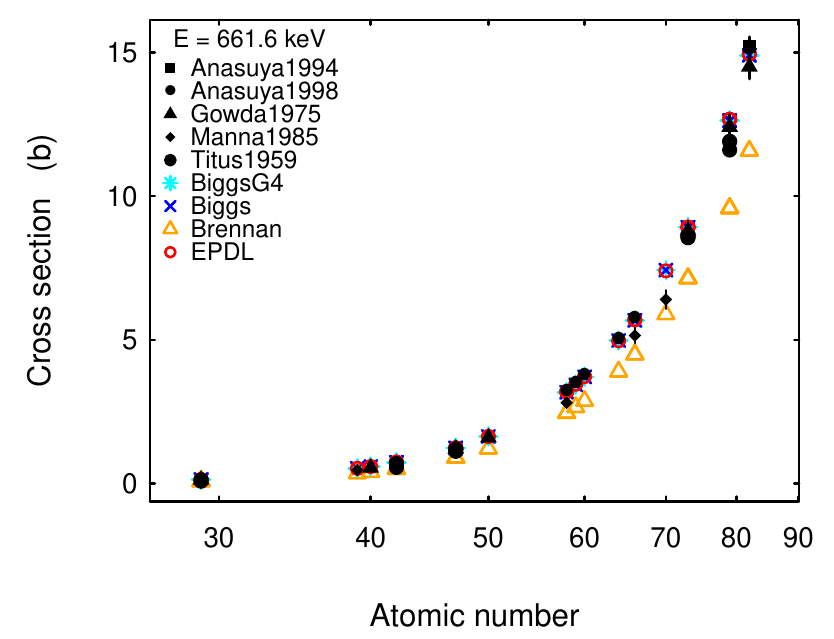}}
\caption{Total photoionization cross section at 661.6~keV as a function of the atomic number Z.}
\label{fig_tote661}
\end{figure}

Figs. \ref{fig_tot1} to \ref{fig_tote661} illustrate calculated and experimental total cross sections.

The validation analysis encompasses various areas of investigation: the
evaluation of possible systematic effects related to the characteristics of
reference data, the evaluation of cross section models covering a wide energy
range, the evaluation of specialized models with limited energy coverage and
the appraisal of the capability of the examined cross section calculations to
describe the photoelectric effect at the low energy end.

\subsection{Evaluation of systematic effects related to reference data}
\label{sec_exptheo}

An analysis was performed prior to the proper validation process to establish
whether semi-empirical cross sections could be used as a reference for comparison without
introducing systematic effects.
This evaluation concerns photon energies greater than 1~keV,
since most of the semi-empirical data collected from the literature
are above this energy;
it involves cross section compilations that are applicable over the
whole energy covered by experimental and semi-empirical data.

The efficiency at reproducing experimental and semi-empirical reference data is 
reported in Table \ref{tab_exptheo} for all compilations covering the selected
energy range.
One observes that it is systematically lower, when semi-empirical data are
considered as a reference in the comparison; the Wald-Wolfowitz test 
rejects the hypothesis of randomness of the sequence of results
associated with experimental and semi-empirical references with 0.01
significance.

\begin{table}[htbp]
  \centering
  \caption{Efficiency calculated with respect to experimental or semi-empirical reference data}
    \begin{tabular}{lccccccc}
    \toprule
     & \multicolumn{3}{c}{ Experimental} && \multicolumn{3}{c}{ Semi-empirical} \\
\cmidrule{2-4}  \cmidrule{6-8} 
        Model  & Pass  & Fail  & Efficiency && Pass  & Fail  & Efficiency \\
    \midrule
    BiggsG4 	& 33    & 9     	& 0.79 $\pm$ 0.06   && 171   & 128   & 0.57 $\pm$ 0.03  \\
    Biggs 		& 33    & 9    	& 0.79 $\pm$ 0.06   && 171   & 128   & 0.57 $\pm$ 0.03 \\
    EPDL  		& 36    & 6    	& 0.86 $\pm$ 0.05   && 180   & 119   & 0.60 $\pm$ 0.03 \\
    Penelope 	& 36    & 6     	& 0.86 $\pm$ 0.05   && 180   & 119   & 0.60 $\pm$ 0.03 \\
    PHOTX 		& 34    & 8     	& 0.81 $\pm$ 0.06   && 176   & 122   & 0.59 $\pm$ 0.03 \\
    Scofield 	& 34    & 8     	& 0.81 $\pm$ 0.06   && 173   & 120   & 0.59 $\pm$ 0.03 \\
    Storm 		& 30    & 12    	& 0.71 $\pm$ 0.07   && 175   & 123   & 0.59 $\pm$ 0.03 \\
    XCOM  		& 35    & 7     	& 0.83 $\pm$ 0.06   && 178   & 120   & 0.60 $\pm$ 0.03 \\
    \bottomrule
    \end{tabular}%
  \label{tab_exptheo}%
\end{table}%

Categorical analysis performed over the compatibility of cross section calculations
with experimental and semi-empirical reference data
confirms that the observed difference is statistically significant
in all cases, with the exception of the Storm and Israel compilation.
The p-values resulting from different tests over contingency tables
are listed in Table \ref{tab_pexptheo}.
The null hypothesis of equivalent compatibility with reference data is rejected
by all tests with 0.01 significance in the comparison involving cross sections
based on Scofield's 1973 calculations (EPDL, PHOTX, XCOM and Scofield's own
tabulations).
For the comparison concerning Biggs-Lighthill cross sections, the null
hypothesis is rejected by all unconditional tests and by Pearson's $\chi^2$
tests, while it is not rejected by Fisher's exact test, which is 
known to be more conservative than unconditional tests.
The insensitivity of the Storm and Israel model to the type of reference data to
which it is compared is related to its overall lower compatibility with
experiment reported in Table \ref{tab_exptheo}.

From these results one can infer that the use of semi-empirical data as a reference
in the comparison with photoelectric cross sections would introduce
systematic effects in the validation process.

All the analyses reported in the following sections concern experimental data
samples only.


\tabcolsep=4pt
\begin{table}[htbp]
  \centering
  \caption{Test of equivalent compatibility of calculated total cross sections with experimental or semi-empirical data}
    \begin{tabular}{lcccccc}
    \toprule
    & \multicolumn{6}{c}{Test} \\
\cmidrule{2-7}
    Model & Fisher & $\chi^2$ & Boschloo & Z-pooled & Santner & Barnard \\   
     \midrule
    Biggs 	& 0.0110 & 0.0081 & 0.0099 & 0.0090 & 0.0090 & 0.0072 \\
    EPDL  	& 0.0010 & 0.0013 & 0.0018 & 0.0023 & 0.0018 & 0.0008 \\
    PHOTX 	& 0.0064 & 0.0066 & 0.0063 & 0.0075 & 0.0080 & 0.0056 \\
    Scofield & 0.0063 & 0.0063 & 0.0061 & 0.0073 & 0.0079 & 0.0053 \\
    Storm	& 0.1319 & 0.1189 & 0.1219 & 0.1223 & 0.1278 & 0.1247 \\
    XCOM	& 0.0034 & 0.0032 & 0.0037 & 0.0041 & 0.0042 & 0.0024 \\
    \bottomrule
    \end{tabular}%
  \label{tab_pexptheo}%
\end{table}%
\tabcolsep=6pt


\subsection{Evaluation of total cross section compilations with wide energy coverage}

Some of the total cross section models considered in this study cover a wide
energy range: those based on Scofield's 1973 non-relativistic calculations
(including EPDL, PHOTX and XCOM compilations), Storm and Israel's compilation and
Biggs-Lighthill's parameterization, both in its original form and in the
modified version used by Geant4.
Their extended applicability has contributed to their extensive use in particle transport codes.

General purpose Monte Carlo codes have traditionally handled photon interactions 
above 1 keV; extensions to lower energies have been included only relatively 
recently in some of them.
The validation process has investigated the ability of these cross section
compilations to reproduce experimental data as a function of energy, with
special attention devoted to characterizing the behaviour at low energies, below
1~keV.
%
%
%

The efficiency of total cross section models applicable from 1 keV up to 
the highest energy measurements included in the experimental sample 
(approximately 1.2 MeV) is reported in Table \ref{tab_exptheo}.
The largest efficiency is achieved by EPDL (which is also
the basis of Penelope's tabulations).
Categorical tests based on contingency tables, summarized in Table
\ref{tab_cont1kev}, show that the differences in compatibility with experiment
between the various models and EPDL are not statistically significant in this
energy range.

One observes some small differences in Table \ref{tab_exptheo} regarding the
efficiencies of cross section compilations derived from Scofield's 1973
calculations.
They are due to differences in the granularity of the energy grid at which
cross sections are tabulated, which affects the precision of interpolation.

\tabcolsep=4pt
\begin{table}[htbp]
  \centering
  \caption{P-values from different tests comparing the compatibility with experiment of total cross section models with extended energy coverage and EPDL, for energies above 1 keV}
    \begin{tabular}{lcccccc}
    \toprule
    & \multicolumn{6}{c}{Test} \\
\cmidrule{2-7}
    Model & Fisher & $\chi^2$ & Boschloo & Z-pooled & Santner & Barnard \\
    \midrule
    BiggsG4 	& 0.570 	& 0.393 	& 0.427 	& 0.427 	& 0.586 	& 0.942 \\
    Biggs 		& 0.570 	& 0.393 	& 0.427 	& 0.427 	& 0.586 	& 0.942 \\
    Penelope 	& 1     	& 1     	& 1     	& 1     	& 1     	& 1 \\
    PHOTX 		& 0.771 	& 0.558 	& 0.608 	& 0.639 	& 0.744 	& 0.997 \\
    Scofield 	& 0.771 	& 0.558 	& 0.608 	& 0.639 	& 0.744 	& 0.997 \\
    Storm 		& 0.183 	& 0.111 	& 0.130 	& 0.122 	& 0.230 	& 0.138 \\
    XCOM  		& 1     	& 0.763 	& 1     	& 0.850 	& 0.913 	& 1 \\
    \bottomrule
    \end{tabular}%
  \label{tab_cont1kev}%
\end{table}%
\tabcolsep=6pt

Only EPDL and Biggs-Lighthill's parameterization cover the whole energy range
corresponding to the experimental data sample, including energies below 1 keV;
their efficiencies are reported in Table \ref{tab_loweeff} for a series of low
energy intervals.
All models exhibit low efficiencies below approximately 100 eV; above this
energy the efficiencies of cross sections based on EPDL and on the original 
Biggs-Lighthill's parameterization appear quite stable (compatible with statistical
uncertainties) and similar, although EPDL ones are always larger.

\begin{table}[htbp]
  \centering
  \caption{Efficiency below 1 keV of total cross section models with extended coverage }
    \begin{tabular}{cclccc}
    \toprule
   E$_{min}$ & E$_{max}$ & Model & Pass & Fail & Efficiency \\
    \midrule
    \multirow{3}{*}{10 eV} & \multirow{3}{*}{1 keV} 	& BiggsG4 & 12    & 31    	& 0.28  $\pm$ 0.07 \\
          								&       & Biggs 	& 13    & 30    	& 0.30  $\pm$ 0.07 \\
          								&       & EPDL  	& 18    & 25    	& 0.42  $\pm$ 0.07 \\
\midrule
    \multirow{3}{*}{100 eV} & \multirow{3}{*}{1 keV}	& BiggsG4 	& 14    & 20    	& 0.41 $\pm$ 0.08 \\
          								&     	& Biggs 	& 22    & 12    	& 0.65 $\pm$ 0.08 \\
          								&      & EPDL  	& 23    & 11    	& 0.68 $\pm$ 0.08 \\
\midrule
    \multirow{3}{*}{150 eV} & \multirow{3}{*}{1 keV} 	& BiggsG4 & 14    & 12   	& 0.54  $\pm$ 0.09 \\
          								&       & Biggs 	& 17    & 9     	& 0.65  $\pm$ 0.09 \\
          								&       & EPDL  	& 19    & 7     	& 0.73  $\pm$ 0.08 \\
 \midrule
    \multirow{3}{*}{250 eV} & \multirow{3}{*}{1 keV} 	& BiggsG4 & 14    & 7     	& 0.67 $\pm$ 0.10 \\
         								&      	& Biggs 	& 14    & 7     	& 0.67 $\pm$ 0.10 \\
          								&     	& EPDL  	& 15    & 6     	& 0.71 $\pm$ 0.09 \\
    \bottomrule
    \end{tabular}%
  \label{tab_loweeff}%
\end{table}%

\begin{figure}
\centerline{\includegraphics[angle=0,width=8.5cm]{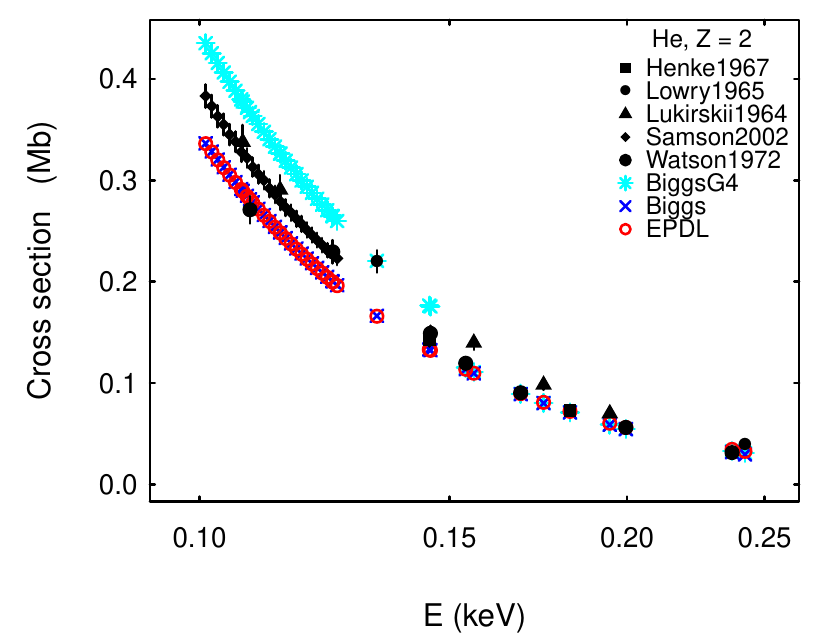}}
\caption{Total photoionization cross section for helium as a function of photon energy, above 100~eV:  original and modified Biggs-Lighthill
parameterizations exhibit different behavior with respect to experimental data.}
\label{fig_biggs2}
\end{figure}

\begin{figure}
\centerline{\includegraphics[angle=0,width=8.5cm]{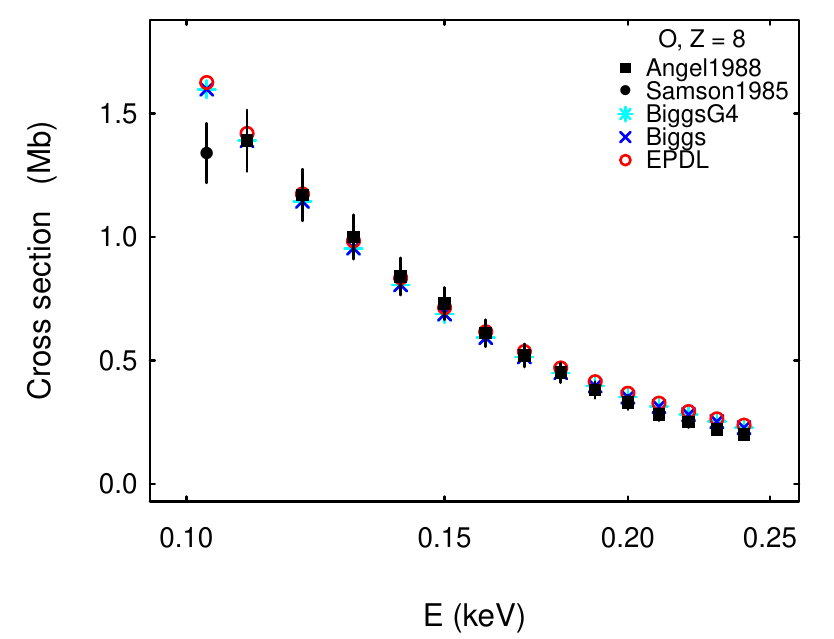}}
\caption{Total photoionization cross section for oxygen as a function of photon energy, above 100~eV:  original and modified Biggs-Lighthill
parameterizations exhibit different behavior with respect to experimental data.}
\label{fig_biggs8}
\end{figure}

\begin{figure}
\centerline{\includegraphics[angle=0,width=8.5cm]{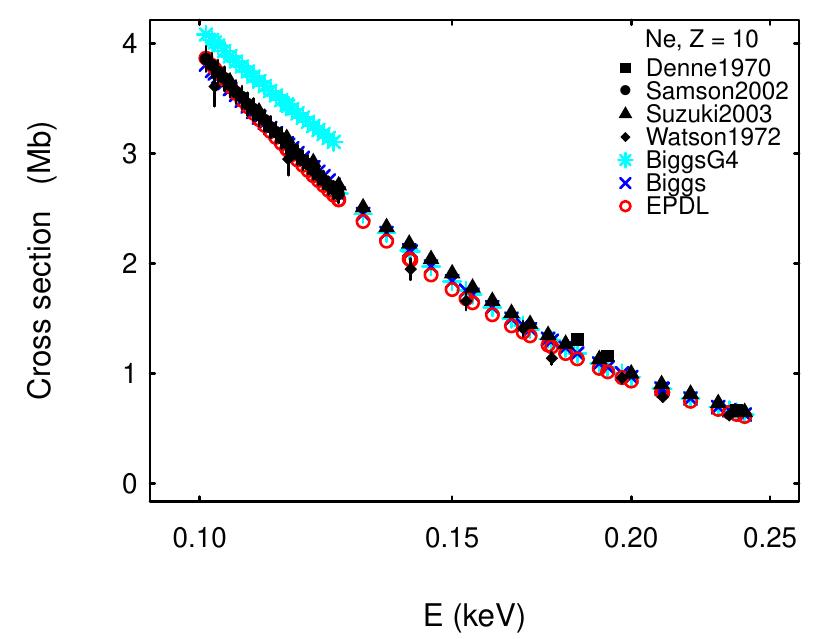}}
\caption{Total photoionization cross section for neon as a function of photon energy, above 100~eV:  original and modified Biggs-Lighthill
parameterizations exhibit different behavior with respect to experimental data.}
\label{fig_biggs10}
\end{figure}

\begin{figure}
\centerline{\includegraphics[angle=0,width=8.5cm]{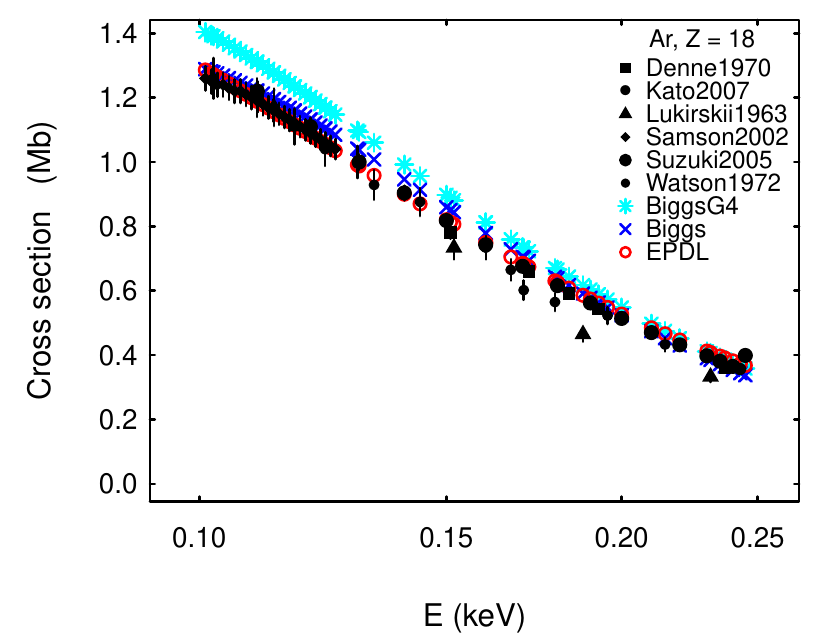}}
\caption{Total photoionization cross section for argon as a function of photon energy, above 100~eV:  original and modified Biggs-Lighthill
parameterizations exhibit different behavior with respect to experimental data.}
\label{fig_biggs18}
\end{figure}

\begin{figure}
\centerline{\includegraphics[angle=0,width=8.5cm]{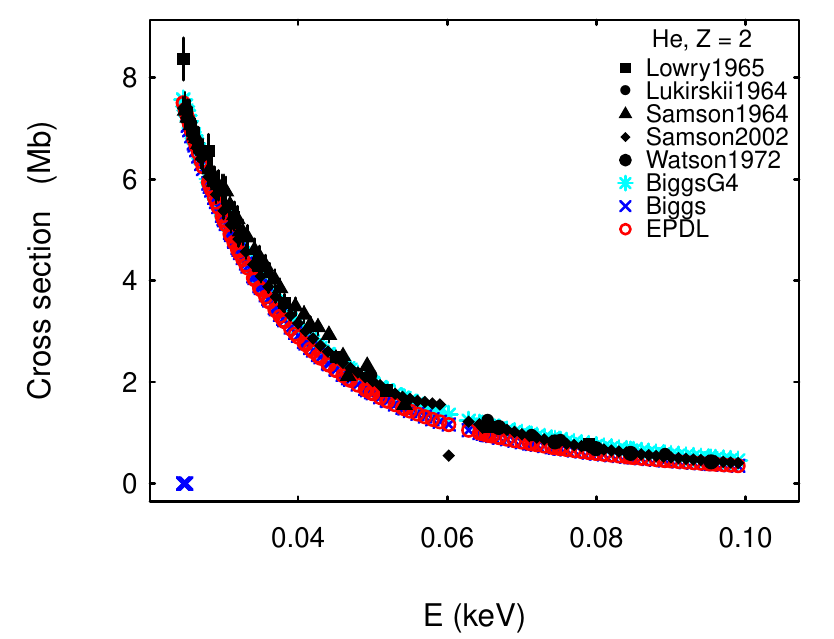}}
\caption{Total photoionization cross section for helium as a function of photon energy, below 100~eV:  original and modified Biggs-Lighthill
parameterizations exhibit different behavior with respect to experimental data.}
\label{fig_biggsLow2}
\end{figure}

\begin{figure}
\centerline{\includegraphics[angle=0,width=8.5cm]{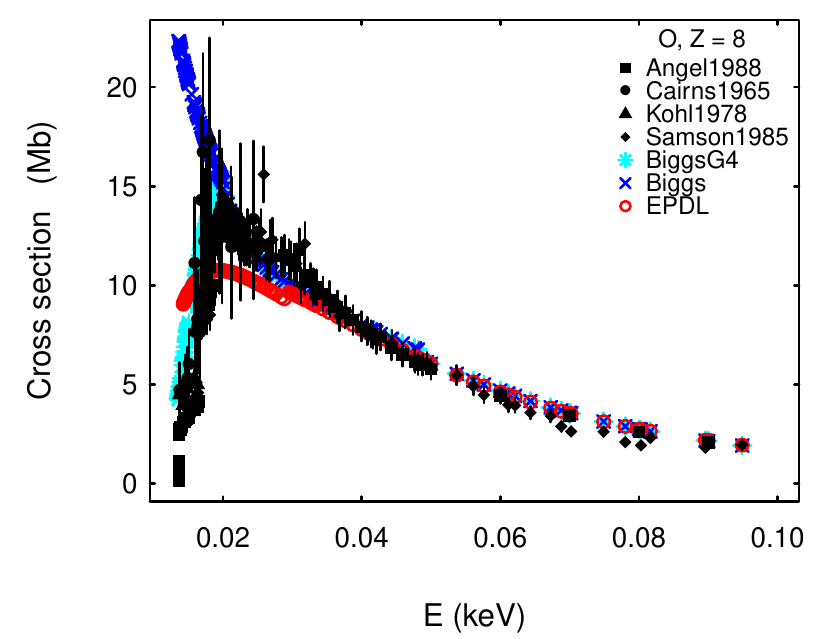}}
\caption{Total photoionization cross section for oxygen as a function of photon energy, below 100~eV:  original and modified Biggs-Lighthill
parameterizations exhibit different behavior with respect to experimental data.}
\label{fig_biggsLow8}
\end{figure}

\begin{figure}
\centerline{\includegraphics[angle=0,width=8.5cm]{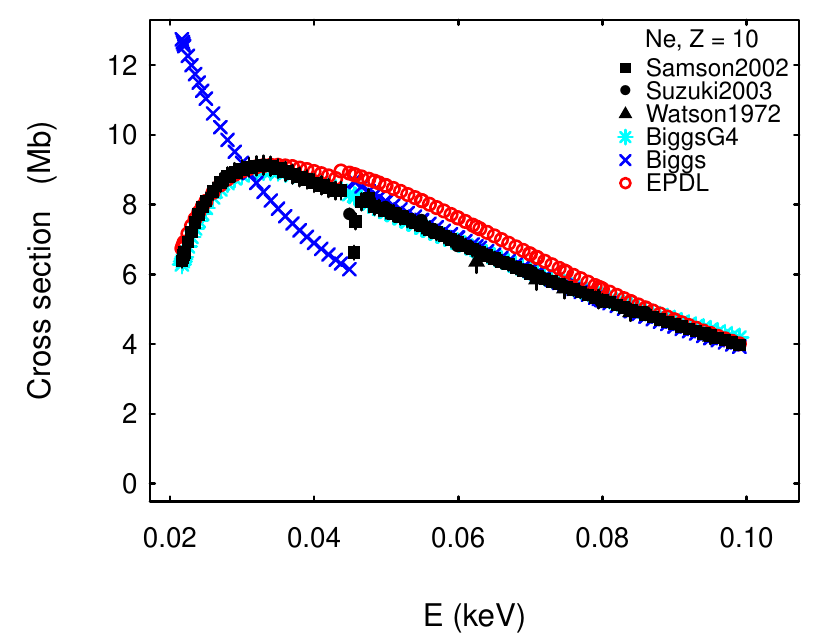}}
\caption{Total photoionization cross section for neon as a function of photon energy, below 100~eV:  original and modified Biggs-Lighthill
parameterizations exhibit different behavior with respect to experimental data.}
\label{fig_biggsLow10}
\end{figure}

\begin{figure}
\centerline{\includegraphics[angle=0,width=8.5cm]{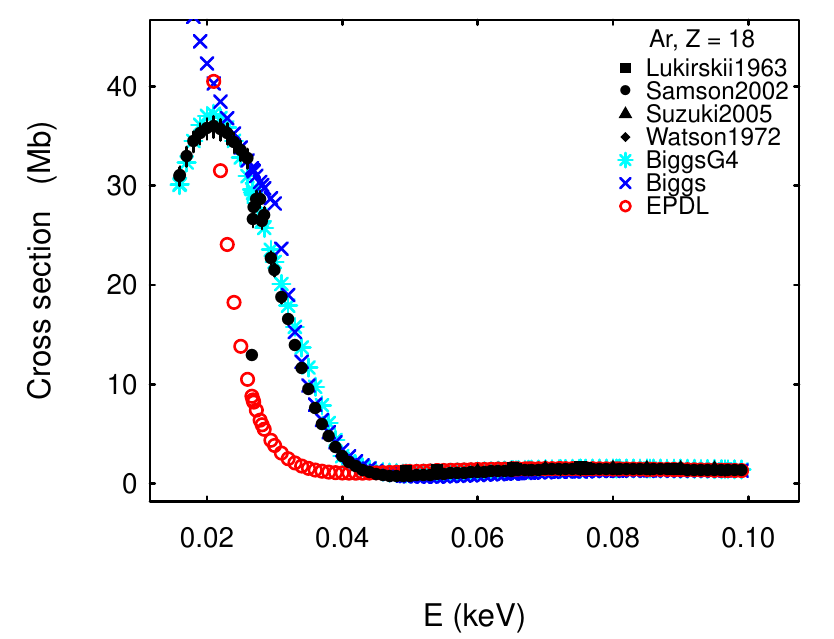}}
\caption{Total photoionization cross section for argon as a function of photon energy, below 100~eV:  original and modified Biggs-Lighthill
parameterizations exhibit different behavior with respect to experimental data.}
\label{fig_biggsLow18}
\end{figure}

The modified coefficients of Biggs and Lighthill's parameterization implemented
in Geant4 do not appear to improve the compatibility with experiment of the
calculated cross sections; discrepancies with respect to experimental data are
qualitatively visible in Figs. \ref{fig_tot1} and \ref{fig_biggs2}-\ref{fig_biggs18}.
Cross sections calculated with the original coefficients appear unable to 
reproduce experimental data consistently in the very low energy range, below a few tens of eV:
a few examples are shown in Figs. \ref{fig_biggsLow2}-\ref{fig_biggsLow18}.

A dedicated statistical analysis was performed to quantify whether the
difference in compatibility with experiment between the two parameterizations is
significant.
For this purpose, the data sample was limited to test cases
where the two cross section calculations produce different values at the same
energy of an experimental measurement; these test cases concern only noble gases, 
oxygen and hydrogen.
Two test cases were considered: one involving energies below 100 eV and 
one concerning measurements above or equal to 100 eV.
The corresponding efficiencies are reported in Table \ref{tab_biggspass}, along
with EPDL and RTAB results: the original Biggs and Lighthill parameterization, as well
as EPDL and RTAB compilations, exhibit low efficiencies in the lower energy range,
while the efficiency of EPDL, RTAB and the original Biggs and Lighthill parameterization 
achieve substantially better compatibility with experiment above 100 eV.
It is worth recalling that this test concerns data below 1~keV, since the coefficients 
of the two parameterizations differ only in the low energy range.

The results of the analysis of these categorical data, summarized in Table 
\ref{tab_biggscont}, show that above 100 eV the null hypothesis of equivalent compatibility 
with experiment of the two parameterizations is rejected with 0.01 significance 
by all the tests applied to the associated contingency table, while it is not rejected below 100~eV.


\tabcolsep=4pt
\begin{table}[htbp]
  \centering
  \caption{Efficiency of total cross sections based on original and modified
Biggs-Lighthill parameterization, limited to the test cases where they differ}
    \begin{tabular}{lccccccc}
    \toprule
         & \multicolumn{3}{c}{E $<$ 100 eV} 	&& \multicolumn{3}{c}{E $\ge$ 100 eV} \\
\cmidrule{2-4} \cmidrule{6-8}
    Model & Pass & Fail & Efficiency  && Pass &Fail & Efficiency \\
\midrule
    BiggsG4 	& 6    & 16    & 0.27 $\pm$ 0.09  	&& 1   	& 13   	& 0.07 $\pm$ 0.07 \\
    Biggs 		& 2    & 20    & 0.09 $\pm$ 0.07  	&& 10   	& 4    	& 0.71 $\pm$ 0.12 \\
    EPDL		& 5	& 17    & 0.23  $\pm$ 0.09		&& 11  	& 3	   	& 0.79 $\pm$ 0.11  \\
    RTAB             & 7	& 15    & 0.32 $\pm$ 0.10		&& 11  	& 3	   	& 0.79 $\pm$ 0.11  \\
  \bottomrule
    \end{tabular}%
  \label{tab_biggspass}%
\end{table}%
\tabcolsep=6pt

\begin{table}[htbp]
  \centering
  \caption{Comparison of the compatibility
with experiment of total cross sections based on original and modified
Biggs-Lighthill parameterization}
    \begin{tabular}{lcc}
    \toprule
    Test 		 	& p-value, E $<$ 100 eV	& p-value, E $\ge$ 100 eV\\
    \midrule
    Fisher  			& 0.240				& 0.0013 \\
    Z-pooled 		& 0.137				& 0.0004 \\
    Boschloo 		& 0.174				& 0.0004 \\
    Santner 			& 0.259				& 0.0004 \\
    Barnard 		& 0.223				& 0.0004 \\
    \bottomrule
    \end{tabular}%
  \label{tab_biggscont}%
\end{table}%

These results suggest reverting to the original coefficients of
Biggs-Lighthill's parameterization in Geant4 for improved accuracy of the
physics models that use cross sections based on it,
at least at energies above a few tens of eV.


\begin{table*}[htbp]
  \centering
  \caption{P-values concerning the comparison of the compatibility with experiment of specific total cross section models and EPDL }
    \begin{tabular}{lrrccccccc}
    \toprule
    				&\multicolumn{2}{c}{ Energy range}	&& \multicolumn{6}{c}{p-values from tests over contingency tables} \\
\cmidrule{2-3} \cmidrule{5-10}
    Model 				& E$_{min}$ & E$_{max}$  		&& Fisher & $\chi^2$ 	& Boschloo 	& Z-pooled 	& Santner & Barnard \\
    \midrule
    \multirow{2}{*}{Brennan}	& 30 eV 	& 700 keV 		&& 0.006 	& 0.004 		& 0.005 	& 0.004 	& 0.004 	& 0.004 \\
    					& 150 eV 	& 700 keV 		&& 0.004	& 0.002		& 0.003	& 0.003	& 0.003	& 0.002 \\
    \multirow{2}{*}{Chantler}	& 10 eV	& 433 keV 		&& 0.0097	& 0.006		& 0.008	& 0.008	& 0.007	& 0.00999 \\
    					& 150 eV	& 433 keV 		&& 0.006	& 0.003		& 0.005	& 0.005	& 0.006	& 0.003 \\
    Ebel 				& 1 keV	& 300 keV 		&& 1		& 0.739		& 0.836	& 0.834	& 0.897	& 0.994 \\
    Elam 				& 100 eV	& 1 MeV			&& 0.124	& 0.084		& 0.093	& 0.093	& 0.133	& 0.087 \\
    \multirow{2}{*}{Henke}	& 10 eV	& 30 keV 			&& 0.151	& 0.101		& 0.109	& 0.110	& 0.128	& 0.108 \\
    					& 150 eV	& 30 keV 			&& 0.049	& 0.027		& 0.031	& 0.031	& 0.032	& 0.030 \\
    McMaster			& 1 keV	& 700 keV			&&  0.041	& 0.033		& 0.033	& 0.040	& 0.071	& 0.039\\
    \multirow{2}{*}{RTAB}	& 10 eV	& 300 keV			&& 1		& 0.843		& 0.828	& 0.917	& 0.920	& 0.997 \\
					& 150 eV	& 300 keV			&& 1		& 0.793		& 1		& 0.880	& 0.903	& 0.870 \\
    Storm-Israel			& 1 keV	& 100 MeV			&& 0.183	& 0.111		& 0.130	& 0.122	& 0.230	& 0.138\\
    Veigele				& 100 eV	& 1 MeV			&&  1	& 0.843		& 0.828	& 0.917	& 0.920	& 0.997\\
    XCOM-DB			& 1 keV	& 500 keV			&& 0.371	& 0.232		& 0.273	& 0.273	& 0.403	& 0.371\\
    \bottomrule
    \end{tabular}%
  \label{tab_contspecific}%
\end{table*}%


\subsection{Evaluation of specific total cross section compilations}

This investigation addresses the issue of whether specialized calculation models
produce more accurate results than extensive cross section compilations 
in the limited energy range they cover and therefore deserve to partially replace 
the extensive cross section compilations currently used in general purpose 
Monte Carlo codes.

The total cross section compilations listed in Table \ref{sec_compilations}
cover different energy ranges, therefore the evaluation of their intrinsic capabilities
requires specific validation tests limited to the energy range of applicability
of each of them.

Efficiencies pertinent to each cross section model are reported in detail in the
following subsections; they are calculated over the subset of the experimental
data sample consistent with the applicability of each model.
The associated tables show the model especially addressed in each
test in italic.
 
The compatibility with experiment of each specialized cross
section model has been compared with that of EPDL, which is the compilation with
the highest efficiency in Table \ref{tab_exptheo}, limited to the energy range 
where the model subject to evaluation is applicable.
The results of the statistical analysis that addresses this issue are summarized
in Table \ref{tab_contspecific};
they are discussed in detail in the following subsections.


\subsubsection{Brennan and Cowan's cross sections}
these cross sections cover the energy range between 30 eV and 700 keV, and are limited to
atomic number greater than 2.
The efficiencies listed in Table \ref{tab_effbrennan} show lower capability of 
Brennan and Cowan's cross sections with respect to other compilations
applicable to this energy range.

According to the results of the categorical analysis listed in Table
\ref{tab_contspecific}, the compatibility with experiment of these cross
sections and EPDL ones is rejected with  0.01 significance, both over the entire energy
interval of applicability and excluding test cases in the lower energy end.


From this analysis one can infer that Brennan and Cowan's calculations 
do not achieve greater accuracy than Scofield's 1973 non-relativistic  ones
tabulated in EPDL.

\tabcolsep=4pt
\begin{table}[htbp]
  \centering
  \caption{Total cross sections: efficiency between 30/150 eV and 700 keV}
    \begin{tabular}{lccccccc}
    \toprule
         			& \multicolumn{3}{c}{E$_{min}$  = 10 eV} 	&& \multicolumn{3}{c}{E$_{min}$ = 150 eV} \\
\cmidrule{2-4} \cmidrule{6-8}
    Model & Pass & Fail & Efficiency  && Pass &Fail & Efficiency \\
    \midrule
    BiggsG4 		& 38    & 23    & 0.62 	$\pm$ 0.06 	&& 37		& 16		& 0.70 $\pm$ 0.06 \\
    Biggs 			& 38    & 23    & 0.62 $\pm$ 0.06 	&& 40		& 13		& 0.75 $\pm$ 0.06 \\
    \textit{Brennan} 	& 26    & 35    & 0.43 $\pm$ 0.06 	&& 27		& 26		& 0.51 $\pm$ 0.07 \\
    EPDL  			& 42    & 19    & 0.69 $\pm$ 0.06 	&& 42		& 11		& 0.79 $\pm$ 0.06 \\
    \bottomrule
    \end{tabular}%
  \label{tab_effbrennan}%
\end{table}%
\tabcolsep=6pt

\subsubsection{Chantler's total cross sections} 
these calculations cover energies between 10 eV and 433 keV.
The efficiency of the cross section models applicable in this energy range is 
reported in Table \ref{tab_effchantler}: it is lower for Chantler's cross sections 
than for other options applicable to this energy interval.
The p-values listed in Table \ref{tab_contspecific} show that the hypothesis of
equivalent compatibility with experiment of Chantler's cross sections with
respect to EPDL ones is rejected by all tests, both over the entire energy
interval and excluding the lower energy end.

\tabcolsep=4pt
\begin{table}[htbp]
  \centering
  \caption{Total cross sections: efficiency between 10/150 eV and 433 keV}
    \begin{tabular}{lccccccc}
    \toprule
         			& \multicolumn{3}{c}{E$_{min}$  = 10 eV} 	&& \multicolumn{3}{c}{E$_{min}$ = 150 eV} \\
\cmidrule{2-4} \cmidrule{6-8}
    Model & Pass & Fail & Efficiency  && Pass &Fail & Efficiency \\
    \midrule
    BiggsG4 		& 34    & 24    & 0.50 $\pm$ 0.06  	&& 36		& 15		& 0.71 $\pm$ 0.06 \\
    Biggs 			& 35    & 33    & 0.51 $\pm$ 0.06  	&& 39		& 12		& 0.76 $\pm$ 0.06 \\
    \textit{Chantler} 	& 24    & 44    & 0.35 $\pm$ 0.06  	&& 27		& 24		& 0.53 $\pm$ 0.07 \\
    EPDL  			& 40    & 28    & 0.59 $\pm$ 0.06  	&& 41		& 10		& 0.80 $\pm$ 0.06 \\
    \bottomrule
    \end{tabular}%
  \label{tab_effchantler}%
\end{table}%
\tabcolsep=6pt

\subsubsection{Ebel's total cross sections}
this parameterization is applicable to energies between 1 keV and 300 keV.
The efficiencies listed in Table \ref{tab_effebel}, which are calculated for all
compilations covering this energy range, show similar compatibility with
experiment for Ebel's and EPDL cross sections, which is confirmed by the
statistical comparison of the two categories in Table \ref{tab_contspecific}.

\begin{table}[htbp]
  \centering
  \caption{Total cross sections: efficiency between 1 keV and 300 keV}
    \begin{tabular}{lccc}
    \toprule
    Model 			& Pass  & Fail  	& Efficiency \\
    \midrule
    BiggsG4 		& 22    & 8     & 0.73 $\pm$ 0.08 \\
    Biggs 			& 22    & 8     & 0.73 $\pm$ 0.08 \\
    Brennan 		& 19    & 11   & 0.63 $\pm$ 0.08 \\
    Chantler 		& 17    & 13   & 0.57 $\pm$ 0.09 \\
    \textit{Ebel}  		& 24    & 6     & 0.80 $\pm$ 0.07 \\
    Elam  			& 23    & 7     & 0.77 $\pm$ 0.08 \\
    EPDL  			& 25    & 5     & 0.83 $\pm$ 0.07 \\
    PHOTX 			& 23    & 7     & 0.77 $\pm$ 0.08 \\
    Scofield 		& 23    & 7     & 0.77 $\pm$ 0.08 \\
    Storm 			& 19    & 11   & 0.63 $\pm$ 0.08 \\
    Veigele 			& 17    & 13   & 0.57 $\pm$ 0.09 \\
    XCOMDB 		& 21    & 9     & 0.70 $\pm$ 0.08 \\
    XCOM  			& 24    & 6     & 0.80 $\pm$ 0.08 \\
    \bottomrule
    \end{tabular}%
  \label{tab_effebel}%
\end{table}%

\subsubsection{Elam's and Veigele's cross sections}
they cover energies between 100 eV and 1 MeV.
The efficiencies of the models applicable in this energy range are listed in Table \ref{tab_effelam}.
The results of the categorical analysis in Table \ref{tab_contspecific} show
that the hypothesis of equivalent compatibility with experiment as for EPDL
cross sections is rejected for Veigele's cross sections, while it is not
rejected for Elam's.

\begin{table}[htbp]
  \centering
  \caption{Total cross sections: efficiency between 100 eV and 1 MeV}
    \begin{tabular}{lccc}
    \toprule
    Model 			& Pass  & Fail  	& Efficiency \\
    \midrule
    BiggsG4 		& 40    & 24    & 0.63  $\pm$ 0.06  \\
    Biggs 			& 48    & 16    & 0.75  $\pm$ 0.05 \\
    \textit{Elam} 		& 40    & 24    & 0.63  $\pm$ 0.06  \\
    EPDL  			& 49    & 15    & 0.77  $\pm$ 0.05 \\
    \textit{Veigele} 	& 30    & 34    & 0.47  $\pm$ 0.06  \\
    \bottomrule
    \end{tabular}%
  \label{tab_effelam}%
\end{table}%

\subsubsection{Henke's cross sections}
this model covers energies between 10 eV and 30 keV.
The efficiencies for the total cross section models applicable to these energies
are listed in Table \ref{tab_effhenke}, which also reports values for photon energies
above 150 eV.
Henke's efficiency is lower;
nevertheless the tests summarized in Table \ref{tab_contspecific} do not reject the 
hypothesis of equivalent compatibility with experiment for Henke's and EPDL cross 
sections with 0.01 significance. 

It is worth remarking that all models appear inadequate at reproducing
experimental measurements in the lower energy range, while they perform better
above 150 eV.
Chantler's calculations remain largely incompatible with experiment even in the 
higher energy interval.

\tabcolsep=4pt
\begin{table}[htbp]
  \centering
  \caption{Total cross sections: efficiency between 10/150 eV and 30 keV}
    \begin{tabular}{lccccrrc}
    \toprule
         			& \multicolumn{3}{c}{E$_{min}$  = 10 eV} 	&& \multicolumn{3}{c}{E$_{min}$ = 150 eV} \\
\cmidrule{2-4} \cmidrule{6-8}
    Model & Pass & Fail & Efficiency  && Pass &Fail & Efficiency \\
    \midrule
    BiggsG4 		& 18    	& 33    	& 0.35 $\pm$ 0.07   		&& 20    & 14    	& 0.59 $\pm$ 0.08 \\
    Biggs 			& 19    	& 32    	& 0.37 $\pm$ 0.07  		&& 23    & 11    	& 0.68 $\pm$ 0.08 \\
    Chantler 		& 7     	& 44    	& 0.14 $\pm$ 0.05  		&& 10    & 24    	& 0.29 $\pm$ 0.08 \\
    EPDL 			& 23    	& 28    	& 0.45 $\pm$ 0.07  		&& 24    & 10    	& 0.71 $\pm$ 0.08 \\
    \textit{Henke} 	& 15    	& 36    	& 0.29 $\pm$ 0.06  		&& 15    & 19    	& 0.44 $\pm$ 0.08 \\
    RTAB 			& 24    	& 27    	& 0.47 $\pm$ 0.07  		&& 23    & 11    	& 0.68 $\pm$ 0.08 \\
    \bottomrule
    \end{tabular}%
  \label{tab_effhenke}%
\end{table}%
\tabcolsep=6pt

\subsubsection{RTAB total cross sections}
this compilation covers energies between approximately 10 eV and 300 keV.
The efficiencies for the total cross section models applicable to these energies
are listed in Table \ref{tab_effRTAB}, which also reports values for photon energies
above 150 eV.
The efficiency for RTAB cross sections is similar to that obtained with EPDL; the 
results of the statistical analysis of the related contingency table 
are consistent with this observation.

\tabcolsep=4pt
\begin{table}[htbp]
  \centering
  \caption{Total cross sections: efficiency between 10/150 eV and 300 keV}
    \begin{tabular}{lccccrrc}
    \toprule
         			& \multicolumn{3}{c}{E$_{min}$  = 10 eV} 	&& \multicolumn{3}{c}{E$_{min}$ = 150 eV} \\
\cmidrule{2-4} \cmidrule{6-8}
    Model & Pass & Fail & Efficiency  && Pass &Fail & Efficiency \\
\midrule
    BiggsG4 	& 29    & 34    & 0.46 $\pm$ 0.06  && 31    & 15    & 0.67 $\pm$ 0.07 \\
    Biggs 		& 30    & 33    & 0.48 $\pm$ 0.06  && 34    & 12    & 0.74 $\pm$ 0.06 \\
    Chantler 	& 19    & 44    & 0.30 $\pm$ 0.06  && 22    & 24    & 0.48 $\pm$ 0.07 \\
    EPDL  		& 35    & 28    & 0.56 $\pm$ 0.06  && 36    & 10    & 0.78 $\pm$ 0.06 \\
    Penelope 	& 37    & 26    & 0.59 $\pm$ 0.06  && 36    & 10    & 0.78 $\pm$ 0.06 \\
    \textit{RTAB} 	& 36    & 27    & 0.57 $\pm$ 0.06  && 35    & 11    & 0.76 $\pm$ 0.06 \\
    \bottomrule
    \end{tabular}%
  \label{tab_effRTAB}%
\end{table}%
\tabcolsep=6pt

It is worth remarking that all models appear inadequate at reproducing
experimental measurements in the lower energy range, while they perform better
above 150 eV.
Chantler's calculations remain largely incompatible with experiment even in the 
higher energy interval.

%
\subsubsection{XCOMBD cross sections}
this compilation, encompassed in the DABAX database,  concerns energies between 1 keV and 500 keV.
The efficiencies of the applicable total cross section models are listed in 
Table \ref{tab_effxcomdb}.
Neither the tests on contingency tables concerning XCOMDB and EPDL
(reported in Table \ref{tab_contspecific}), nor those concerning 
 XCOMDB and the standard XCOM tabulation reject the hypothesis of
equivalent compatibility with experiment.

The different efficiency for cross sections based on XCOMDB and standard XCOM
tabulations could derive from a different energy grid of the two tabulations,
which affects the cross section values calculated by interpolation, or from a
different version of XCOM used as a basis for XCOMDB.
The XCOM version used for creating the XCOMDB tabulations could not
be retrieved in the DABAX documentation.

\begin{table}[htbp]
  \centering
  \caption{Total cross sections: efficiency between 1 keV and 500 keV}
    \begin{tabular}{lccc}
    \toprule
    Model 			& Pass  & Fail  	& Efficiency \\
    \midrule
    BiggsG4 		& 27    & 8     	& 0.77 $\pm$ 0.07 \\
    Biggs 			& 27    & 8     	& 0.77 $\pm$ 0.07 \\
    Brennan			& 22    & 13   	& 0.63 $\pm$ 0.08 \\
    Elam  			& 28    & 7     	& 0.80 $\pm$ 0.07 \\
    EPDL  			& 30    & 5     	& 0.86 $\pm$ 0.06 \\
    PHOTX 			& 28    & 7    	& 0.80 $\pm$ 0.07 \\
    Scofield 		& 28    & 7     	& 0.80 $\pm$ 0.07 \\
    Storm 			& 24    & 11    	& 0.69 $\pm$ 0.08 \\
    Veigele 			& 22    & 13    	& 0.63 $\pm$ 0.08 \\
    \textit{XCOMDB} 	& 26    & 9     	& 0.74 $\pm$ 0.07 \\
    XCOM  			& 29    & 6     	& 0.83 $\pm$ 0.06 \\
    \bottomrule
    \end{tabular}%
  \label{tab_effxcomdb}%
\end{table}%

The quantitative validation analysis documented here shows that none 
of the specialized cross section models provides better accuracy than
EPDL tabulations: some of them exhibit significantly worse compatibility with 
experiment, while others are at most statistically equivalent to EPDL 
at reproducing experimental measurements.
Therefore their use in their specific range of applicability at the place of
EPDL would not improve the accuracy of general purpose Monte Carlo codes.

\subsection{Evaluation of total cross sections at low energy}

The results documented in the previous sections hint that 
neither the models applicable to an extended energy range nor
those covering specific energies appear capable of reproducing
cross sections measurements at low energies.


$\chi^2$ tests performed over a data sample with energies 
between 10 eV and 150 eV quantify the capability of all models
applicable in that energy range to reproduce experimental data.
The resulting efficiencies are documented in Table \ref{tab_effverylow};
they confirm that none of the cross section compilation is suitable 
for accurate simulation of the photoelectric effect in this
energy range.

\tabcolsep=4pt
\begin{table}[htbp]
  \centering
  \caption{Efficiency of total cross section models in the low energy range}
    \begin{tabular}{lccccccc}
    \toprule
          & \multicolumn{3}{c}{10-100 eV} &       & \multicolumn{3}{c}{10-150 eV} \\
    \cmidrule{2-4} \cmidrule{6-8}
    Model & Pass  & Fail  & Efficiency &    & Pass  & Fail  & Efficiency \ \\
    \midrule
    BiggsG4 	& 9     & 22    & 0.29  $\pm$ 0.08  &       & 7      & 28    & 0.20  $\pm$ 0.07 \\
    Biggs 		& 5     & 26    & 0.16  $\pm$ 0.07  &       & 8      & 27    & 0.23  $\pm$ 0.07  \\
    Chantler 	& 4     & 27    & 0.13  $\pm$ 0.08  &       & 5      & 30    & 0.14  $\pm$ 0.06  \\
    EPDL  		& 8     & 23    & 0.26  $\pm$ 0.06  &       & 10    & 25    & 0.29  $\pm$ 0.07  \\
    Henke 		& 9     & 22    & 0.29  $\pm$ 0.08  &       & 11    & 24    & 0.31  $\pm$ 0.08  \\
    Penelope 	& 10   & 21    & 0.32  $\pm$ 0.08  &       & 12    & 23    & 0.34  $\pm$ 0.08  \\
    RTAB  		& 10   & 21    & 0.32  $\pm$ 0.08  &       & 12    & 23    & 0.34  $\pm$ 0.08  \\
    \bottomrule
    \end{tabular}%
  \label{tab_effverylow}%
\end{table}%
\tabcolsep=6pt



\section{Results of Partial Cross Section Validation}
\label{sec_resultshell}

\begin{figure}
\centerline{\includegraphics[angle=0,width=8.5cm]{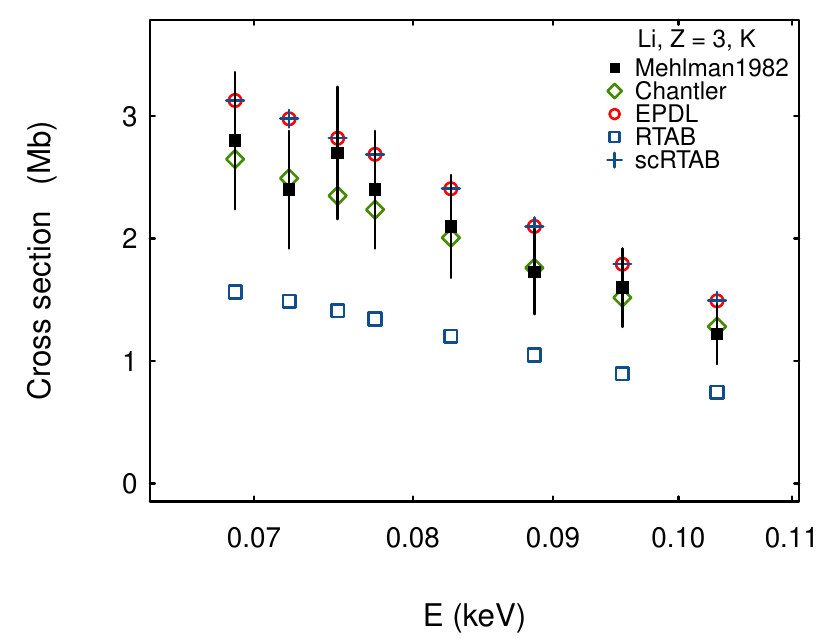}}
\caption{Cross section for the photoionization of the K shell of lithium as a function of photon energy.}
\label{fig_K3}
\end{figure}

\begin{figure}
\centerline{\includegraphics[angle=0,width=8.5cm]{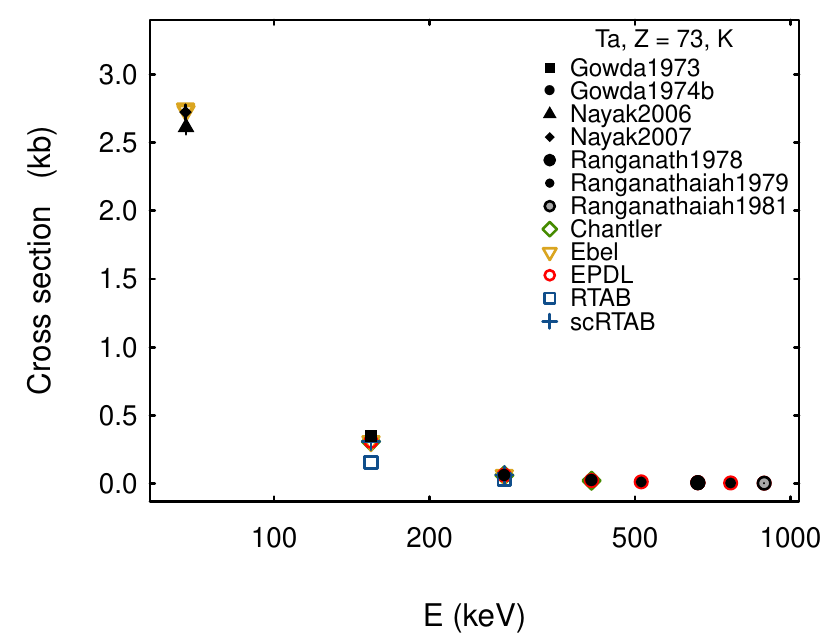}}
\caption{Cross section for the photoionization of the K shell of tantalum as a function of photon energy.}
\label{fig_K73}
\end{figure}

\begin{figure}
\centerline{\includegraphics[angle=0,width=8.5cm]{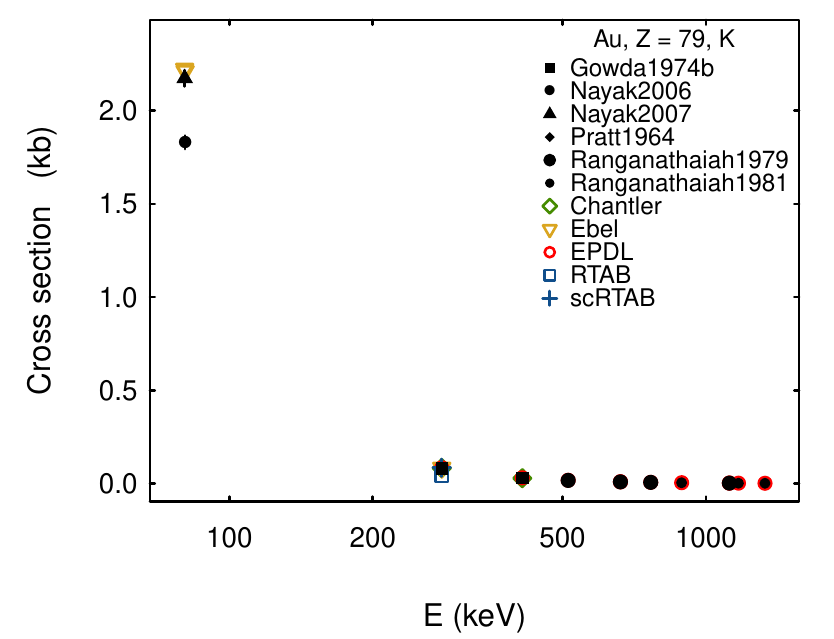}}
\caption{Cross section for the photoionization of the K shell of gold as a function of photon energy.}
\label{fig_K79}
\end{figure}

\begin{figure}
\centerline{\includegraphics[angle=0,width=8.5cm]{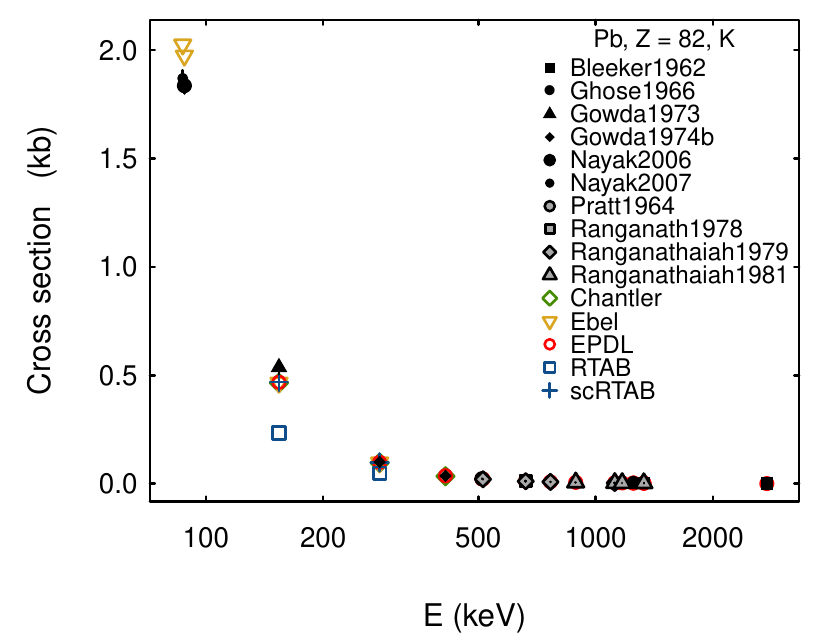}}
\caption{Cross section for the photoionization of the K shell of lead as a function of photon energy.}
\label{fig_K82}
\end{figure}

\begin{figure}
\centerline{\includegraphics[angle=0,width=8.5cm]{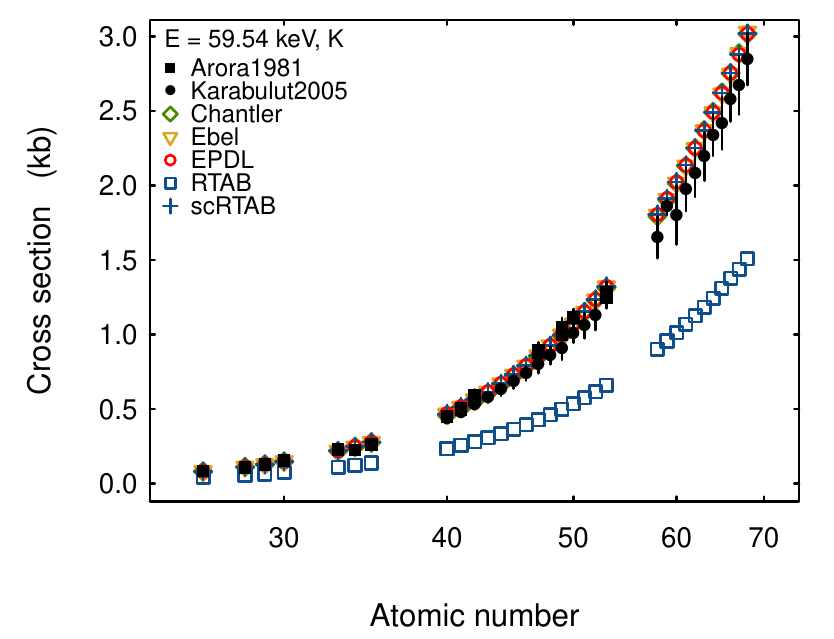}}
\caption{Cross section for the photoionization of the K shell at 59.54 keV as a function of the atomic number. }
\label{fig_K_e59}
\end{figure}

\begin{figure}
\centerline{\includegraphics[angle=0,width=8.5cm]{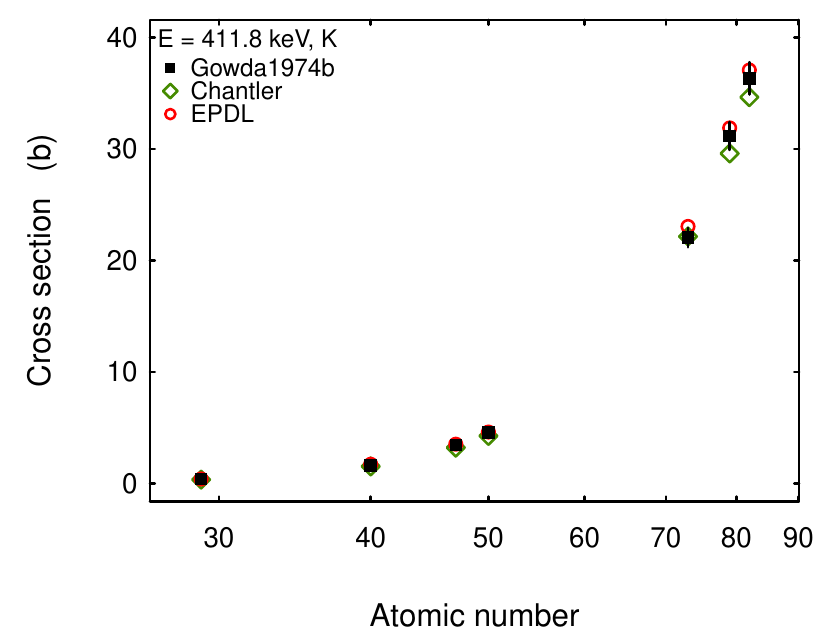}}
\caption{Cross section for the photoionization of the K shell at 411.8 keV as a function of the atomic number. }
\label{fig_K_e411}
\end{figure}

\begin{figure}
\centerline{\includegraphics[angle=0,width=8.5cm]{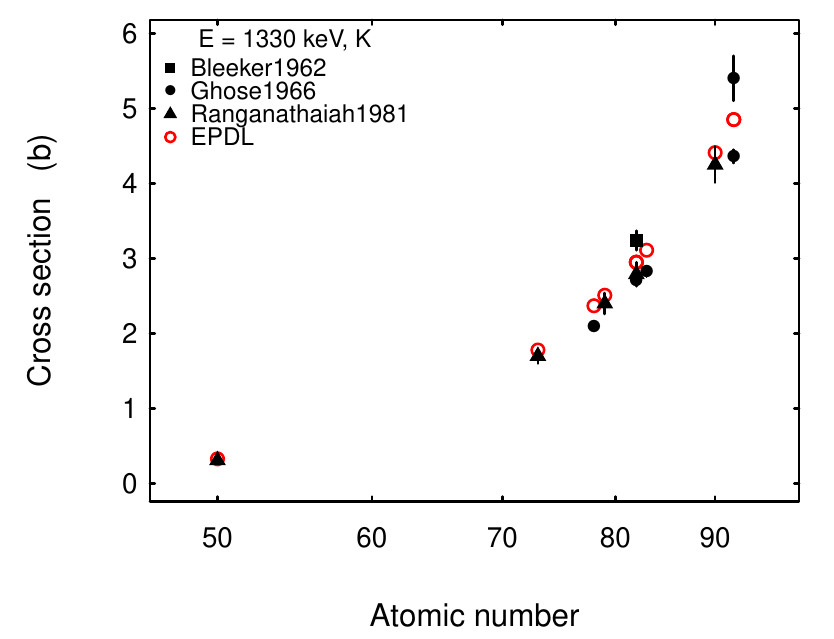}}
\caption{Cross section for the photoionization of the K shell at 1.33 MeV as a function of the atomic number. }
\label{fig_K_e1330}
\end{figure}

\begin{figure}
\centerline{\includegraphics[angle=0,width=8.5cm]{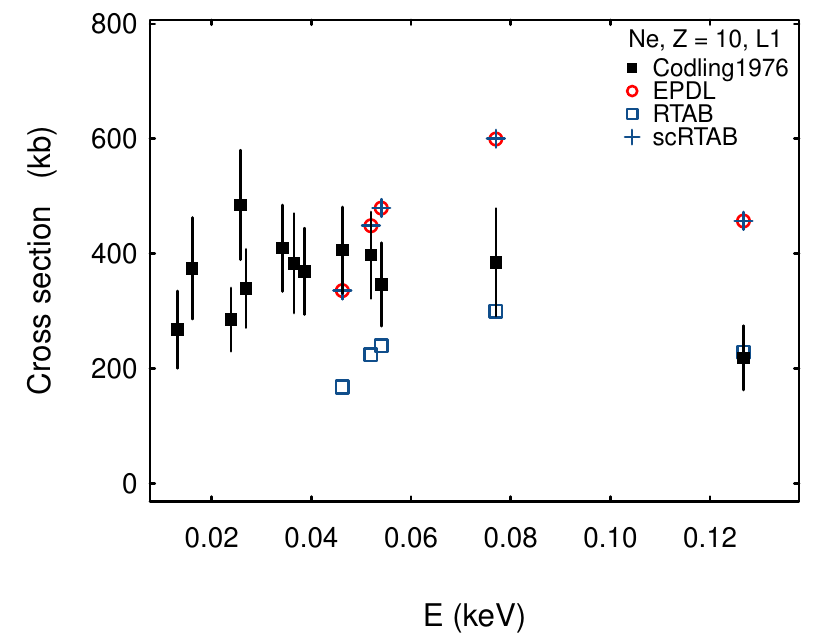}}
\caption{Cross section for the photoionization of the L$_1$ subshell of neon as a function of photon energy.}
\label{fig_L1_10}
\end{figure}

\begin{figure}
    \centering
    \subfigure[L$_1$ subshell]
    {
        \includegraphics[width=8.5cm]{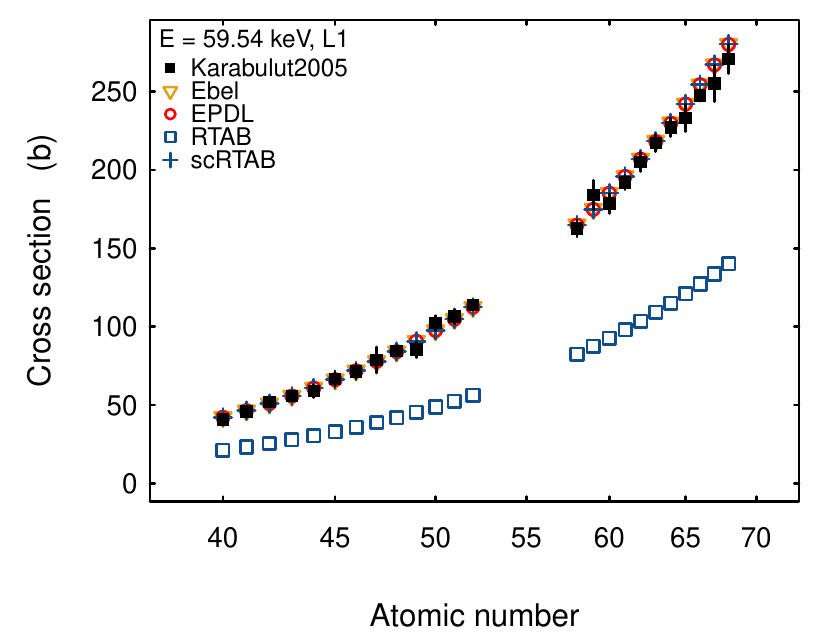}
        \label{fig_L1e59}
    }
    \subfigure[L$_2$ subshell]
    {
        \includegraphics[width=8.5cm]{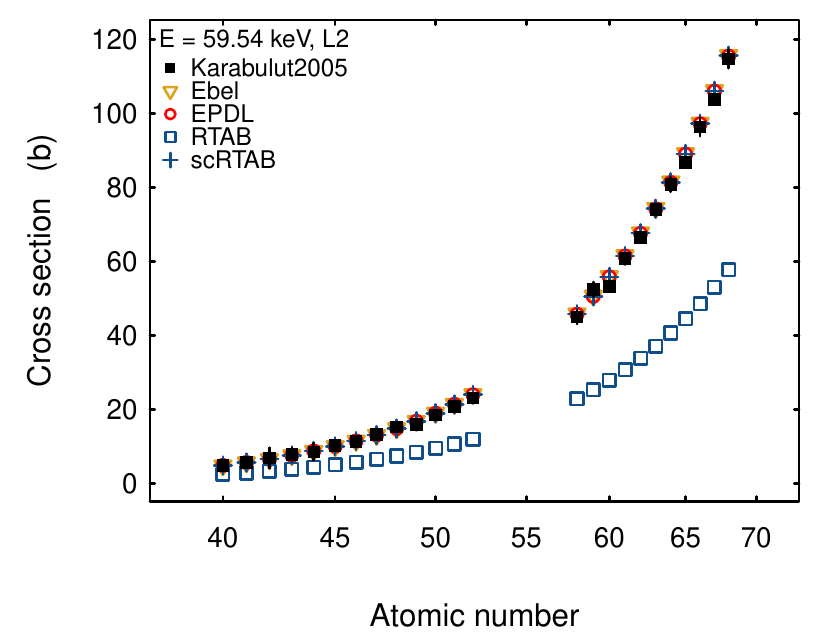}
        \label{fig_L2e59}
    }
    \subfigure[L$_3$ subshell]
    {
        \includegraphics[width=8.5cm]{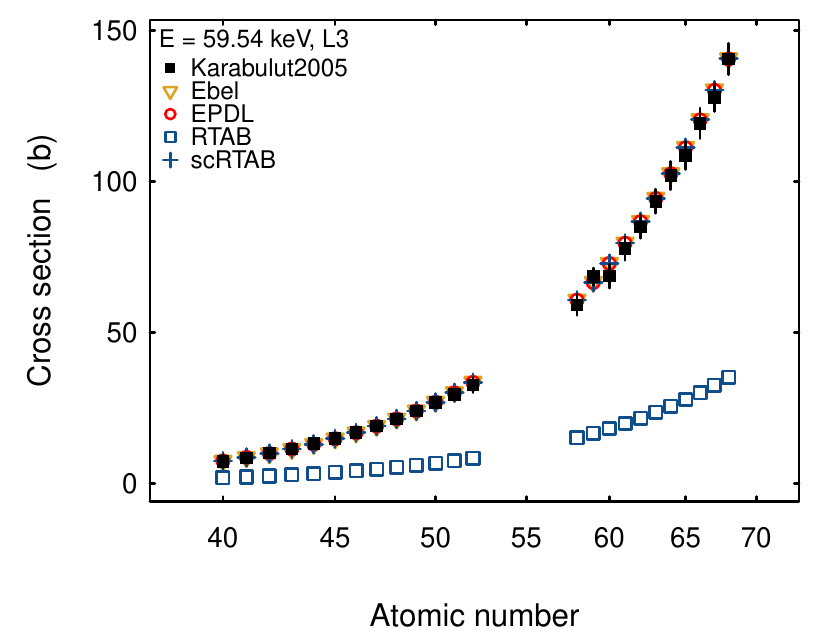}
        \label{fig_L3e59}
    }
    \caption{Cross section for the photoionization of the L shell at 59.54 keV as a function of the atomic number.}
    \label{fig_Le59}
\end{figure}

\begin{figure}
    \centering
    \subfigure[Argon (Z=18)]
    {
        \includegraphics[width=8.5cm]{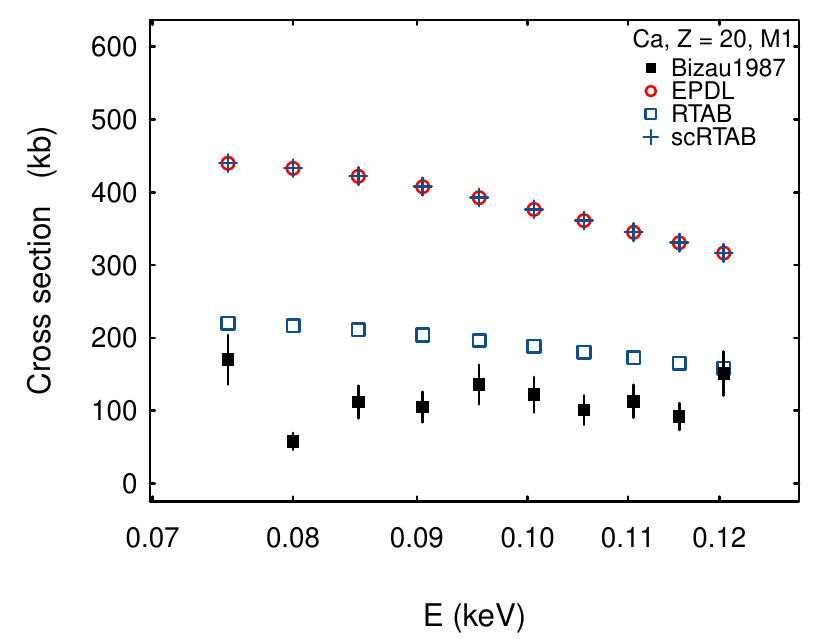}
        \label{fig_M1_18}
    }
    \subfigure[Calcium (Z=20)]
    {
        \includegraphics[width=8.5cm]{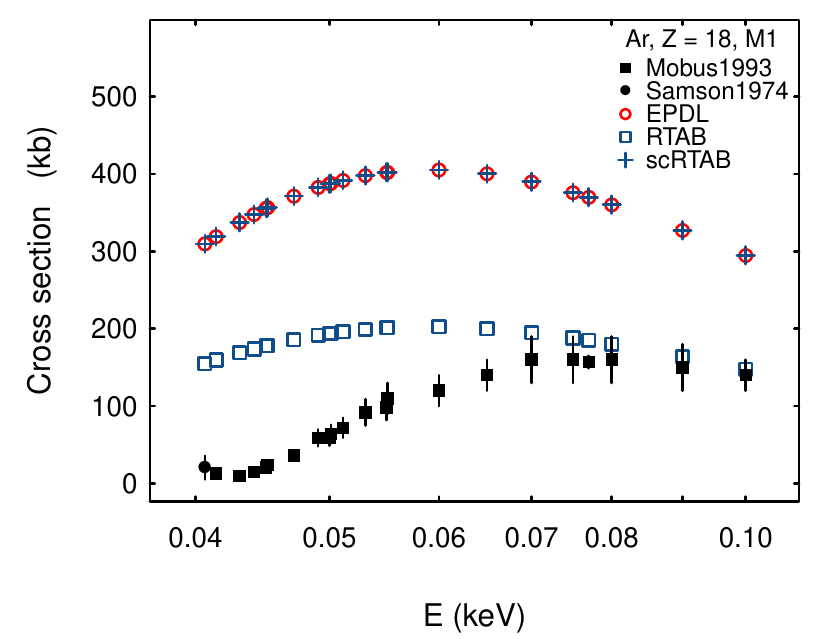}
        \label{fig_M1_20}
    }
      \subfigure[Manganese (Z=25)]
    {
        \includegraphics[width=8.5cm]{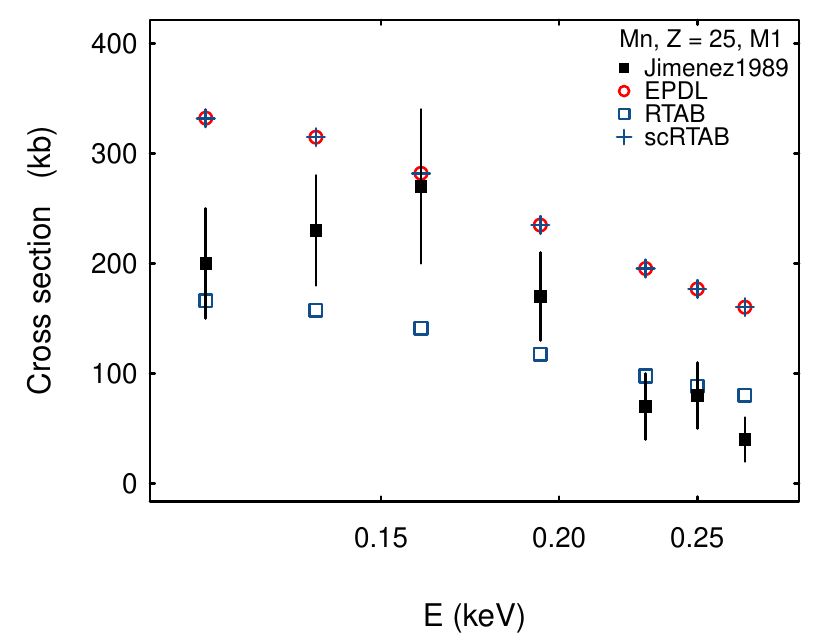}
        \label{fig_M1_25}
    }
  \caption{Cross section for the photoionization of the M$_1$ subshell of argon, calcium and manganese as a function of energy.}
    \label{fig_M1}
\end{figure}

\begin{figure}
\centerline{\includegraphics[angle=0,width=8.5cm]{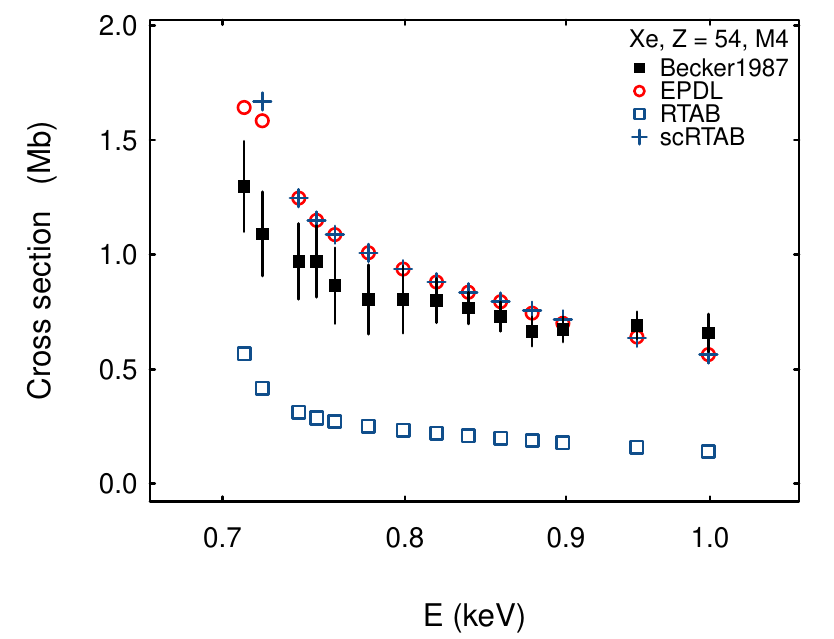}}
\caption{Cross section for the photoionization of the M$_4$ subshell of xenon as a function of photon energy.}
\label{fig_M4_54}
\end{figure}

\begin{figure}
\centerline{\includegraphics[angle=0,width=8.5cm]{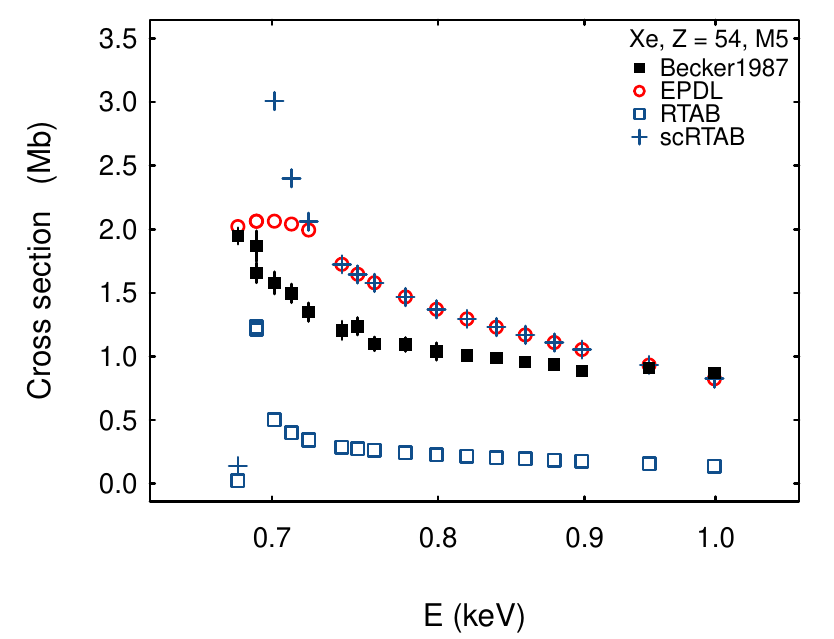}}
\caption{Cross section for the photoionization of the M$_5$ subshell of xenon as a function of photon energy.}
\label{fig_M5_54}
\end{figure}

\begin{figure}
\centerline{\includegraphics[angle=0,width=8.5cm]{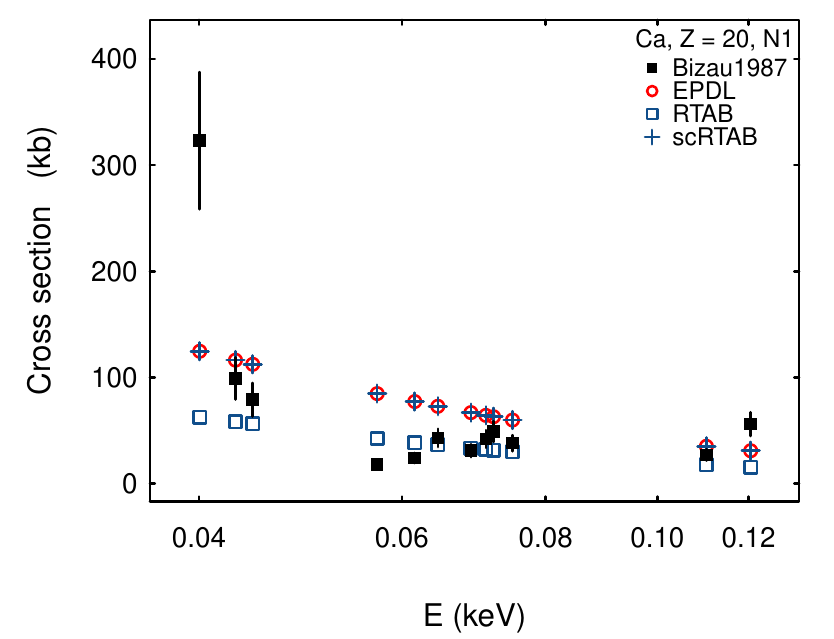}}
\caption{Cross section for the photoionization of the N$_1$ subshell of calcium as a function of photon energy.}
\label{fig_N1_20}
\end{figure}

\begin{figure}
\centerline{\includegraphics[angle=0,width=8.5cm]{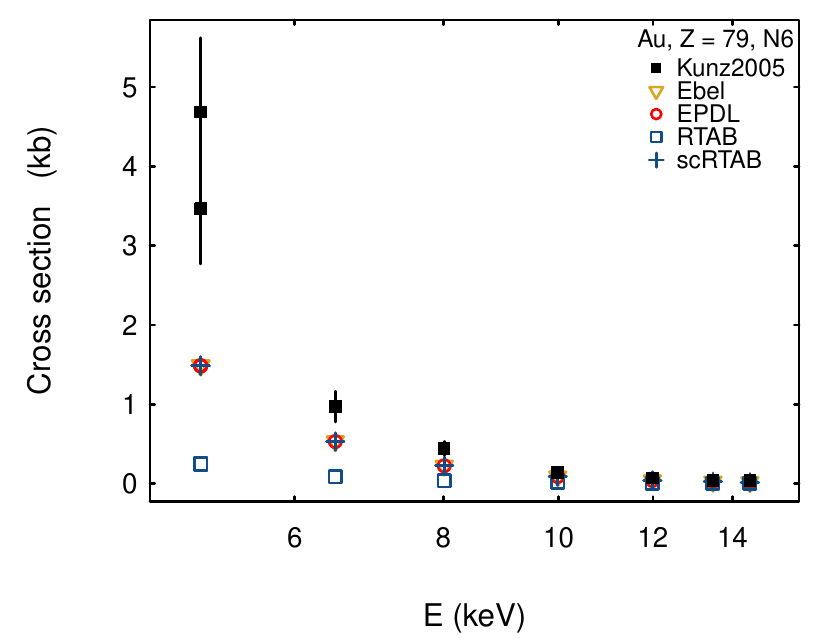}}
\caption{Cross section for the photoionization of the N$_6$ subshell of gold as a function of photon energy.}
\label{fig_N6_79}
\end{figure}

\begin{figure}
\centerline{\includegraphics[angle=0,width=8.5cm]{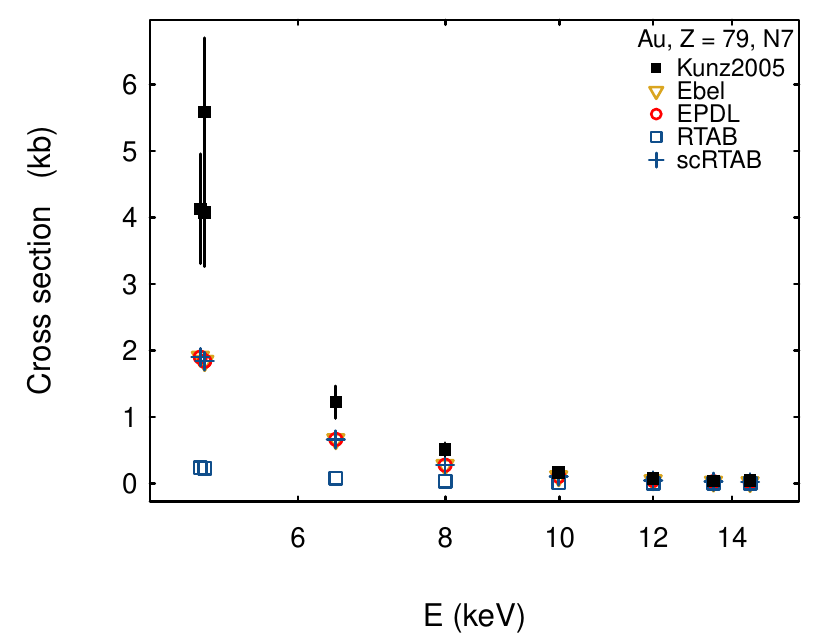}}
\caption{Cross section for the photoionization of the N$_7$ subshell of gold as a function of photon energy.}
\label{fig_N7_79}
\end{figure}

\begin{figure}
\centerline{\includegraphics[angle=0,width=8.5cm]{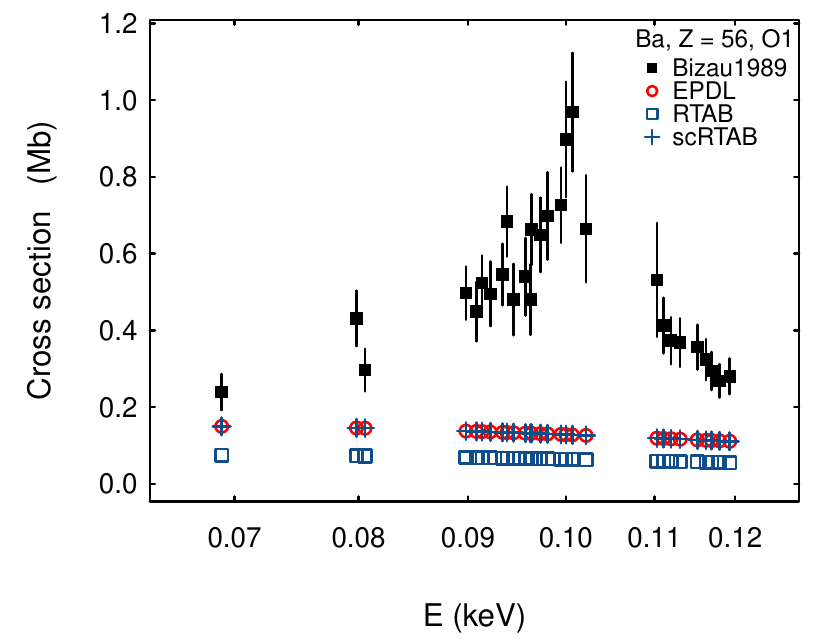}}
\caption{Cross section for the photoionization of the O$_1$ subshell of barium as a function of photon energy.}
\label{fig_O1_56}
\end{figure}

\begin{figure}
\centerline{\includegraphics[angle=0,width=8.5cm]{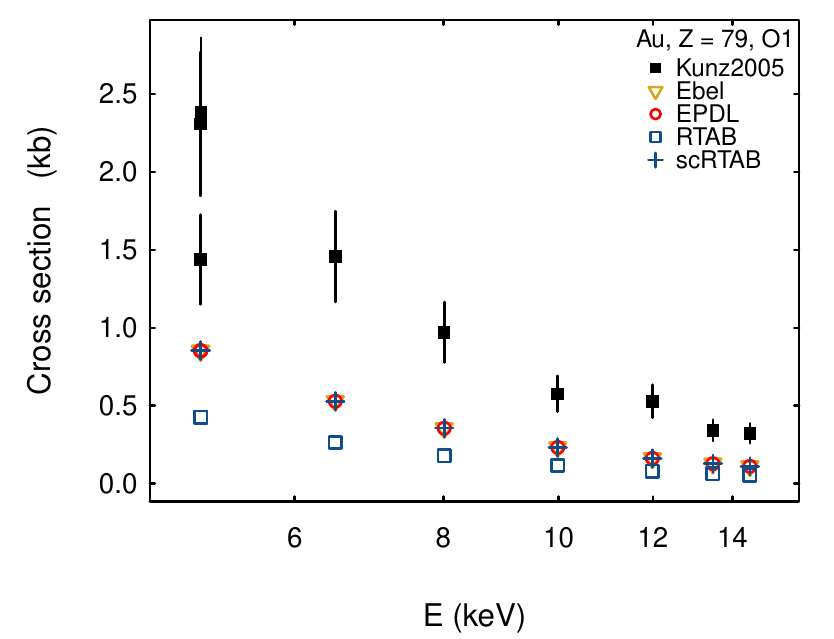}}
\caption{Cross section for the photoionization of the O$_1$ subshell of gold as a function of photon energy.}
\label{fig_O1_79}
\end{figure}

\begin{figure}
\centerline{\includegraphics[angle=0,width=8.5cm]{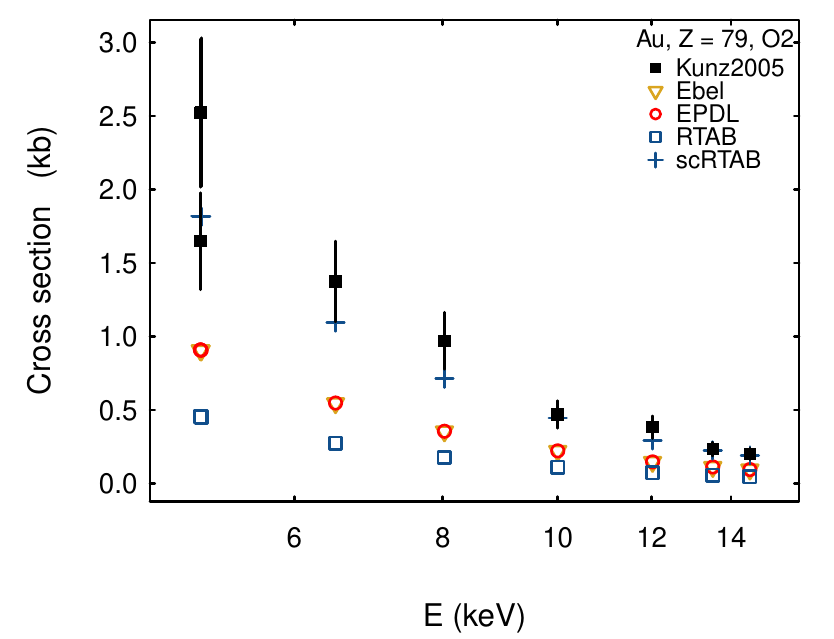}}
\caption{Cross section for the photoionization of the O$_2$ subshell of gold as a function of photon energy.}
\label{fig_O2_79}
\end{figure}

\begin{figure}
\centerline{\includegraphics[angle=0,width=8.5cm]{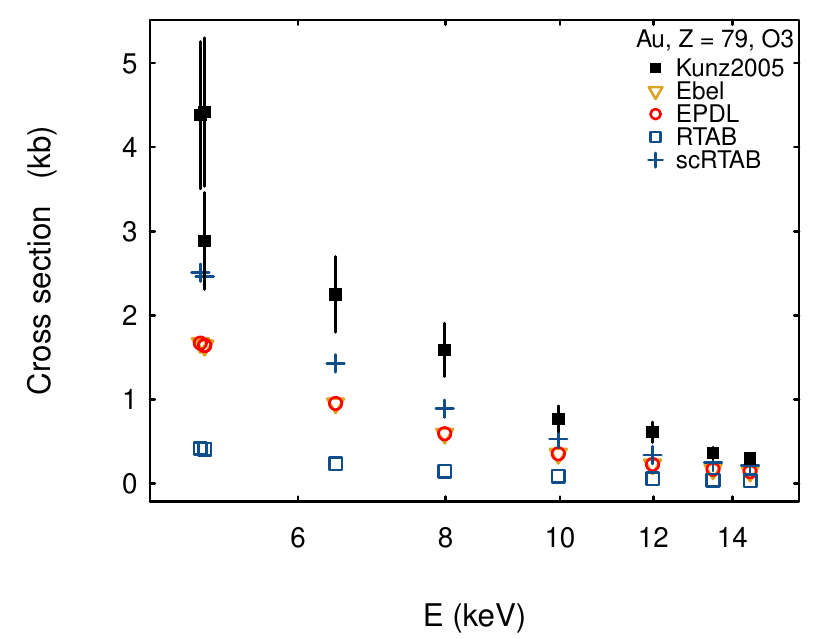}}
\caption{Cross section for the photoionization of the O$_3$ subshell of gold as a function of photon energy.}
\label{fig_O3_79}
\end{figure}

Figs. \ref{fig_K3} to \ref{fig_O3_79} illustrate some examples of calculated and
experimental cross sections for inner and outer shell photoionization.

A systematic discrepancy of RTAB shell cross sections with respect to
experimental data is observed, which hints at a missing multiplicative
factor in the tabulated values.
RTAB cross sections scaled by the presumed missing factor
are identified in figures and in the following tables as ``scRTAB''.

%
%
%


\subsection{K shell}

All but one of the experimental data sets concern energies above 1~keV.

Only the compilations based on Scofield's 1973 non-relativistic calculations
cover the whole energy range of the experimental data sample above 1~keV;
their efficiencies are listed in Table \ref{tab_Kscofield}.
The lower efficiency obtained with PHOTX is related to the 
characteristics of the energy grid over which cross sections are tabulated.

\begin{table}[htbp]
  \centering
  \caption{K shell cross section efficiency of models based on Scofield 1973 non-relativistic calculations}
    \begin{tabular}{lccc}
    \toprule
    Model 		& Pass & Fail   & Efficiency \\
    \midrule
    EPDL  		& 62    & 17    & 0.78 $\pm$ 0.05 \\
    Penelope 	& 62    & 17    & 0.78  $\pm$ 0.05 \\
    PHOTX 		& 52    & 27    & 0.66  $\pm$ 0.05 \\
    Scofield 	& 62    & 17    & 0.78  $\pm$ 0.05 \\
    \bottomrule
    \end{tabular}%
  \label{tab_Kscofield}%
\end{table}%

Chantler's, EPDL and scRTAB models are applicable to the single set of
measurements below 1~keV retrieved from the literature; according to the outcome
of the $\chi^2$ test, their calculations are compatible with the experimental
data set with 0.01 significance.

The capabilities of Chantler's, Ebel's and scRTAB models, which are limited to a 
specific energy range, are evaluated by restricting the $\chi^2$ test of comparison
with experiment to the data sample consistent with their coverage.
The resulting efficiencies, calculated within 1~keV and 300~keV, which correspond to the
coverage of Ebel's model, are reported in Table \ref{tab_Kebel}.
Since the experimental data sample includes no additional measurements 
between 300 keV and 433 keV, which is the highest energy tabulated by
Chantler's calculations, the results in Table \ref{tab_Kebel} express
the outcome of the validation test for this model too.

\begin{table}[htbp]
  \centering
  \caption{Efficiency of K shell cross section models for energies between 1 keV and 300 keV}
    \begin{tabular}{lccc}
    \toprule
    Model & Pass  & Fail  & Efficiency \\
    \midrule
    Chantler 	& 45    & 13    & 0.78  $\pm$ 0.05\\
    Ebel  		& 50    & 8      & 0.86  $\pm$ 0.05\\
    EPDL  		& 45    & 13    & 0.78  $\pm$ 0.05\\
    Penelope 	& 45    & 13    & 0.78  $\pm$ 0.05\\
    PHOTX 		& 47    & 11    & 0.81  $\pm$ 0.05\\
    scRTAB 		& 45    & 13    & 0.78  $\pm$ 0.05\\
    Scofield 	& 45    & 13    & 0.78  $\pm$ 0.05\\
    \bottomrule
    \end{tabular}%
  \label{tab_Kebel}%
\end{table}%

Ebel's parameterization appears more efficient than EPDL tabulations at
reproducing experimental K shell cross sections in the energy range between 1
and 300 keV; nevertheless, the hypothesis of equivalent compatibility with
experiment for these K shell cross section models is not rejected by any of the
tests applied to the analysis of the associated contingency table with 0.01
significance.
The failures of EPDL-based cross sections at reproducing experimental data 
mostly concern test cases close to absorption edges.
Issues about the accuracy of related calculations of atomic binding energies
collected in EADL (Evaluated Atomic Data Library) \cite{eadl} were documented in
\cite{tns_binding}; as discussed in section \ref{sec_exp}, experimental
measurements in these regions are delicate, as realistic estimates of
experimental uncertainties are prone to be affected by systematic effects
deriving from the experimental environment.


\subsection{L shell}

All L subshell measurements included in the experimental data sample 
concern energies between 1~keV and 300~keV, with the exception of one
set of L$_1$ cross section data, which encompasses lower energy measurements.

Efficiencies above 1~keV are reported in Table \ref{tab_Leff} for all models
dealing with L subshells.
All cross section models are equally capable of reproducing the measurements in the
experimental data sample.

\begin{table}[htbp]
  \centering
  \caption{Efficiency of L shell cross section models above 1 keV}
    \begin{tabular}{clccc}
    \toprule
    Subshell & Model & Pass  & Fail  & Efficiency \\
    \midrule
    \multirow{6}[0]{*}{L$_1$} 	& Ebel  		& 24    & 0     & 1.00 - 0.04 \\
          				& EPDL  		& 24    & 0     & 1.00 - 0.04  \\
          				& Penelope 	& 24    & 0     & 1.00 - 0.04  \\
          				& PHOTX 		& 24    & 0     & 1.00 - 0.04  \\
          				& scRTAB 		& 24    & 0     & 1.00 - 0.04  \\
          				& Scofield 		& 24    & 0     & 1.00 - 0.04  \\
\midrule
    \multirow{6}[0]{*}{L$_2$} 	& Ebel  		& 23    & 1     & 0.96 $\pm$ 0.04 \\
          				& EPDL  		& 23    & 1     & 0.96 $\pm$ 0.04 \\
          				& Penelope 	& 23    & 1     & 0.96 $\pm$ 0.04 \\
         				& PHOTX 		& 23    & 1     & 0.96 $\pm$ 0.04 \\
          				& scRTAB 		& 23    & 1     & 0.96 $\pm$ 0.04 \\
         			 	& Scofield 		& 23    & 1     & 0.96 $\pm$ 0.04 \\
\midrule
    \multirow{6}[0]{*}{L$_3$} 	& Ebel  		& 27    & 0     & 1.00 - 0.03 \\
          				& EPDL  		& 27    & 0     & 1.00 - 0.03 \\
         				& Penelope 	& 27    & 0     & 1.00 - 0.03 \\
         				& PHOTX 		& 27    & 0     & 1.00 - 0.03 \\
          				& scRTAB 		& 27    & 0     & 1.00 - 0.03 \\
          				& Scofield 		& 27    & 0     & 1.00 - 0.03 \\
    \bottomrule
    \end{tabular}%
  \label{tab_Leff}%
\end{table}%

The hypothesis of compatibility with the single experimental data set including
measurements below 1~keV is rejected for all models subject to evaluation.
Nevertheless, no conclusions regarding the capabilities of the cross section
models at lower energies can be inferred from such a limited test; the failure
over a single experimental data set could be due either to deficiency of the
models or to systematic effects affecting the measurements, which cannot be
investigated in terms of consistency with other, independent experimental
measurements.

\subsection{Outer shells}

The cross sections for outer shells appear incompatible with experiment;
the $\chi^2$ test rejects the hypothesis of compatibility between 
calculated and experimental cross sections in most test cases.
Nevertheless, caution should be exercised in drawing general conclusions 
from these results, as 
 the experimental data sample
for the validation of outer shells is small and for a given 
test case the available data often originate from a single experimental source.
This test scenario is prone to be affected by systematic effects,
which would hinder the reliability of the outcome of the statistical analysis.

Firm conclusions about the accuracy of cross section calculations for
outer shells would require more extensive experimental data samples.



\section{Qualitative Evaluation of Photoelectron Angular Distribution}
\label{sec_angular}

A preliminary appraisal of photoelectron angular distribution models
has been performed, limited to the different options implemented in Geant4 and 
GEANT 3.
Inconsistencies in the analytical formula of the latter appearing in the software 
documentation have been corrected in the implementation used for this
evaluation.

Reports of direct measurements of photoelectron angular distribution are
scarce in the literature; the narrow scope of the experimental data
sample we could retrieve from the literature and the lack of adequate detail on
the characteristics of the reported measurements prevent a meaningful
statistical study for the validation of the models used in general purpose Monte
Carlo codes.

\begin{figure}
\centerline{\includegraphics[angle=0,width=8.5cm]{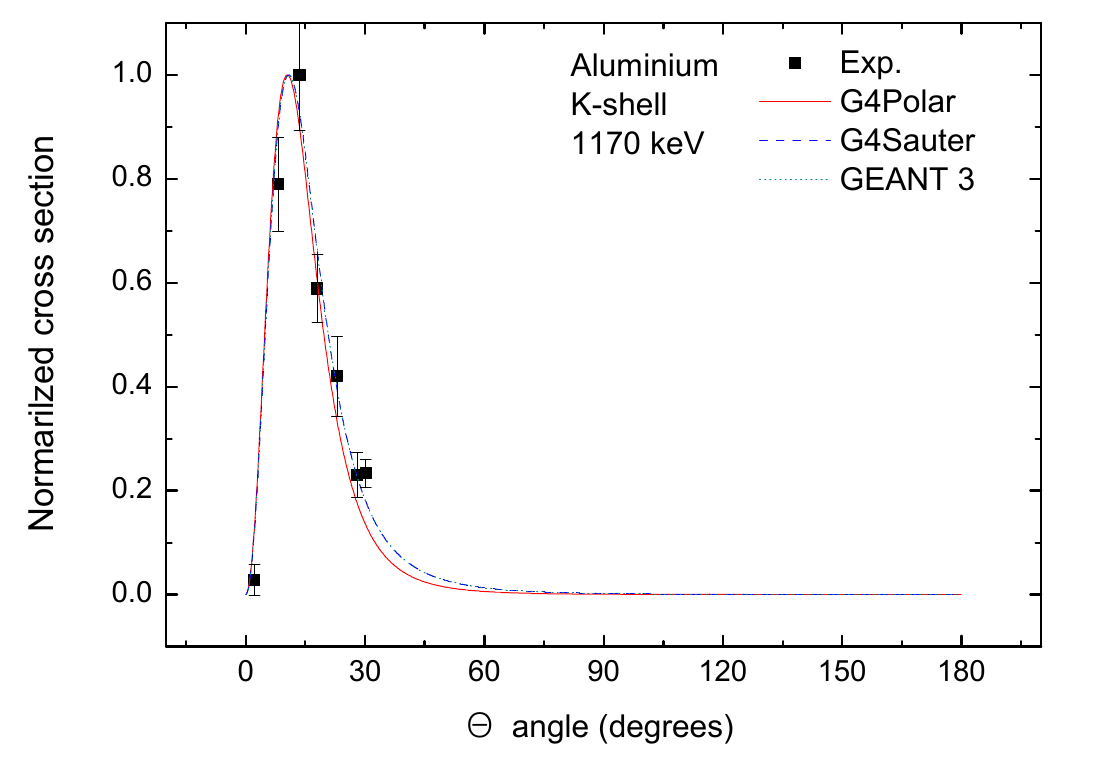}}
\caption{Photoelectron angular distribution for aluminium, K shell, at 1.17 MeV.}
\label{fig_angK}
\end{figure}

\begin{figure}
\centerline{\includegraphics[angle=0,width=8.5cm]{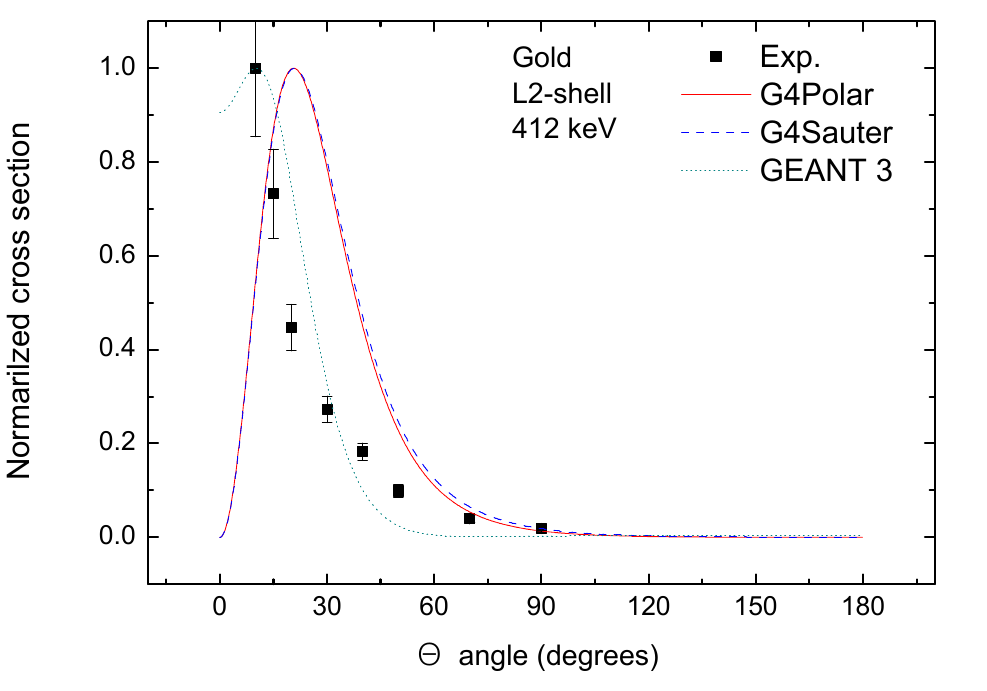}}
\caption{Photoelectron angular distribution for gold, L$_2$ subshell, at 412 keV.}
\label{fig_angL2}
\end{figure}

Figs. \ref{fig_angK} and \ref{fig_angL2} show qualitative comparisons of
photoelectron angular distribution models implemented in Geant4 and GEANT 3 with
experimental data; discrepancies are visible, both between calculations and
experimental data and between the outcome of different model implementations,
which suggest that further quantitative investigation would be useful.
In the scenarios corresponding to the limited experimental sample retrieved from
the literature, Geant4 angular distribution models exhibit similar behaviour;
the corrected GEANT 3 model appears in some cases different from the others and
qualitatively closer to measurements.
Nevertheless, no conclusions can be drawn from such a narrow, qualitative test.

An indirect investigation of photoelectric differential cross section models
could be pursued through testing the so-called ``asymmetry parameter''
\cite{starace_1983}; due to its complexity, this study should be considered as
the subject of a dedicated paper.

%


\section{Conclusion}

An extensive set of cross section models for the simulation of photoionization
has been quantitatively evaluated regarding their accuracy at reproducing
experimental measurements.
Since the validation analysis documented in this paper concerns
cross section data libraries and parameterization methods, which are (and can
be) used by any Monte Carlo codes, rather than specific implementations of
the photoelectric effect, its results, and the recommendations
which derive from them, are of general relevance for the simulation of the
photoelectrict effect.

Statistical tests against a large sample of total and partial cross section
measurements demonstrate that tabulations based on Scofield's 1973
non-relativistic calculations, which cover an extended range of energies
starting from a few electronvolts, achieve the best overall capability to
reproduce experiment among the models subject to test.
Small differences in the results of goodness-of-fit tests are observed across
different compilations (EPDL, Penelope database, PHOTX, XCOM) based on the same
theoretical calculation method; they appear related to the granularity of the
tabulations.

Some discrepancies with experiment of EPDL-based K~shell cross sections are
observed close to absorption edges, which could be related to intrinsic
limitations of the calculations, also highlighted in the evaluation of related
calculations of atomic binding energies; nevertheless, the possibility of
systematic effects in these delicate measurements suggests caution in drawing
conclusions about the reliability of EPDL in the proximity of absorption edges.

No other photoelectric cross section calculation methods, among those considered
in this study, demonstrate statistically better accuracy than EPDL tabulations:
neither theoretical calculations, including relativistic ones, nor empirical or
theory-driven analytical parameterizations.
The validation tests documented in this paper demonstrate that the compatibility
with experiment of some of them is significantly worse than that of EPDL, 
while it is statistically equivalent for others,
although restricted to the energy range they cover.
Only K-shell cross sections based on Ebel's parameterization produce more accurate
results than EPDL in some test cases close to absorption edges, although the difference
in compatibility with experiment with respect to EPDL is not statistically significant.

Special attention has been devoted to the evaluation of total cross sections
based on Biggs and Lighthill's parameterization, which is extensively used in
Geant4.
While the original set of parameterization coefficients produces cross sections
that are statistically equivalent to EPDL regarding compatibility with
experiment, the modified coefficients implemented in Geant4 significantly
worsen the resulting cross section compatibility with experimental
measurements above a few tens of electronvolt.
The original and modified parameterization produce statistically equivalent
compatibility with experiment below a few tens of electronvolt; both 
parameterizations, as well as other total cross section models, exhibit low
efficiency in that energy range.
As a result of this validation test, it is recommended that future versions of
Geant4 revert to using the original coefficients of Biggs and Lighthill
parameterization above a few tenso of electronvolt, unless a new, more accurate
parameterization is developed and quantitatively demonstrated to improve
compatibility with experiment over the original one.
The suggestion to implement this modification in Geant4 version 10.2, which
is in preparation at the time of writing this paper, has been conveyed to the 
maintainers of the related Geant4 code.
Given that EPDL photoelectric cross sections are already used in other Geant4
models \cite{lowe_e}, the burden of maintaining alternative cross sections
based on Biggs-Lighthill's parameterization should be considered as well, as this
validation test has demonstrated that even the original parameterization is
not statistically superior to EPDL  in reproducing experimental data.

The validation tests show that in the low energy end cross sections based on
Scofield's non-relativistic calculations produce results consistent with
experiment with 0.01 significance down to approximately 150~eV.
Nevertheless, the limited representativity of the experimental data sample,
which only includes noble gases and hydrogen, suggests caution at extrapolating
this result to experimental scenarios involving solid targets.

No model, among those considered in this validation study, is able to 
reproduce experimental data consistently below approximately 150~eV.
Their failure at low energies reflects the inadequacy of theoretical
calculations in independent particle approximation and of the assumption of
isolated atoms adopted in general purpose Monte Carlo codes, in conditions where
effects related to the structure of the interacting material should be taken into
account.

The quantitative results of the validation tests documented in this paper allow
Monte Carlo code developers and experimental users to base the selection of
physics models in their simulations on sound scientific grounds.

%
%
%

\section*{Acknowledgment}
The authors are grateful to Lynn Kissel for providing a copy of the RTAB database.
The authors thank Sergio Bertolucci for support, Vladimir Grichine and 
Georg Weidenspointner for valuable discussions, and
Anita Hollier for proofreading the manuscript.
The CERN Library has provided helpful assistance and essential reference 
material for this study.


\end{document}